\documentclass[longauth]{aa} 

\usepackage{graphicx}
\usepackage[varg]{txfonts}
\usepackage[T1]{fontenc}
\usepackage{lscape}
\usepackage{subfig}
\usepackage{amssymb,natbib,supertabular}
\usepackage{defs}
\usepackage{appendix}
\usepackage{color}
\usepackage{longtable}
\usepackage{float}
\usepackage[breaklinks, colorlinks, citecolor=blue]{hyperref}

\newcommand{\jwst}{{\rm JWST}}
\newcommand{\0}{\phantom{0}}
\begin{document}

   \title{JWST's PEARLS: Resolved study of the stellar and dust components in starburst galaxies at cosmic noon}

   \titlerunning{ \jwst\ starburst galaxies}
  \authorrunning{Polletta et al.}
  
   \author{M. Polletta\inst{1}
          }

   \institute{INAF - Istituto di Astrofisica Spaziale e Fisica cosmica (IASF), via A. Corti 12, 20133 Milan, Italy\\
             }
\author{M. Polletta\inst{\ref{inst1}} 
\and B.~L.~Frye\inst{\ref{inst2}} 
\and N. Garuda\inst{\ref{inst2}}
\and S.~P. Willner\inst{\ref{inst3}}
\and S. Berta\inst{\ref{inst4}}
\and R. Kneissl\inst{\ref{inst5},\ref{inst5a}}
\and H. Dole\inst{\ref{inst6}}
\and R.~A. Jansen\inst{\ref{inst7}}
\and M.~D. Lehnert\inst{\ref{inst8}}
\and S.~H. Cohen\inst{\ref{inst7}}
\and J. Summers\inst{\ref{inst7}}
\and R.~A. Windhorst\inst{\ref{inst7}}
\and J.~C.~J. D'Silva\inst{\ref{inst9},\ref{inst10}}
\and A.~M. Koekemoer\inst{\ref{inst11}}
\and D. Coe\inst{\ref{inst11},\ref{inst12},\ref{inst13}}
\and C.~J. Conselice\inst{\ref{inst14}}
\and S.~P. Driver\inst{\ref{inst9}}
\and N.~A. Grogin\inst{\ref{inst11}}
\and M.~A. Marshall\inst{\ref{inst10},\ref{inst15}}
\and M. Nonino\inst{\ref{inst16}}
\and R. Ortiz~III\inst{\ref{inst7}}
\and N. Pirzkal\inst{\ref{inst11}}
\and A. Robotham\inst{\ref{inst9}}
\and R.~E. Ryan, Jr.\inst{\ref{inst11}}
\and C.~N.~A. Willmer\inst{\ref{inst2}}
\and H. Yan\inst{\ref{inst17}}
\and V. Arumugam\inst{\ref{inst4}}
\and C. Cheng\inst{\ref{inst18},\ref{inst19}}
\and H.~B. Gim\inst{\ref{inst20}}
\and N.~P. Hathi\inst{\ref{inst11}}
\and B. Holwerda\inst{\ref{inst21}}
\and P. Kamieneski\inst{\ref{inst7}}
\and W.~C. Keel\inst{\ref{inst22}}
\and J. Li\inst{\ref{inst9}}
\and M. Pascale\inst{\ref{inst23}}
\and H. Rottgering\inst{\ref{inst24}}
\and B.~M. Smith\inst{\ref{inst7}}
\and M.~S. Yun\inst{\ref{inst25}}
}
  \offprints{M. Polletta\\ \email{maria.polletta@inaf.it}}

\institute{INAF – Istituto di Astrofisica Spaziale e Fisica Cosmica Milano,  Via A. Corti 12, I-20133 Milano, Italy\label{inst1}
\and Steward Observatory, University of Arizona, 933 N Cherry Ave, Tucson, AZ, 85721-0009, USA\label{inst2}
\and Center for Astrophysics \textbar\ Harvard \& Smithsonian, 60 Garden Street, Cambridge, MA 02138, USA\label{inst3}
\and Institut de Radioastronomie Millimétrique, 300 rue de la piscine, F-38406 Saint-Martin-d'Hères, France\label{inst4}
\and European Southern Observatory, ESO Vitacura, Alonso de C\'ordova 3107, Vitacura, 19001, Casilla, Santiago, Chile\label{inst5}
\and Joint ALMA Observatory, Alonso de C\'ordova 3107, Vitacura 763-0355, Santiago, Chile\label{inst5a}
\and Universit\'e Paris-Saclay, CNRS, Institut d'Astrophysique Spatiale, 91405, Orsay, France\label{inst6}
\and School of Earth and Space Exploration, Arizona State University, Tempe, AZ 85287-1404, USA\label{inst7}
\and Univ Lyon, Univ Lyon1, Ens de Lyon, CNRS, Centre de Recherche Astrophysique de Lyon UMR5574, F-69230 Saint-Genis-Laval, France\label{inst8}
\and International Centre for Radio Astronomy Research (ICRAR) and the International Space Centre (ISC), The University of Western Australia, M468, 35 Stirling Highway, Crawley, WA 6009, Australia\label{inst9}
\and ARC Centre of Excellence for All Sky Astrophysics in 3 Dimensions (ASTRO 3D), Australia\label{inst10}
\and Space Telescope Science Institute, 3700 San Martin Drive, Baltimore, MD 21218, USA\label{inst11}
\and Association of Universities for Research in Astronomy (AURA) for the European Space Agency (ESA), STScI, Baltimore, MD 21218, USA\label{inst12}
\and Center for Astrophysical Sciences, Department of Physics and Astronomy, The Johns Hopkins University, 3400 N Charles St. Baltimore, MD 21218, USA\label{inst13}
\and Jodrell Bank Centre for Astrophysics, Alan Turing Building, University of Manchester, Oxford Road, Manchester M13 9PL, UK\label{inst14}
\and National Research Council of Canada, Herzberg Astronomy \& Astrophysics Research Centre, 5071 West Saanich Road, Victoria, BC V9E 2E7, Canada\label{inst15}
\and INAF-Osservatorio Astronomico di Trieste, Via Bazzoni 2, 34124 Trieste, Italy\label{inst16}
\and Department of Physics and Astronomy, University of Missouri, Columbia, MO 65211, USA\label{inst17}
\and Chinese Academy of Sciences South America Center for Astronomy, National Astronomical Observatories, CAS, Beijing 100101, China\label{inst18}
\and CAS Key Laboratory of Optical Astronomy, National Astronomical Observatories, Chinese Academy of Sciences, Beijing 100101, China\label{inst19} 
\and Department of Physics, Montana State University, Bozeman, MT 59717, USA\label{inst20}
\and Department of Physics and Astronomy, University of Louisville, 102 Natural Science Building, Louisville KY 40292, USA\label{inst21}
\and Dept. of Physics and Astronomy, University of Alabama, Box 870324, Tuscaloosa, AL 35404, USA\label{inst22}
\and Department of Astronomy, University of California, 501 Campbell Hall \#3411, Berkeley, CA 94720, USA\label{inst23}
\and Leiden Observatory, PO Box 9513, 2300 RA Leiden, The Netherlands\label{inst24}
\and Department of Astronomy, University of Massachusetts at Amherst, Amherst, MA 01003, USA\label{inst25}
}

   \date{Received 09 May 2024/ Accepted 29 June 2024}

\abstract{Dusty star-forming galaxies (DSFGs) significantly contribute to the stellar buildup in galaxies during ``cosmic noon,'' the peak epoch of cosmic star formation. Major mergers and gas accretion are often invoked to explain DSFGs' prodigious star formation rates (SFRs) and large stellar masses.  We conducted a spatially resolved morphological analysis of the rest-frame ultraviolet/near-infrared ($\sim$0.25--1.3$\mu$m) emission in three DSFGs at $z\simeq2.5$. Initially discovered as carbon monoxide (CO) emitters by NOrthern Extended Millimeter Array (NOEMA) observations of a bright ($S_{\rm 350\mu m}=111\pm10$\,mJy) \herschel\ source, we observed them with the {\it James Webb} Space Telescope/NIRCam as part of the PEARLS program.  The NIRCam data reveal the galaxies' stellar populations and dust distributions on scales of 250\,pc.  Spatial variations in stellar mass, SFR, and dust extinction are determined in resolved maps obtained through pixel-based spectral energy distribution fitting.  The CO emitters are massive ($M_{\rm star}\simeq(3-30)\times10^{10}$~\msun), dusty starburst galaxies with SFRs ranging from 340 to 2500\,\msun\,yr$^{-1}$, positioning them among the most active star-forming galaxies at $2<z<3$.  Notably, they belong to the $\sim$1.5\%\ of the entire \jwst\ population with extremely red colors. Their morphologies are disk like (S\'ersic index $n\simeq1$), with effective radii of 2.0--4.4\,kpc, and exhibit substructures such as clumps and spiral arms.  The galaxies have dust extinctions up to \av=5--7 mag extending over several kiloparsecs with asymmetric distributions that include  off-center regions resembling bent spiral arms and clumps. The near-infrared dust-attenuation curve in these sources deviates from standard laws, possibly implying different dust--star geometries or dust grain properties than commonly assumed in starburst galaxies. The proximity ($<$5\arcsec) of galaxies with consistent redshifts, strong color gradients, an overall disturbed appearance, asymmetric dust obscuration, and widespread star formation collectively favor interactions (minor mergers and flybys) as the mechanism driving the CO galaxies' exceptional SFRs. The galaxies' large masses and rich environment hint at membership in two proto-structures, as initially inferred from their association with a \planck-selected high-$z$ source.
}



   \maketitle
%

\section{Introduction}\label{sec:intro}
The star formation rate (SFR) density of the Universe reached its maximum about 10\,Gyr ago or at $2\la z\la3$ \citep[also called "Cosmic Noon;"][]{madau14}. This epoch marks the theoretical transition between different growth regimes for galaxies, the peak of supermassive black holes' (SMBHs') accretion rate \citep{delvecchio14}, and the collapse of cosmic structures \citep{dekel09}.  At $z\,{\gtrsim}\,$2 galaxies go through a rapid growth phase during which stars are formed ``in situ'' within the galaxy from infalling cold gas. This early growth is followed by a slower phase during which the stars that formed ``ex situ'' are accreted through merger events \citep[e.g.,][]{oser10}.  In galaxy protoclusters, environment-related quenching mechanisms are likely efficient by $z\sim2$ producing substantial quenched galaxy populations and forming the first true galaxy clusters \citep{alberts22}.
 
The main contributors ($\sim$85\%) to the cosmic SFR density at $z\,{\sim}\,$2--3 are dusty galaxies with extreme SFRs---hundreds to thousands of solar masses per year \citep{dunlop17}. These galaxies are usually identified as bright submillimeter (submm) sources and are frequently referred to as dusty star-forming galaxies (DSFGs) or submm galaxies (SMGs). SMGs were discovered in the late 1990s as bright ($>$1--10\,mJy) sources at  450 and 850\,$\mu$m \citep[][]{blain93,clements93,smail97}. They are usually massive galaxies ($M_{\rm star}\sim 10^{11}$\,\msun) with large molecular gas reservoirs \citep{frayer98,greve05,bothwell13}. 

In spite of the importance of DSFGs and decades of study, it is still unclear which mechanisms power their intense SFRs and how they acquire the gas to fuel these high rates.  There is a substantial controversy in interpreting the observations. On the one hand, there is evidence for rapid growth through major merger-driven events, where interactions with similarly massive and gas-rich galaxies can trigger a short-lived and intense burst of star formation  \citep{engel10,casey16,ginolfi20,gomez_guijarro18}. On the other hand, DSFGs are often large, gas-rich disk galaxies, implying secular growth through gas inflows and disk instabilities \citep[e.g.,][]{hodge12,gullberg19,hodge19,rizzo20,jimenez_andrade20,amvrosiadis23,liu24,gillman24}. 
In this case, the galaxy interstellar medium (ISM) is continuously fed and star formation is sustained  for a long time (1--10\,Gyr) by gas accretion from the circumgalactic medium (CGM), the intergalactic medium (IGM), or other galaxies through flybys or minor mergers \citep{davis16,camps_farina23,narayanan15}.  Current cosmological simulations (EAGLE, \citealt{crain15}; Illustris-TNG, \citealt{nelson18,pillepich18}; SIMBA, \citealt{dave19}; Romulus, \citealt{tremmel17}; Auriga \citealt{dave19}) are tuned to reproduce the evolutionary history of stellar masses and galaxy sizes as well as galaxy scaling-relations and their scatter \citep[see ][for a review]{crain23}. Even though independent simulations produce similarly realistic galaxies, they are often based on different prescriptions, in particular regarding the
mechanisms regulating star formation, such as supernovae or feedback from an active galactic nucleus (AGN)\null. These different prescriptions result in different gas accretion rates onto the galaxy and feedback-driven outflow rates in the inner regions of massive galaxies' haloes ($<$200\,kpc) \citep{crain23}. The study of the gas and star formation properties of massive star-forming galaxies (SFGs) is thus fundamental to providing observational constraints needed to refine theoretical models of galaxy evolution \citep{baugh05,dave10,hayward13,lagos20}.

Determining the triggering and fueling mechanism of star formation in DSFGs has important implications on our understanding of this important galaxy population as well as of the growth of massive galaxies.  
Because environmental-related processes, such as interactions and gas feeding through cold streams \citep{dekel09}, may affect the star formation triggering and fueling mechanisms, it is also important to consider the environment in which a galaxy is evolving when studying its star formation. 
DSFGs are considered to be progenitors of  the population of massive galaxies that is predominant in local galaxy clusters  \citep{lilly99,chapman05,simpson14} and signposts of matter density peaks and galaxy protoclusters \citep{oteo18,miller18,polletta21,calvi23}.
The study of their star formation and gas properties in connection with the environment is thus necessary to assess the role environmental density plays in their growth and evolution. An overdense environment at high redshift is characterized by a higher merger probability and a larger gas reservoir than the field or an overdensity at low redshifts.  These conditions might play a significant and perhaps decisive role in both driving and establishing the timing of the extreme SFRs in massive galaxies by triggering star formation and regulating the gas accretion rates onto galaxies. 

The various driving mechanisms are expected to leave different imprints on the molecular gas properties, the gas kinematics, the star formation efficiency (SFE), the gas-depletion timescale, and in the stellar morphologies \citep{hopkins13,boogaard24}. 
In order to establish whether these intense star formers grow through major mergers or through gas accretion, we thus need to map their stellar, dust, and cold gas components. The maps require sufficient spatial resolution to reveal the components' relative distribution. Resolved (with $<$0\farcs1 angular resolution) studies of these sources, at both near-infrared (NIR), where the effects of dust obscuration are reduced, and at millimeter (mm) wavelengths, are necessary to measure the spatial extent of the stellar and dust components, establish the formation of disks and bulges, identify mergers, and trace gas fueling and ejection processes \citep{dekel09,krumholz18}. At low resolution, minor mergers can mimic smooth rotating disks \citep[e.g., ][]{rizzo22}, and deviations from circular motion due to gas inflow or turbulent motions throughout the disk can be misinterpreted as signs of merging. Heavy obscuration in the galaxy center, if not accounted for, can create artificial peaks of stellar mass erroneously leading to a merger scenario \citep{sun24}. On the other hand, identifying a major merger depends on the spatial resolution and the source separation, and the latter depends on the merger stage. 

At mm wavelengths, facilities such as the NOrthern Extended Millimeter Array (NOEMA) or the Atacama Large Millimeter Array (ALMA) offer the sensitivity, resolution, and frequency coverage to map the molecular gas and the dust continuum, and constrain the gas mass and kinematics with sufficient resolution to test these different scenarios. These observations have also enabled the identification and study of their visible and NIR counterparts and the possibility to carry out statistical multiwavelength studies of large samples \citep[e.g.,][]{danielson17,aravena19,as2uds20,donevski20,birkin21,berta23,liao24}.

The advent of the {\it James Webb} Space Telescope \citep[\jwst;][]{Gardner2006,Rieke2005, Beichman2012, Windhorst2008}, with its unprecedented sensitivity and angular resolution at NIR wavelengths, offers the opportunity to explore the rest-frame visible emission of even the dustiest galaxies at $z\sim2$. It is now possible to study their internal structure on  scales of 10-–100\,pc or even less in lensed systems \citep[e.g.,][]{smail23,kamieneski23,kamieneski24a,liu24} and estimate  their physical properties such as stellar and dust mass, SFR, and gas fraction. The  \jwst\ spatial resolution permits identifying signs of interaction and mapping their stellar structure and morphology and the close environment. Since its launch, several studies have been carried out with \jwst\ on DSFGs, often combined with high resolution mm data.  The sub-arcsec resolution \jwst\ images of DSFGs at $1.4<z<10$ show a diversity of structures and sizes and a wide range of morphologies \citep{gillman23,gillman24}. Some DSFGs are gas-rich isolated massive disks \citep{huang23,liu24}, others show disturbances such as arms, bars, tidal features \citep{smail23,wu23,chen22,huang23,sugahara24}. The majority exhibit a compact stellar mass distribution \citep{chen22,smail23} with a core mass fraction that increases at lower redshift possibly due to the growth of a bulge \citep{lebail23}.  Star formation activity is often spread over galaxy-scales \citep{kamieneski24b,crespo_gomez24}, but quiescent regions are sometimes present in the core or in the disk \citep{lebail23,kamieneski24b,cheng23}. Many live in an overdense environments (groups or protoclusters) \citep{smail23,wu23,kokorev23,frye24}.  In general they are heavily obscured \citep{smail23,alvarez_marquez23}, with dust obscuration more concentrated in the center \citep{price23,sun24}, but not exclusively \citep{kamieneski24b}. The emerging picture is consistent with inside-out galaxy evolution, with the growth with time of a centrally concentrated older stellar populations over more extended, younger and obscured star-forming regions that dominate the mass budget \citep{chen22}. Overall, these results suggest that star formation in DSFGs at high-$z$ is driven by gas accretion, rather than by major mergers \citep[but see][]{sugahara24}. All these studies are revolutionizing our understanding of this emblematic galaxy population, but evidence in favor of the accretion scenario is still sparse and the amount of information to be delivered to modelers needs to be parsed and standardized. Last, these studies cover a broad range of redshifts and luminosities and the environment is rarely well characterized. Further progress is clearly needed on the study of DSFGs at high resolution. The premises are very promising, facilities such as ALMA, NOEMA and \jwst\ hold the keys to unveil the mechanisms at the origin of the vigorous growth of these extreme star formers and to finally untangle such a fascinating and enigmatic population.  

This paper presents  \jwst/Near-Infrared Camera \citep[NIRCam;][]{Rieke2003} and NOEMA observations of three DSFGs at $z\sim2.5$ with the goal of establishing the mechanisms that regulate their star formation activity.  The target selection is summarized in Sect.~\ref{sec:targets}. Sect.~\ref{sec:obs} describes NOEMA, LOFAR, and \jwst/NIRCam observations and data reduction.  Sect.~\ref{sec:cigale} presents the integrated and the resolved spectral energy distributions (SEDs) and derived properties, including stellar masses, dust extinctions, and SFRs\null.  The DSFGs' morphological properties are analyzed in Sect.~\ref{sec:morph}. Sect.~\ref{sec:SF_mode} presents their star formation  and molecular gas properties and compares them with samples from the literature and with scaling relations. Sect.~\ref{sec:dust_properties} discusses dust properties and presents near-IR (NIR) attenuation curves. Sect.~\ref{sec:comparison} compares our sources with other red  \jwst\ sources from the literature. Sect.~\ref{sec:discussion} discusses the mechanisms that might be at the origin of the DSFGs' prodigious star formation activity and the link with their environment, and Sect.~\ref{sec:summary} summarizes our conclusions.

Throughout this work we adopt a \citet{chabrier03} initial mass function (IMF) and a flat $\Lambda$ cold dark matter ($\Lambda$CDM) model with cosmological parameters 
$\Omega_{\Lambda}\,=\,0.685$; $\Omega_{\mathrm{M}}\,=\,0.315$;
$H_{{0}}\,=\,67.4$\,\kms\,Mpc$^{-1}$ \citep{planck_cosmo18}. 
Magnitudes are given in the AB system \citep{OkeGunn1983}. The prefix `p' in front of distance units given in kpc means physical, and `c' means comoving.

\section{Selected targets}\label{sec:targets}
The selected sources were discovered as carbon monoxide (CO) emitters in NOEMA follow-up observations (Sect.~\ref{sec:noema}) of the brightest \herschel\ 350\,$\mu$m source \citep{planck15} in a 20\arcmin$\times$20\arcmin\ field centered on the \planck\ high-$z$ source \citep[known as PHz;][]{planck16} \g191\ (G191 hereinafter).  The \herschel\ source, known as \herschel\ ID 01 (H01, hereinafter), is a known CO-emitter at $z=2.55$ from previous observations carried out with the Eight MIxer Receiver (EMIR) on the Institut de radioastronomie millim\'etrique (IRAM) 30-m telescope \citep{polletta22}.  The total star formation rate (SFR) of H01 is $\sim$2300\,\msun\,yr$^{-1}$.  This is derived from the total infrared (IR) luminosity ($L_{\rm IR}$) and assuming the relation in  \citet{kennicutt12} corrected for a Chabrier IMF (i.e., ${\rm SFR}\,=\,1.40\times 10^{-10}\times L_{\rm IR}$)\footnote{The SFR conversion factor for different IMFs, as derived by \citet{madau14}, is SFR$_{\rm Chabrier}$\,=\,SFR$_{\rm Kroupa}$/1.063.  This value was obtained assuming the flexible stellar population synthesis (FSPS) models of \citet{conroy10} and a constant SFR and is valid for various metallicities but for an age of 10\,Gyr.  At younger ages the conversion factor is slightly smaller, e.g., SFR$_{\rm Chabrier}$\,=\,SFR$_{\rm Kroupa}$/1.028 at an age of 6\,Gyr.  Because conversion factors at younger ages are not available, we adopted 1.063, which is the most commonly used in the literature, even when considering high redshift objects.}.  The factor adopted to derive the SFR from the total IR luminosity can differ by up to 50\% even for the same IMF\null.  The IR luminosity, $\log (L_{\rm IR}$/\lsun)\,=\,13.21$\pm$0.21, was estimated by fitting the \herschel/SPIRE fluxes (Table~\ref{tab:herschel_fluxes}) with a modified black body and assuming the CO redshift.
\begin{table} 
\caption{\label{tab:herschel_fluxes}\herschel/SPIRE fluxes of the selected targets.} 
\centering 
\renewcommand{\arraystretch}{1.2}
\setlength{\tabcolsep}{3.0pt}
\begin{tabular}{c c c r r r c l} 
\hline\hline
\herschel\ & $\alpha_{Herschel}$ & $\delta_{Herschel}$    & $S_\mathrm{250\mu m}$& $S_\mathrm{350\mu m}$ & $S_\mathrm{500\mu m}$ \\
 ID        &  (h:m:s) & (\deg:\arcmin:\arcsec)  &    (mJy)             &        (mJy)          &    (mJy)              \\
\hline
    01     & 10:44:38.5  & 33:51:05.9 &    89$\pm$11         &  111$\pm$10           &    85$\pm$11          \\
\hline
    01a    & \nodata   &  \nodata  &    54$\pm$11         &   67$\pm$10           &    51$\pm$11          \\
    01b    & \nodata   &  \nodata  &    35$\pm$11         &   44$\pm$10           &    34$\pm$11          \\
    01c    & \nodata   &  \nodata  &    $<$35             &   $<$44               &    $<$34              \\
\hline                                                                                            
\end{tabular}\\
\end{table}

The G191 field (RA\,=\,161.223\,deg, Dec\,=\,$+$33.8482\,deg) was included in the target list of the \jwst\ Prime Extragalactic Areas for Reionization and Lensing Science (PEARLS) GTO Program \citep[ID 1176, PI: Windhorst; ][]{windhorst23}, as a protocluster candidate.  The source was first identified as one of the \planck\ high-$z$ (PHz) sources with the largest SFR \citep[i.e., SFR$\,\gtrsim\,$40000\,\msun\,yr$^{-1}$;][]{planck15}.  The PHz sources are galaxy protocluster candidates at $z$\,$\simeq$\,2--4.  The interest in this field grew after observations with \herschel\ in the submm and with \spitzer\ in the mid-IR (MIR) revealed a significant overdensity of both red SPIRE sources \citep[significant at 6.2$\sigma$; ][]{planck15}, and red \spitzer\ sources \citep{martinache18}.  Such overdensities are expected in the case of highly star-forming protoclusters at $z\gtrsim2$ \citep{chiang13,contini16}.  Nine \herschel\ sources were observed with the IRAM 30-m telescope with the goal of detecting a CO line and determining their redshift, but a CO line was detected in only three sources with redshifts 1.89, 2.55 and 2.62.  Because the undetected sources are all fainter at submm wavelengths than the detected ones, it is not possible to rule out that there might be several \herschel\ sources at the same redshift, but at the moment the protocluster nature of this PHz source is not spectroscopically confirmed.  In this work, we study in detail the brightest \herschel\ source in the G191 field, H01, for which a redshift of 2.55 was confirmed through the detection of three CO line transitions \citep[][]{polletta22}.  

The \herschel\ source H01 in G191 was observed with EMIR \citep{carter12}
at the IRAM/30-m telescope with multiple tunings at 2, and 3\,mm with the
goal of detecting the CO emission line and thus constraining the
source redshift and molecular gas properties.  These observations yielded
significant ($>6\sigma$) detections of the \COthree, and \COfour\
transitions, and a tentative detection of the \COfive\ transition at an
average redshift of $\langle z\rangle =2.55132\pm 0.0009$ \citep[see details in
][]{polletta22}.  Surprisingly, the lines become fainter at higher
transitions, implying a CO spectral line energy distribution (CO SLED) with
peak at $J_{\rm up}=3$.  The line ratio $I_{\rm CO(4-3)}/I_{\rm CO(3-2)} =
0.41\pm 0.10$ is much lower than observed in SMGs (i.e.,
1.40$\pm$0.02 in \citealt{bothwell13}, and 0.96$\pm$0.21 in \citealt{birkin21}), and in the Milky Way \citep[i.e., 1.12;
][]{carilli13}.  Since the size of the EMIR beam decreases with frequency (the EMIR half power beam width varying from 27\arcsec\ at 91\,GHz to 16\arcsec\ at 145\,GHz)\footnote{https://publicwiki.iram.es/Iram30mEfficiencies}, such a peculiar CO SLED could be artificially produced if multiple CO-emitters at the same redshift contribute to the large beam at short frequencies, but not all in the small beam at higher frequencies.  It is also possible that there is only one CO-emitter but a wrong or inaccurate pointing might have caused signal loss in the higher frequency observations. Another possibility is that optically thick dust suppresses the high-$J$ CO lines \citep{papadopoulos10}. Finally, gravitational lensing, a phenomenon that affects some bright DSFGs and PHz \citep{negrello10,canameras15}, can also produce a larger magnification of the diffuse gas emission seen in the low transitions than that coming from the compact gas traced by higher transitions, artificially producing a low peak in the CO SLED \citep[see ][]{hezaveh12}.  To test all these possibilities, we carried out CO observations with NOEMA at a higher spatial resolution.  In this work, we present the follow-up observations carried out in the millimeter (mm) with NOEMA and in the NIR with  \jwst/NIRCam.
\begin{figure*}[t!]
\centering\includegraphics[width=0.47\linewidth]{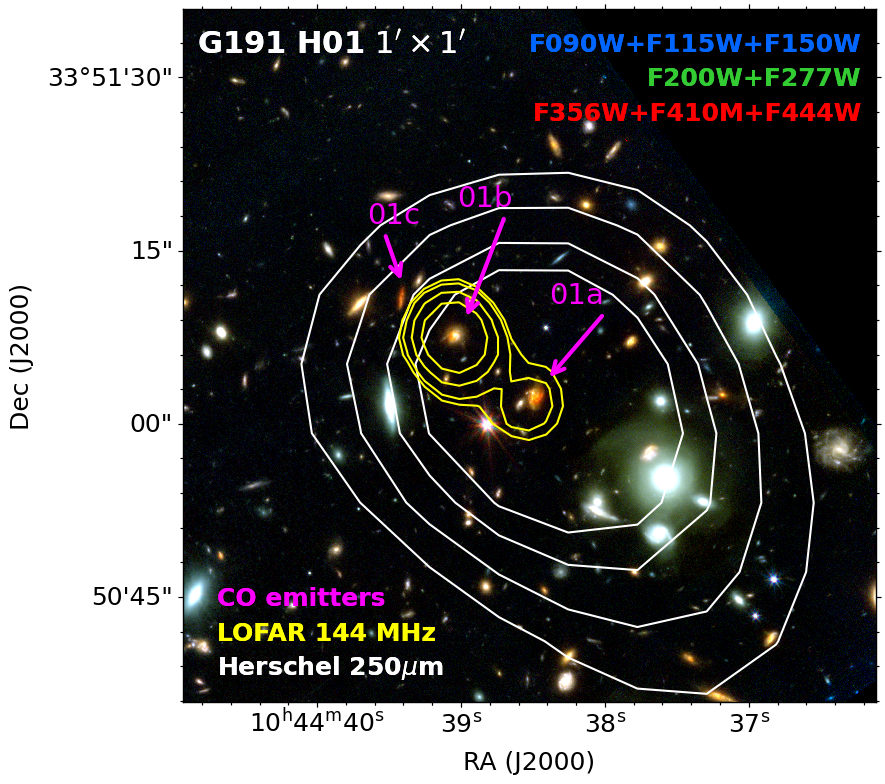}
\includegraphics[width=0.51\linewidth]{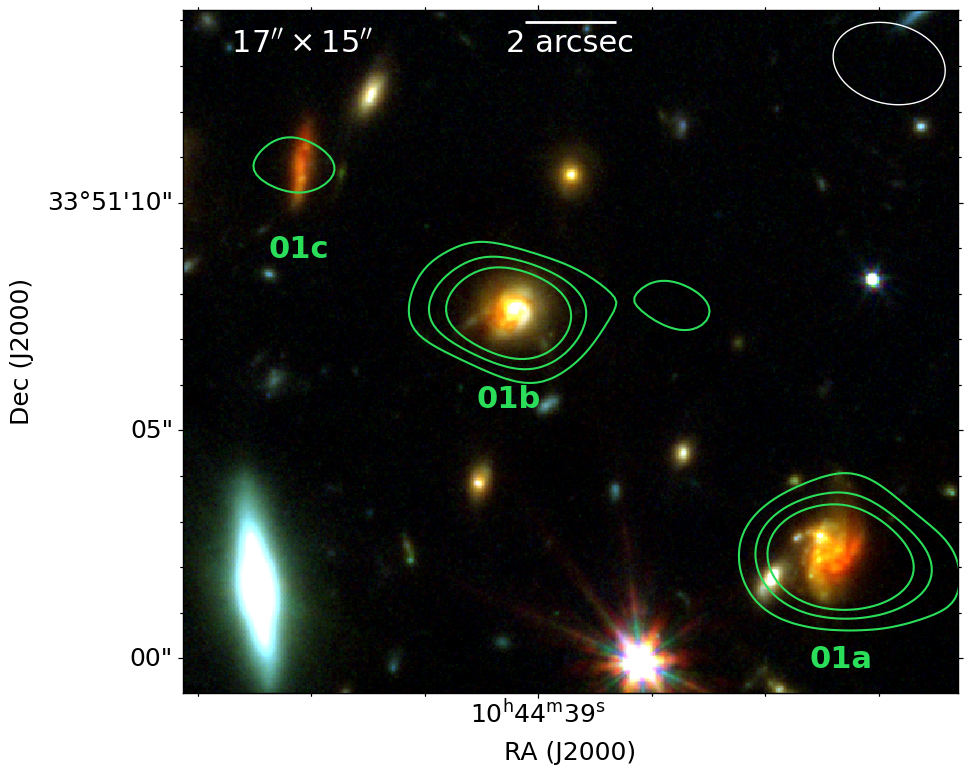}
\caption{{\small Multicolor (red: F356W+F410M+F444W, green: F200W+F277W,
and blue: F090W+F115W+F150W) \jwst/NIRCam images. {\it Left panel:}
1$^{\prime}{\times}{1}^{\prime}$ image centered on the \herschel\
source H01. The \herschel\ contours at 250\,$\mu$m are in white, the radio LOFAR 144\,MHz contours in yellow, and the IRAM/NOEMA CO-detected sources are
indicated by magenta arrows and labeled. 
{\it Right panel:} 17\arcsec$\times$15\arcsec\ image showing the CO contours starting from 3$\sigma$ in steps of 3$\sigma$ (green) and the NOEMA beam (white ellipse in the top right corner).}}
\label{fig:g191_01_image} 
\end{figure*}

\section{Observations and data}\label{sec:obs}

\subsection{NOEMA observations}\label{sec:noema}

We observed H01 in the G191 field with the NOEMA interferometer (project ID: W21DA, PI: Polletta) to assess whether multiplicity affected the \COthree\ emission previously detected with EMIR\null.  The observations were carried out with 11 antennas with 24--368\,m baseline range on 2022 March~26 and~27  for a total observing time of $\sim$5\,hr including $\sim$2.6\,hr  on-source.  The weather conditions were good with precipitable water
vapor (PWV) $\sim$2\,mm, average system temperature $T_{\rm sys}\sim120$\,K, and
wind speed  $\sim$5\,m/s.  The quasar 3C~84 was used as receiver bandpass
(RF) calibrator, the emission line star LkH$\alpha$~101 and the BL Lac 2010$+$723 as flux calibrators, and the Seyfert 1 galaxy 1030$+$415 and the Blazar 1100$+$305
as phase and amplitude calibrators.  The targeted \COthree\ emission line at rest-frame frequency $\nu_{\rm rest}$\,=\,345.796\,GHz is redshifted to $\nu_{\rm obs}$\,=\,97.37\,GHz at $z$\,=\,2.55.  A single spectral tuning was used with the NOEMA {\sf PolyFiX} correlator in dual polarization mode in the 3\,mm band providing coverage over two frequency bandwidths, 79--86\,GHz and 94.2--102\,GHz with spectral resolution 2\,MHz (corresponding to 6\,\kms).  The spectral tuning covered the \COthree\ line in the upper sideband and the continuum centered at 3.1\,mm in the lower sideband.  To search for multiple CO emitters and solve confusion, we chose the C configuration to achieve angular resolution  $\sim$2\arcsec\ at the observed frequency.

Calibration and imaging of the $uv$ data were carried out using the
GILDAS\footnote{http://www.iram.fr/IRAMFR/GILDAS/} package {\sf clic} with
assistance from the IRAM staff.  In the 3\,mm band, the absolute flux
calibration is accurate at the 10\% level.  We resampled the $uv$ tables to
20\,MHz ($\sim$60\,\kms).
With the calibrated visibility data, \COthree\ and dust-continuum 
maps were created using the software GILDAS/MAPPING using natural 
weighting and the Hogbom cleaning algorithm \citep{hogbom74} with a threshold at 1$\sigma$,
where $\sigma\!\sim\!0.5$\,mJy\,beam$^{-1}$ is the noise after
deconvolution.  The continuum $uv$ table was created from the calibrated
visibilities by averaging the line-free channels using the MAPPING task {\sf
uv\_continuum} with line-emission channels filtered with the {\sf uv\_filter} task. 
The continuum emission was subtracted from the line $uv$ data by fitting a
linear baseline for each visibility with the {\sf uv\_baseline}
task.  The final synthesized beam size is 2\farcs5$\times$1\farcs8
with position angle (PA) 75\deg.  The NOEMA primary beam (field of view) for
NOEMA observations at the \COthree\ frequency is 52\farcs3.  The
resulting background root mean square (rms) noise level of the
natural-weighted image cube is $\sigma\sim$\,0.5\,mJy\,beam$^{-1}$ per
60\,\kms\ channel and $\sim$0.03\,mJy\,beam$^{-1}$ for the continuum image. 
A \COthree\ intensity map was created by averaging the flux over the line
velocity range $\sim$1200\,\kms\ and applying a primary-beam correction.

\subsubsection{CO measurements}\label{sec:CO_measurements}

\begin{figure*}[t!]
\centering
\includegraphics[height=3.7cm]{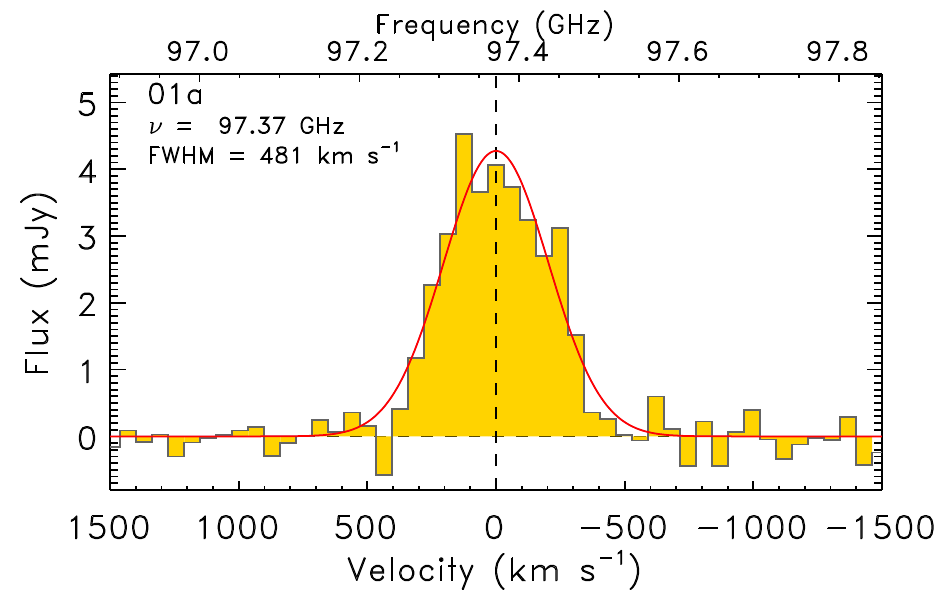}
\includegraphics[height=3.7cm]{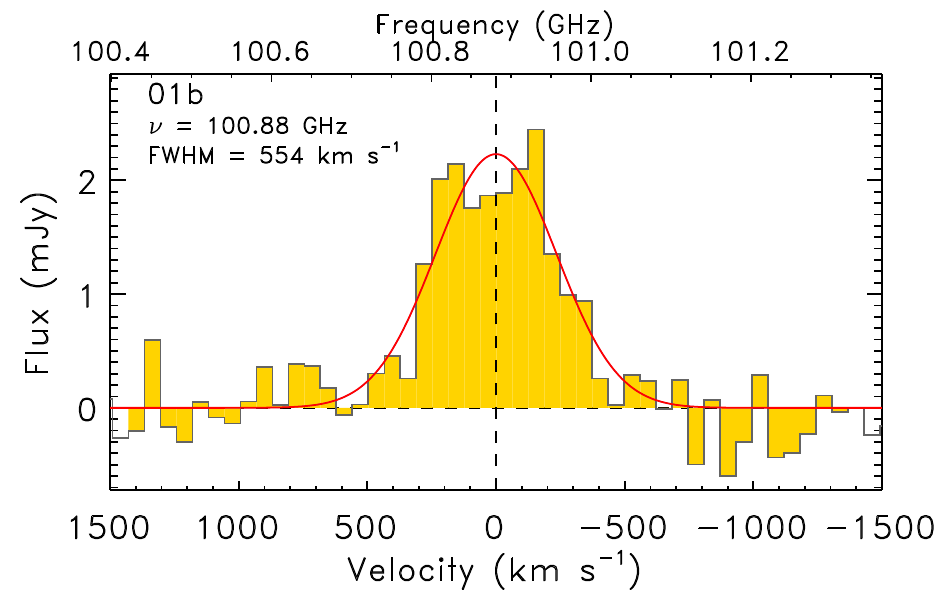}
\includegraphics[height=3.7cm]{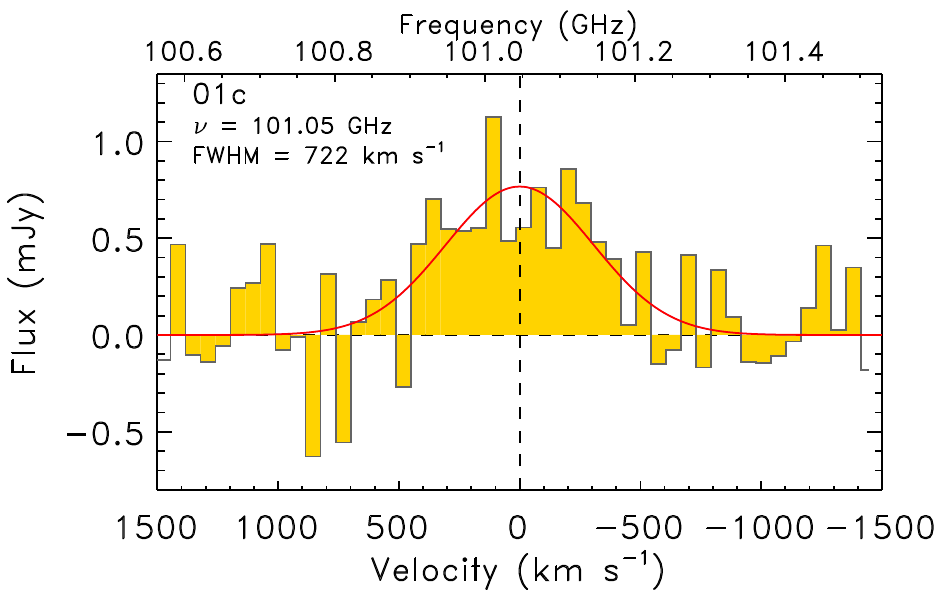}
\caption{{\small NOEMA \COthree\ emission (gray histogram) from 01a at $z=2.5513$ (left panel), 01b at
$z=2.4278$ (middle panel), and 01c at $z=2.4222$ (right panel) and Gaussian bestfit model
(red line). The source ID, line observed frequency and FWHM are annotated in
each panel.}}
\label{fig:noema_line_srcs} 
\end{figure*}

The observations show three emission-line sources within 15\arcsec\ of the image center. One, here called 01a, is at the same frequency as the previous EMIR detection, and two more, designated 01b and 01c, are at a higher frequency corresponding to $z{\sim}$2.42, assuming the observed line is \COthree. Alternative transitions would imply redshifts ($z\sim1.28$, or 3.56) that are inconsistent with the observed spectral energy distributions (SEDs) (see Sect.~\ref{sec:cigale}). The position of each line emitter, measured using the GILDAS/MAPPING GILDAS software, is
reported in Table~\ref{tab:co_data}, and Fig.~\ref{fig:g191_01_image} shows their locations
with respect to the \herschel\ 250\,$\mu$m emission. The \COthree\ line spectra, extracted with the GILDAS {\sf uv\_fit} task, are shown in Fig.~\ref{fig:noema_line_srcs}. Table~\ref{tab:co_data} gives the line intensities, central frequencies, and
widths (FWHM) measured using the Cube Analysis and Rendering Tool for Astronomy \citep[CARTA;][]{comrie21}.\footnote{CARTA is an open-source software package developed by the National Radio Astronomy Observatory (NRAO) for analyzing, visualizing, and manipulating astronomical
data cubes (https://cartavis.org/).} After discovering the CO emission from 01b and 01c in the NOEMA image, we reanalyzed the existing EMIR spectra in the spectral regions where the CO lines from 01b and 01c were expected and were able to identify a line at a frequency consistent with the CO line frequencies of 01b and 01c, and with S/N=5.6. The parameters of the EMIR line associated with 01a, and with 01b and 01c, derived from a single Gaussian fit, are also reported in Table~\ref{tab:co_data}. Because the lines emitted by 01b and 01c do not overlap in frequency with the line from source 01a, we can rule out their contamination to the previous EMIR detection. On the other hand, these two lines overlap in the EMIR spectrum and even if their peaks are shifted by $\sim$490\,\kms\ and the two sources are $\sim$5.7\arcsec\ apart, without sufficient spatial resolution it would be difficult to identify the two separate sources. A search for other line emitters in the NOEMA primary beam yielded no additional detections. 

\begin{table*}[!ht]
\centering
\caption{\COthree\ line measurements.\label{tab:co_data}}
\setlength{\tabcolsep}{3.pt}
\begin{tabular}{lrccccc}
\hline \hline
 ID          & Frequency & Redshift       & $I_\mathrm{CO}$     & FWHM         &  $L^{\prime}_\mathrm{CO(3-2)}$ & $M_{\rm gas}$\tablefootmark{a} \\
             &    (GHz)  &                &  (Jy\,km\,s$^{-1}$) &(km\,s$^{-1}$)& (10$^{9}$\,K\,\kms\,pc$^2$)    & (10$^{10}$\,\msun)  \\
\hline
\noalign{\vspace{2pt}}
\multicolumn{7}{c}{EMIR\tablefootmark{b}}\\
 01a       &   97.383$\pm$0.010   &  2.55089$\pm$0.00071 &  2.75$\pm$0.33  &   \0497$\pm$63\0  &    96.6$\pm$11.6  &   16.7$\pm$2.0 \\
 01b$+$01c &  100.887$\pm$0.029   &  2.42756$\pm$0.00098  &  1.85$\pm$0.47  &   \0643$\pm$172 &    59.7$\pm$15.2  &   10.3$\pm$2.6 \\ 
 \hline
\multicolumn{7}{c}{NOEMA}\\
 01a       &   97.3709$\pm$0.0032 &  2.55133$\pm$0.00012 &  2.19$\pm$0.09  &   \0481$\pm$23\0  &    77.2$\pm$3.2   &   13.3$\pm$0.6 \\
 01b       &  100.8809$\pm$0.0054 &  2.42777$\pm$0.00018 &  1.31$\pm$0.08  &   \0554$\pm$39\0  &    42.6$\pm$2.6   &    \07.4$\pm$0.5 \\
 01c       &  101.0463$\pm$0.0238 &  2.42215$\pm$0.00081 &  0.59$\pm$0.12  &   \0722$\pm$174 &    19.0$\pm$4.0   &    \03.3$\pm$0.7 \\
\hline
\end{tabular}
\tablefoot{
\tablefoottext{a}{\small The gas mass is derived from the \LpCOone\ luminosity as  
$M_{\rm gas}=1.36\alpha_{\rm CO}$\LpCOone\ and assuming 
$\alpha_\mathrm{CO}$\,=\,0.8\,\msun\,pc$^{-2}$\,(K\,\kms)$^{-1}$ as
typically assumed in starburst galaxies. The \LpCOone\ luminosity is derived
from the \LpCOthree\ luminosity assuming the brightness-temperature ratio 
measured in SMGs \citep[$r_{3,2}$\,=\,\LpCOthree/\LpCOone \,=\,0.63;][]{birkin21}.}
\tablefoottext{b}{\small Line parameters obtained from the IRAM 30-m telescope/EMIR spectra \citep{polletta22}, and assuming the same $r_{3,2}$ value and the $M_{\rm gas}-$\LpCOone\ relation adopted for the NOEMA data.}
}
\end{table*}

\subsubsection{Dust continuum measurements}
We detected only 01a and 01b in the 3\,mm continuum map. They are  unresolved at a resolution of 2\arcsec$\times$3\arcsec.  We measured their flux by fitting a point like source using the MAPPING task {\sf uv\_fit} in the $uv$ plane. Table~\ref{tab:noema} gives the source coordinates and continuum flux densities. 
It is quite likely that multiple sources might be associated with a single \herschel\ source, especially at bright flux densities. For example, \citet{montana21} and \citet{scudder16} found that 9--23\% of \herschel\ sources with $S_{\rm 500\mu m}>$35--80\,mJy are multiple.  It is also common to find that some of these sources are at the same redshift \citep[e.g.,][]{ivison13,Wang2016,oteo18,coogan18,zhou24}. In PLCK\,G073.4$-$57.5, the only \planck-selected field for which high spatial resolution mm observations are available, \citet{kneissl19} found between one and four ALMA continuum detections associated with each of the eight observed \herschel\ sources. 

The NOEMA observations demonstrate that source H01 in G191 suffers from multiplicity effects and that its submm continuum emission is likely due to the combined emission of all three CO emitters. Because the \herschel\ observations have insufficient spatial resolution ($FWHM=25$\arcsec\ at 350$\mu$m) to resolve the emission from each source, we estimate their submm flux by assuming that it is proportional to the NOEMA mm flux. This assumption is likely valid because the sources are at similar redshifts, and they presumably have similar dust temperatures and SEDs. Because 01c was not detected in the continuum by NOEMA, we assigned an upper limit to its submm flux. The measured and scaled \herschel\ flux densities are listed in Table~\ref{tab:herschel_fluxes}.

\begin{table} 
\caption{\label{tab:noema}NOEMA measurements} 
\centering 
\begin{tabular}{c c c c} 
\hline\hline
Source & $\alpha_{\rm NOEMA}$ & $\delta_{\rm NOEMA}$ & $S_\mathrm{\rm 3mm}$ \\
 ID    &  (h:m:s)    & (\deg:\arcmin:\arcsec) &          (mJy)        \\
\hline
   01a & 10:44:38.47 & 33:51:02.17 &  0.149$\pm$0.026 \\
   01b & 10:44:39.05 & 33:51:07.57 &  0.098$\pm$0.037 \\
   01c & 10:44:39.43 & 33:51:10.81 &  $<$0.042        \\ 
\hline                                                                                            
\end{tabular}
\end{table}

\subsection{LOFAR observations}\label{sec:lofar}
The G191 field was  observed with LOFAR at 144\,MHz as part of the
mosaic P160$+$35. The LOFAR data from the LOFAR LoTSS second data
release \citep[DR2;][]{shimwell17}  include images at an angular
resolution of 6\,arcsec\ and rms noise 0.074\,mJy\,beam$^{-1}$ and a radio source list.  Two LOFAR sources fall within
the H01 beam and  match the two CO emitters 01a and 01b (Fig.~\ref{fig:g191_01_image}), but source 01c was not detected. 
The sources' LOFAR positions and flux densities are listed in Table~\ref{tab:lofar}.
\begin{table} 
\caption{\label{tab:lofar}LOFAR measurements} 
\centering 
\begin{tabular}{c c c c} 
\hline\hline
Source & $\alpha_{\rm LOFAR}$ & $\delta_{\rm LOFAR}$ & $S_\mathrm{\rm 144 MHz}$ \\
 ID    &      (h:m:s)           &    (\deg:\arcmin:\arcsec)             &         (mJy)        \\
\hline
  01a  &  10:44:38.51 &  33:51:01.51 &  1.26$\pm$0.25  \\
  01b  &  10:44:39.05 &  33:51:07.56 &  2.01$\pm$0.15  \\
  01c  &  \nodata    &  \nodata   &  $<$0.37\tablefootmark{a}        \\
\hline                                                   
\end{tabular}
\tablefoot{
\tablefoottext{a}{5$\sigma$ upper limit}                  }
\end{table}

\subsection{ \jwst\ observations}\label{sec:jwst_obs}

\jwst/NIRCam multiband observations of the G191 field were obtained as part
of the PEARLS GTO Program \citep[ID 1176, PI: Windhorst; ][]{windhorst23}. 
Observations were carried out on UT 2023 April 24 in eight filters: F090W,
F115W, F150W, F200W of the short wavelength (SW) module and F277W, F356W, F410M,
and F444W of the long wavelength (LW) module.  Exposure times ranged
from 1.9\,ks to 2.5\,ks giving 5$\sigma$ point-source limits 28.32, 28.50, 28.52, 28.79, 28.49, 28.49, 27.99, and 28.29 in order of increasing wavelength for the eight NIRCam bands \citep{windhorst23}.

The data were obtained from the Mikulski Archive for Space Telescopes
(MAST) and processed with the STScI pipeline v1.7.2 \citep{bushouse22} with reference files specified by the context file jwst\_0995.pmap.  Raw images
were corrected for $1/f$ noise, flat field, and cosmic-ray hits using the
prescription of
C.~Willott.\footnote{\url{https://github.com/chriswillott/jwst.git}}  We
then ran the ProFound code \citep{ProFound} for the sky subtraction to 
correct remaining residuals, detector-level offsets, 
snowballs, and wisps \citep{robotham23}.  The frames were
then corrected for image distortions, aligned to the GAIA DR3 reference
catalog \citep{GAIADR3}\footnote{The comparison between the position of $\sim$2500 sources in common between ground-based $i$-band images and GAIA DR3 after applying twice a 3$\sigma$ clipping yields a median $\Delta\alpha\sim0.5\pm4$\,mas, and a median $\Delta\delta\sim-1\pm5$\,mas.}, and drizzled into final science mosaics with
resampled pixel size of 30\,mas, following similar methodology to that first
described by \cite{Koekemoer2011} updated to use the \jwst\
pipeline\footnote{\url{https://github.com/spacetelescope/jwst}}.   \citet{windhorst23} gave more
details on the observations, data reduction, source detection, and photometry.

\begin{figure*}[h!]
\centering
\includegraphics[width=0.33\textwidth]{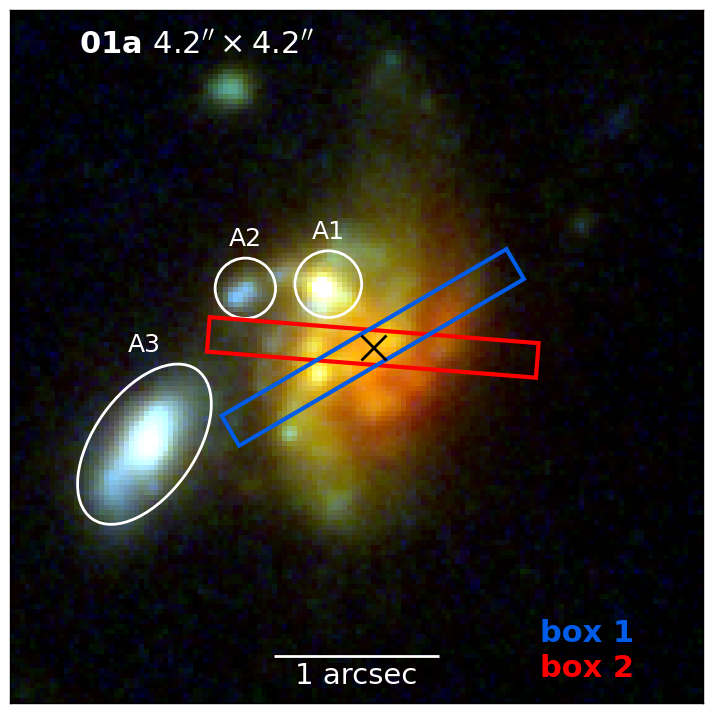}
\includegraphics[width=0.33\textwidth]{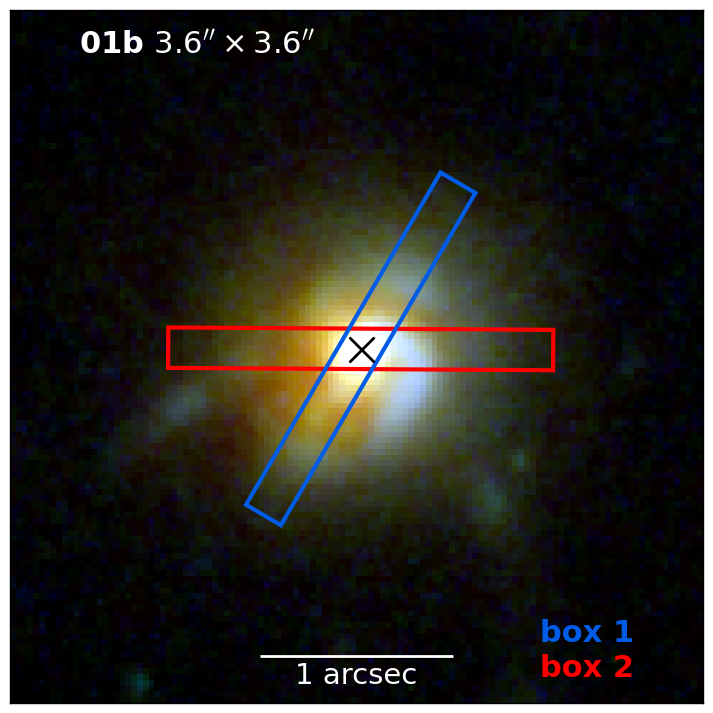}
\includegraphics[width=0.33\textwidth]{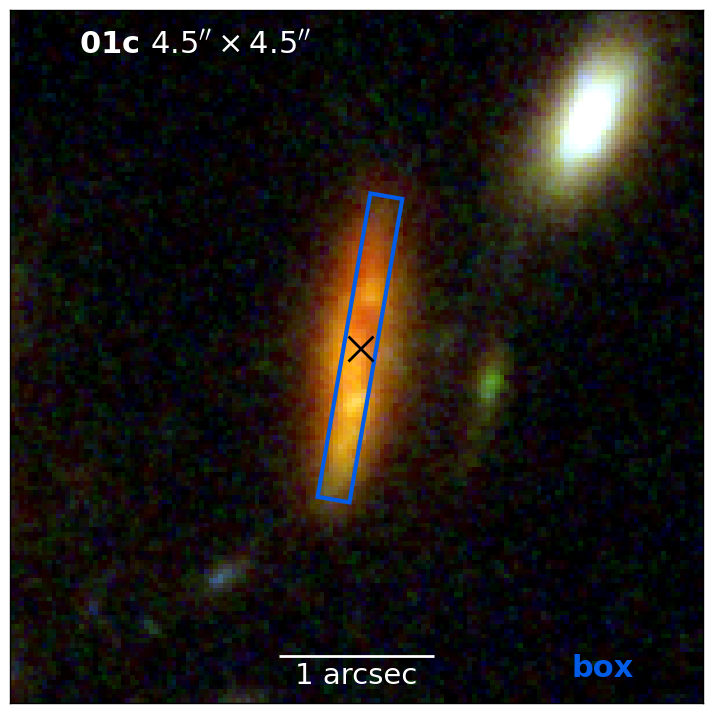}\\
\caption{\small  Multicolor (red: F356W+F410M+F444W, green: F200W+F277W, and
blue: F090W+F115W+F150W) \jwst/NIRCam images of individual CO counterparts.  Panels left to right show 01a, 01b, and 01c. The size of each panel is noted at top, and north is up,  east to the left. The white outlined regions in 01a indicate likely distinct sources. The center of each galaxy is marked with a black cross. Rectangular boxes mark regions used to analyze how the key physical parameters vary across each galaxy (Sect.~\ref{sec:resolved_seds}).} 
\label{fig:targets}
\end{figure*}

\begin{figure*}[h!]
\centering
\includegraphics[width=\textwidth]{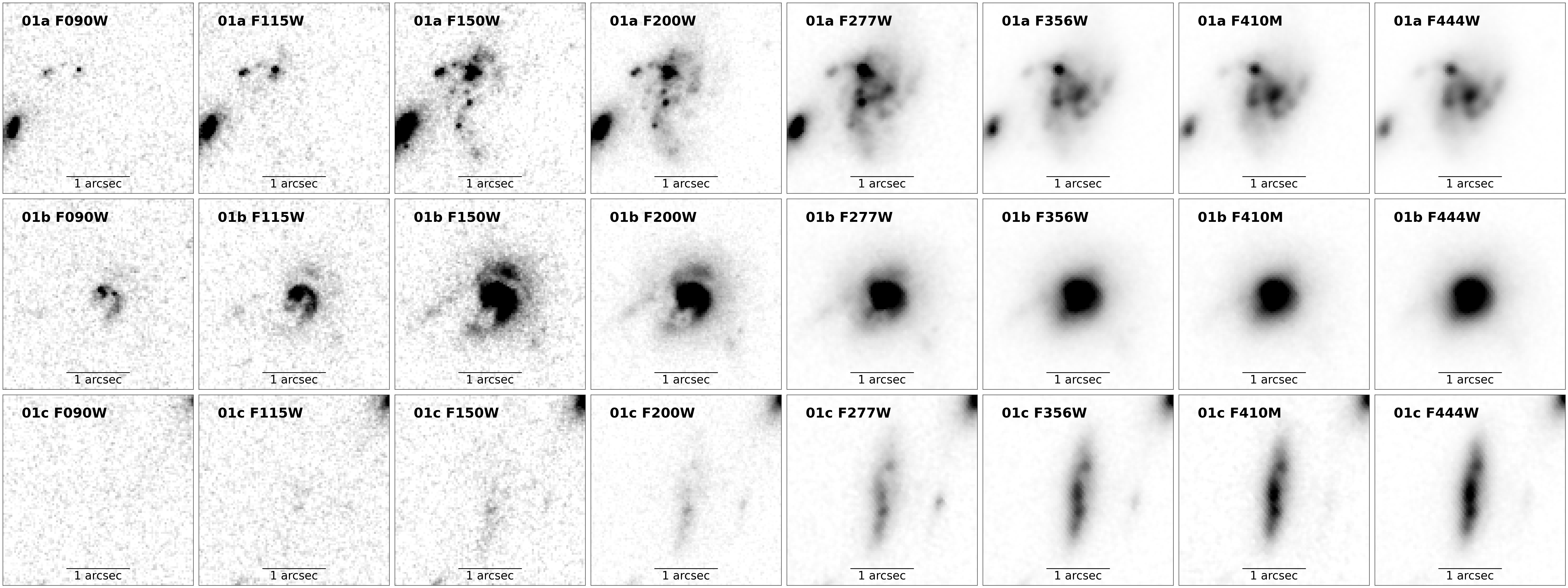}
\caption{\small Negative 3\arcsec$\times$3\arcsec\ thumbnail \jwst/NIRCam images.  Panels top to bottom show 01a, 01b, and 01c and left to right show increasing wavelengths as labeled. The images grayscale stretches from 0\,nJy to 3.2\,nJy in the first three bandpasses, and to 6.3\,nJy in F200W for all three sources. The maximum values adopted in the LW filters increase with wavelength (from F277W to F444W) and
differ for each source (from 10.6 to 42\,nJy in 01a, from 10.6 to 21\,nJy in 01b, and from 6.3 to 11.4\,nJy in 01c). North is up and east to the left.}
\label{fig:cutouts} 
\end{figure*}

The single and multiband images of the three CO emitters are shown in Figs.~\ref{fig:targets} and~\ref{fig:cutouts}. To measure the flux densities of the three galaxies, we  PSF-matched all the images to that of F444W, which has the coarsest resolution.  
The PSFs came from \textsc{WebbPSF} models publicly provided by STScI,\footnote{\url{https://stsci.app.box.com/v/jwst-simulated-psf-library/folder/174723156124}}
rotated to match the orientations of the observed images. Convolution kernels were created for each F444W--other band PSF pair following the convolution theorem by computing the ratio of the Fourier transforms and inverse transforming \citep[e.g.,][]{aniano11,boucaud16}. This approach suffers at high spatial frequencies, where the Fourier transform of the PSF is small and dominated by noise \citep{aniano11}. To eliminate spurious high frequency modes induced by the noise, we applied a simple cosine-bell window function where 80\% of the array values are tapered in the kernel creation. Each image was then convolved using the appropriate kernel following the procedure described by \citet{pascale22}. The PSF-matched images were then stacked, and a segmentation map was created from the stacked image by selecting all pixels
within a specific radius per source, from 22 to 30 pixels (equivalent to 0\farcs66--0\farcs9), with a value above a threshold.  The threshold was derived from a wide region close to the selected sources where we computed the 3$\sigma$-clipped mean value
($\mu_{\rm bck}$) and the standard deviation ($\sigma_{\rm bck}$).  The
threshold value was defined as $\mu_{\rm bck}+10\times\sigma_{\rm bck}$. 
Nearby sources were flagged and removed from the segmentation mask.  For 01a, we identified two clumps (called A1 and A2) whose association with the main galaxy is
dubious in spite of their proximity because of the clumps' much bluer colors (Fig.~\ref{fig:targets}).  Their spectral energy distributions and photometric redshifts (Appendix~\ref{app:01a_fgd_srcs}) indicate that they are distinct sources,  and therefore they were masked.

The total flux density of each source per band was obtained as the sum of the signal in each image of all associated
pixels in the segmentation map.  The associated  uncertainty per source and 
band was computed from the variance defined as
\begin{equation}\label{eq:variance}
\sigma_{\rm src}^2 = S_{\rm src} - \bar{B} + \left(N_S + \frac{N_S^2}{N_B} \right)\quad,
\end{equation}
where $S_{\rm src}$ is the total source flux density, $\bar{B}$ the average background value obtained from seven clean circular regions with 0\farcs72 radius around each source, $N_S$ is the number of source pixels, and $N_B$  the number of background pixels.\footnote{https://wise2.ipac.caltech.edu/staff/fmasci/ApPhotUncert.pdf}  
The  \jwst\ flux densities are listed in Table~\ref{tab:jwst}.
\begin{table} 
\caption{\label{tab:jwst} \jwst/NIRCam measurements.} 
\centering 
\setlength{\tabcolsep}{3.0pt}
\begin{tabular}{l c c c}
\hline\hline
Source ID                  &          01a  &  01b  &  01c  \\
\hline
$\alpha_{\rm  \jwst}$        &    10:44:38.48  &  10:44:39.05 &  10:44:39.42 \\
$\delta_{\rm  \jwst}$        &    33:51:02.27  &  33:51:07.62 &  33:51:10.82 \\
$S_\mathrm{\rm F090W}$     &   0.24$\pm$0.01   &  0.86$\pm$0.02   &  0.09$\pm$0.01 \\
$S_\mathrm{\rm F115W}$     &   0.40$\pm$0.02   &  1.47$\pm$0.04   &  0.18$\pm$0.01 \\
$S_\mathrm{\rm F150W}$     &   1.24$\pm$0.04   &  4.02$\pm$0.07   &  0.31$\pm$0.02 \\
$S_\mathrm{\rm F200W}$     &   3.05$\pm$0.06   &  6.20$\pm$0.08   &  0.65$\pm$0.03 \\
$S_\mathrm{\rm F277W}$     &   7.57$\pm$0.09   & 10.14$\pm$0.10   &  1.49$\pm$0.04 \\
$S_\mathrm{\rm F356W}$     &  14.49$\pm$0.12   & 15.28$\pm$0.12   &  2.89$\pm$0.05 \\
$S_\mathrm{\rm F410M}$     &  18.67$\pm$0.14   & 18.59$\pm$0.14   &  3.97$\pm$0.06 \\
$S_\mathrm{\rm F444W}$     &  21.85$\pm$0.15   & 20.33$\pm$0.14   &  4.67$\pm$0.07 \\
\hline                                                                                            
\end{tabular}\\
\tablefoot{$\alpha_{\rm  \jwst}$ and $\delta_{\rm  \jwst}$ are J2000 equatorial
coordinates in h:m:s, and \deg:\arcmin:\arcsec\, respectively, and $S_\mathrm{filter}$ are flux densities in the
 NIRCam filters in $\mu$Jy.}
\end{table}

\section{Spectral energy distribution modeling}\label{sec:cigale}

\subsection{Whole galaxy SED fits}
To determine the physical properties of the three CO emitters, we modeled their SEDs comprising the \jwst/NIRCam data, the submm--mm data from \herschel\ and NOEMA, and the radio LOFAR data.
\begin{figure}[ht!]
  \centering
 \includegraphics[width=0.49\textwidth]{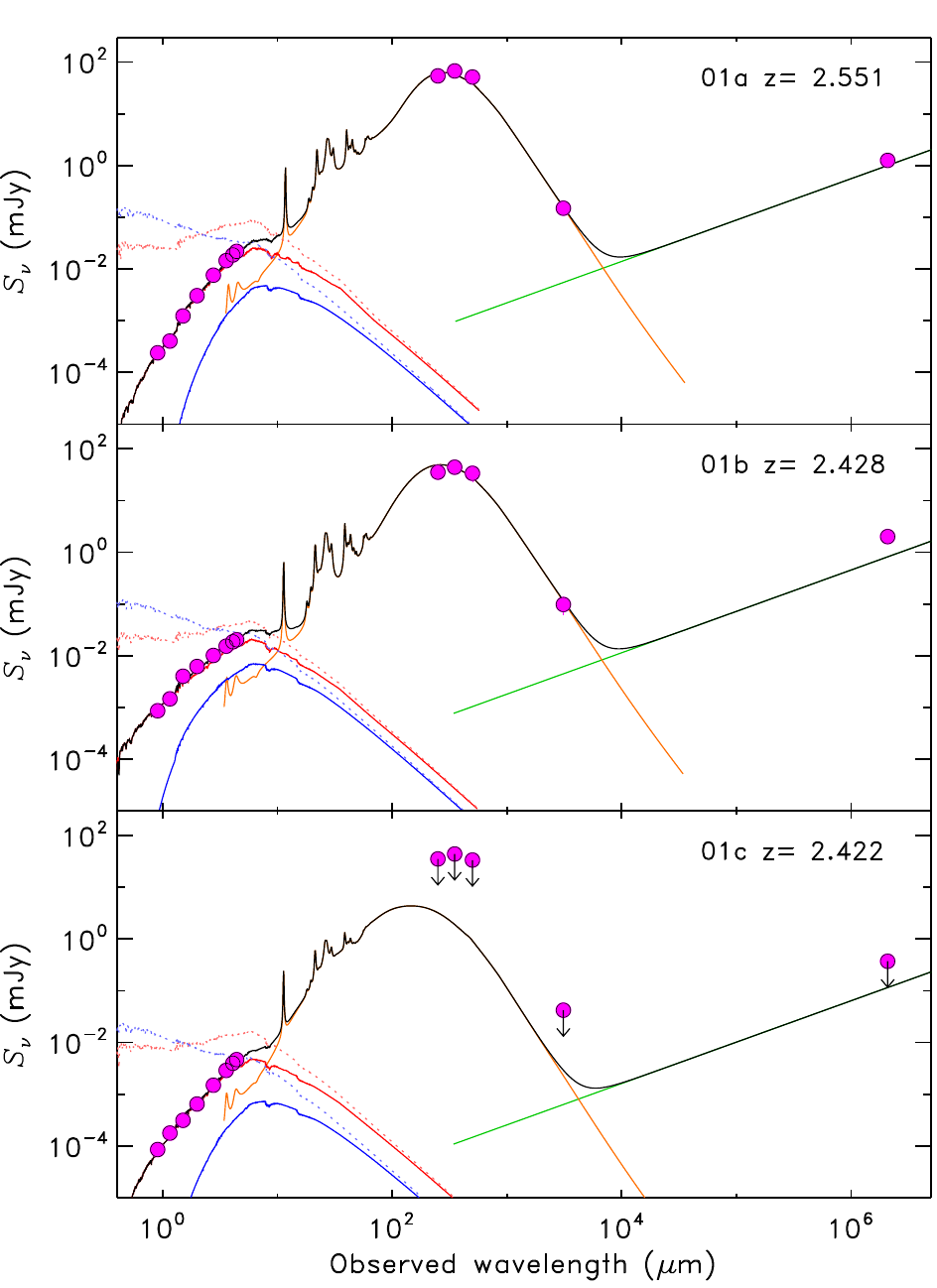}
\caption{SEDs of three CO emitters (filled magenta circles): 01a (top panel), 01b (middle panel), and 01c (bottom panel).  Downward arrows represent 5$\sigma$ upper limits.  The \texttt{CIGALE} bestfit model is shown with a solid black line. Dotted lines show the stellar light (red for the older population and blue for the younger) before dust attenuation, and the solid red and blue lines show the attenuated stellar light. The orange line shows the dust component, and the green line shows the radio synchrotron emission.}
   \label{fig:cigale_sed}
\end{figure}
The SED modeling was carried out with the Code Investigating Galaxy Emission \citep[\texttt{CIGALE};][]{boquien19}.  The code provides estimates of the stellar mass, the dust mass, the dust extinction, and the SFR and constraints on the stellar-population age. We fixed the redshifts at  
\COthree\ values and the metallicities to the solar
value {$Z$}\,=\,0.02.  Following \citet{osborne24}, we used the BC03 \citep{bruzual03} stellar
population models and chose a star formation history (SFH) modeled
with a double exponential with a recent burst with constant SFR ($\tau_{\rm
burst}=20$\,Gyr).  These choices produce the most accurate stellar masses
and SFRs among all those that are possible within
\texttt{CIGALE} \citep{michalowski14,osborne24}.  Dust emission was modeled
by \citet{draine14} models and dust attenuation by the
two-component model of \citet[][CF00 hereinafter]{charlot00}. The CF00 model assumes a different attenuation law and strength for the birth clouds and the interstellar medium (ISM) as it might be the case when the visible--NIR and FIR emissions are not co-spatial \citep[e.g.,][]{chen15,hodge16,smail23}.
Finally, we included a radio component to model the synchrotron emission produced by the interaction of relativistic electrons from supernovae with the local magnetic field.  This radio component is related to the galaxy star formation activity, and modeled as a power law with spectral index $\alpha_{\rm radio}=0.8$ \citep[$F_{\nu}\propto\nu^{-\alpha_{\rm radio}}$;][]{brienza17}. 
Its strength is bound to the estimated SFR through the radio-FIR correlation
\citep{helou85}. We did not include an AGN component in the model because there is no bright central source as expected in case of an AGN \citep{ortiz24} and the CO spectral-line energy distribution (SLED) in 01a peaks at a low transition, contrary to what is observed in AGN \citep{carilli13}. While an unobscured AGN is ruled out by the
shape of the rest-frame visible--NIR SED, an obscured AGN might instead be present and contribute to the mid-IR and radio emission\null.  No nebular component was
included because all the major emission lines are outside the NIRCam filter bandpasses.
The best fit was determined as the template with the lowest $\chi^2$, and
the bestfit parameters and associated uncertainties are the
likelihood-weighted means and standard deviations, respectively.  To assess the parameters' reliability and choose the best model, we compared the results obtained with different
SFHs (i.e., delayed and double exponential, with and without a burst),  extinction curves (i.e., CF00 and the \citet{calzetti00} law), and dust models. A burst greatly improves the fits (it halves the reduced-$\chi^2$), while the choice of underlying SFH, of extinction curves, and of dust model produce only minor variations. The parameters derived assuming a double exponential or delayed SFHs are all consistent. The main parameters derived from the best fit model obtained with a double exponential SFH are listed in Table~\ref{tab:cigale_params}, and the SEDs and bestfit models are shown in Fig.~\ref{fig:cigale_sed}.  
Based on the SED modeling, the three galaxies are massive, highly star-forming, and heavily obscured. Interestingly, the SED of source 01b exhibits a small excess of radio emission with respect to that predicted by the best fit model and based on the estimated SFR\null.  This excess is a factor of $\sim$4 and might be an indication of synchrotron emission from a radio jet powered by an active galactic nucleus (AGN). This possibility will be discussed in Sect.~\ref{sec:sfr_assessment}\null.
\begin{table*}[ht!]
\caption{\texttt{CIGALE} physical properties.\label{tab:cigale_params}}
\begin{center}
\renewcommand{\arraystretch}{1.4}
\begin{tabular}{c c c c c c c c c c}
\hline 
  ID   &     age   & age$_{\rm burst}$ &       \av       & SFR$_{\rm 100\,Myr}\tablefootmark{a}$& SFR\tablefootmark{b}     & $M_{\rm star}$   & $M_{\rm dust}$   &MS Offset\tablefootmark{c} & $\chi^2$ \\
       &  (Gyr) & (Myr) &    (mag)      &  (\msun\,yr$^{-1}$) & (\msun\,yr$^{-1}$) & (10$^{10}$\,\msun) &   (10$^{9}$\,\msun)  & & \\
\hline 
  01a & 1.5$\pm$0.1 & 20$\pm$7  & 3.8$\pm$0.1    &     541$\pm$18     &   2457$\pm$123 & 31$\pm$2    & 2.4$\pm$0.4    & 16$\pm$1 &  0.8 \\ 
  01b & 1.1$\pm$0.2 & 20$\pm$4  & 2.9$\pm$0.1    &     417$\pm$21     &   1918$\pm$191 &  9$\pm$2    & 1.6$\pm$0.5    & 19$\pm$3 &  2.8 \\ 
  01c & 1.3$\pm$0.2 & 50$\pm$6  & 3.6$\pm$0.1    &     171$\pm$ 9     &    343$\pm$ 53 & 2.6$\pm$0.4 & 0.05$\pm$0.002 &  6$\pm$1 &  4.8 \\ 
\hline                                                                            
\end{tabular}
\tablefoot{\texttt{CIGALE} fits are for a double-exponential SFH with an additional burst.
\tablefoottext{a}{SFR averaged over the past 100~Myrs.}
\tablefoottext{b}{Instantaneous SFR.}
\tablefoottext{c}{Ratio between the estimated instantaneous SFR and the SFR expected for a galaxy with the same stellar mass and redshift on the main sequence \citep{popesso23}.}
}
\end{center}
\end{table*}

\subsection{Resolved SED modeling}\label{sec:resolved_seds}

\begin{figure*}[h!]
\centering
\includegraphics[width=0.3\textwidth]{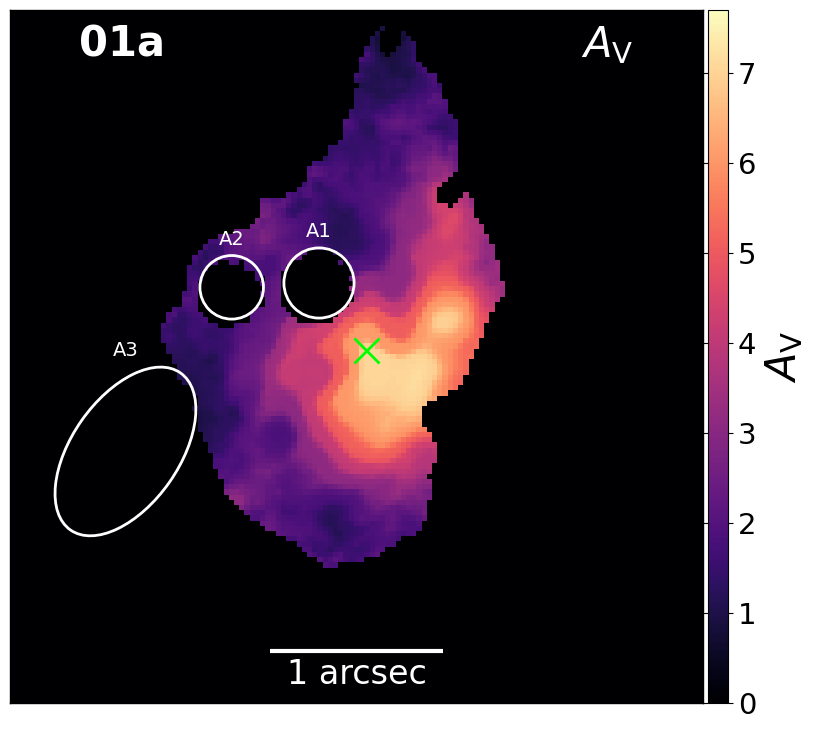}           \includegraphics[width=0.3\textwidth]{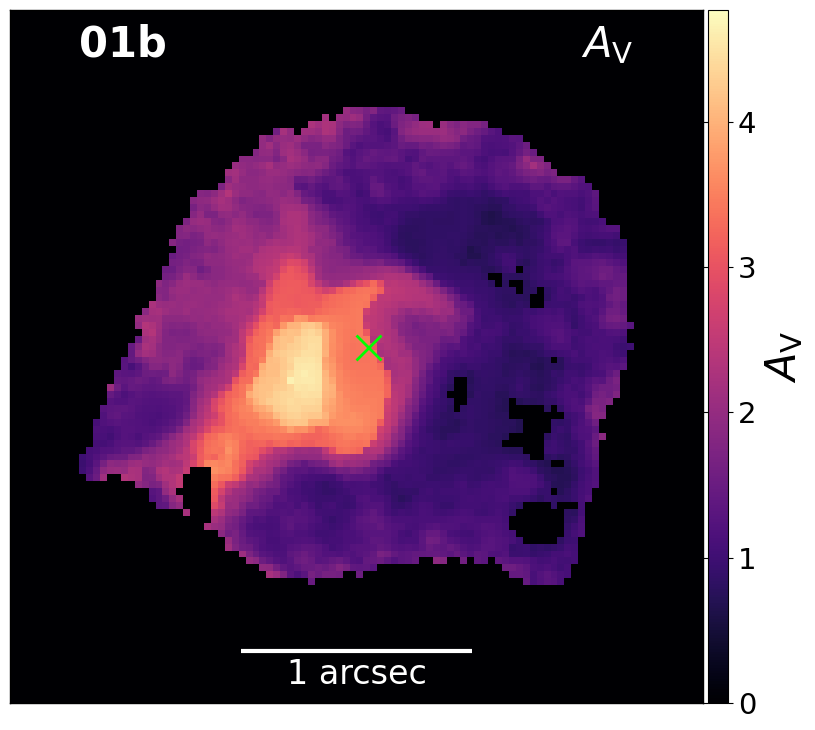}             \includegraphics[width=0.3\textwidth]{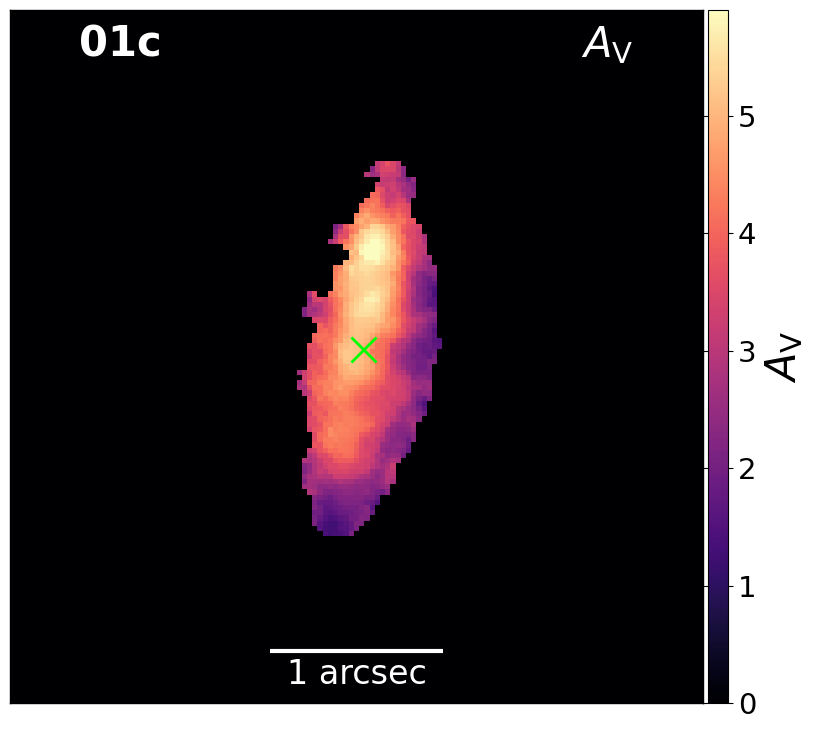}\\
\includegraphics[width=0.3\textwidth]{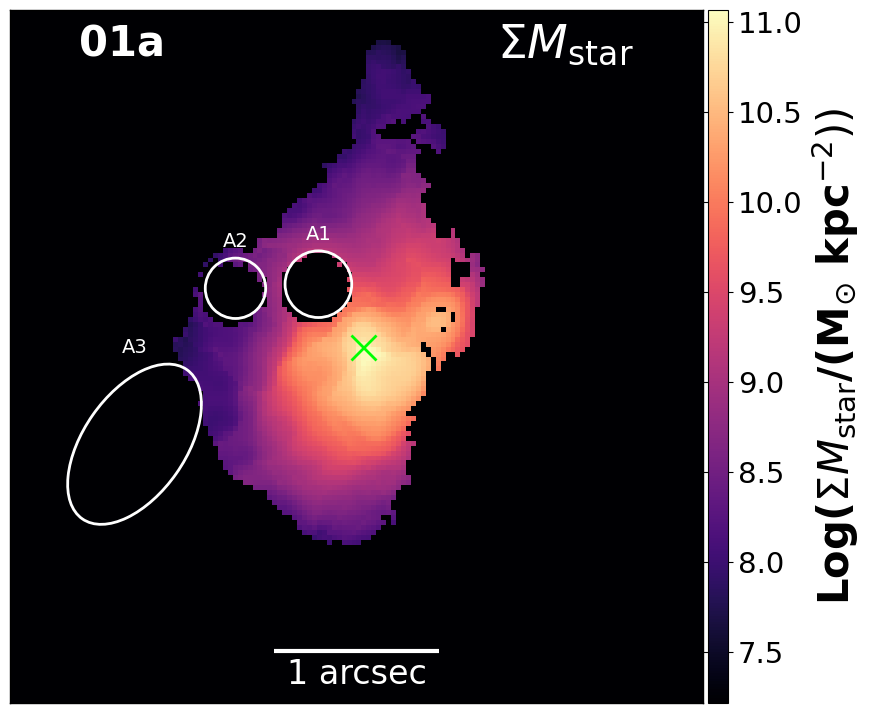}       \includegraphics[width=0.3\textwidth]{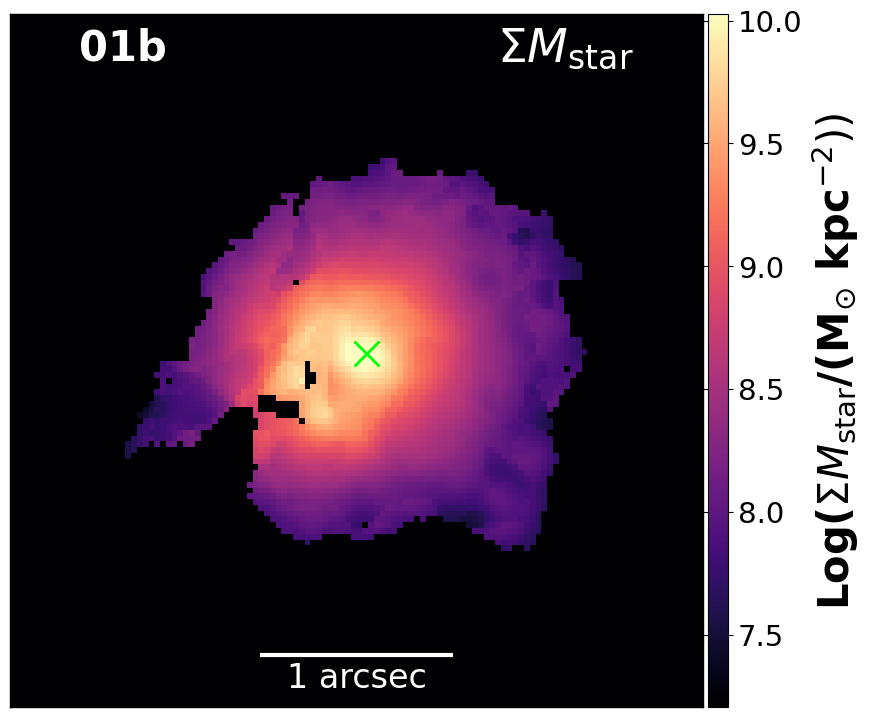}         \includegraphics[width=0.3\textwidth]{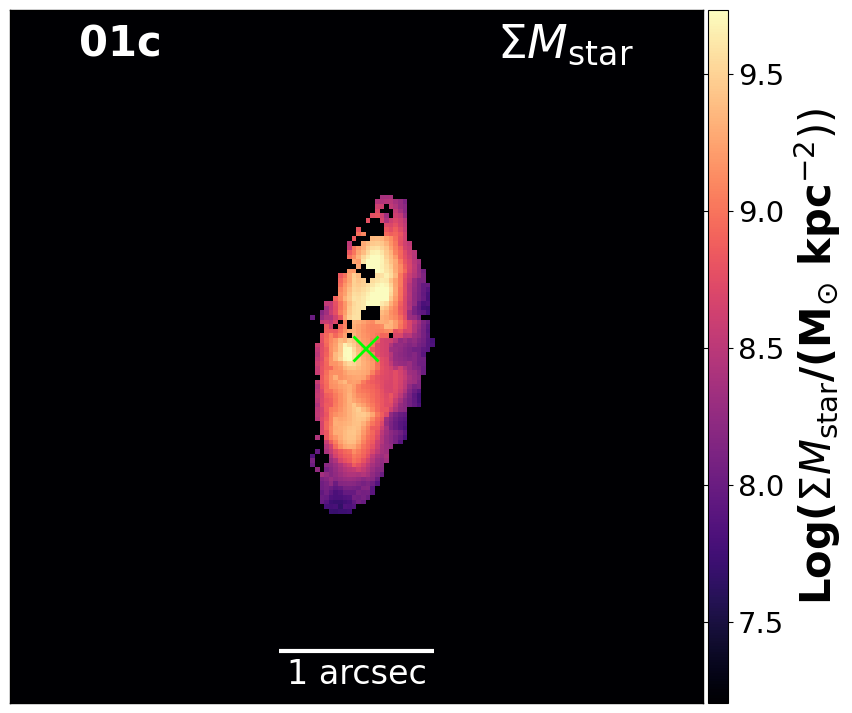}\\
\includegraphics[width=0.3\textwidth]{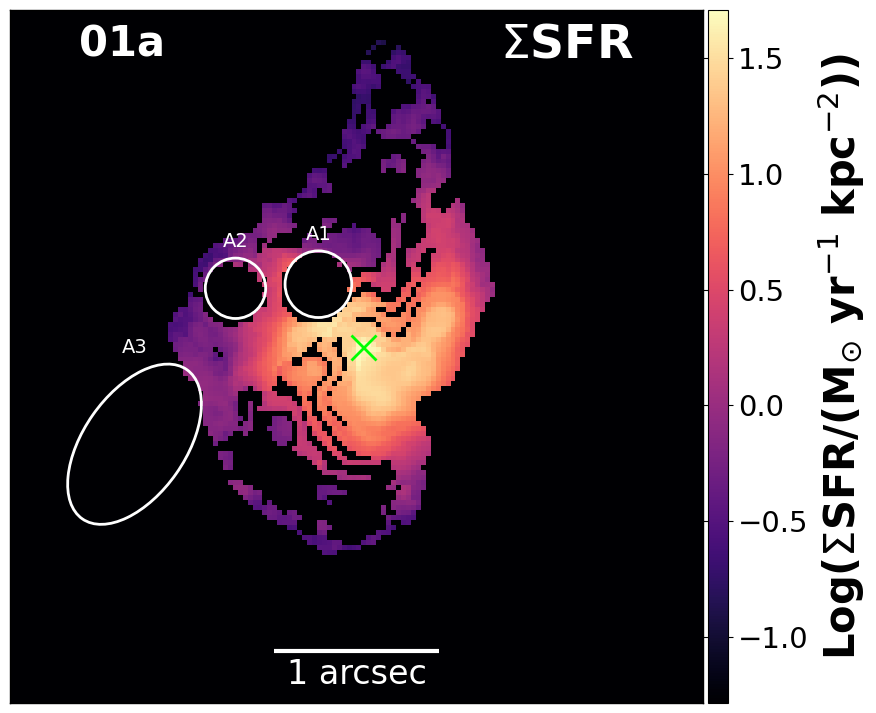}      \includegraphics[width=0.3\textwidth]{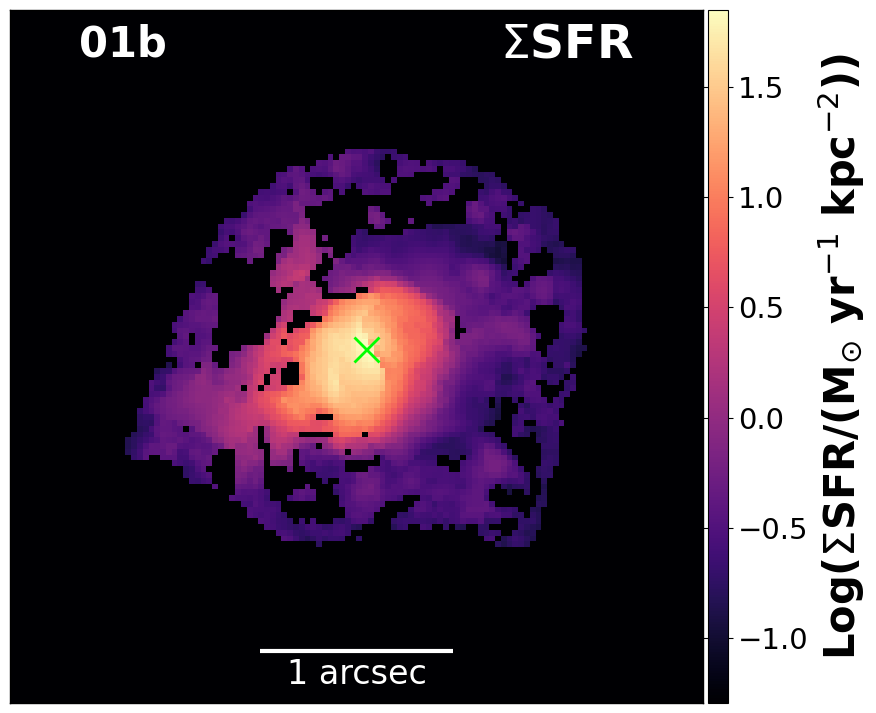}        \includegraphics[width=0.3\textwidth]{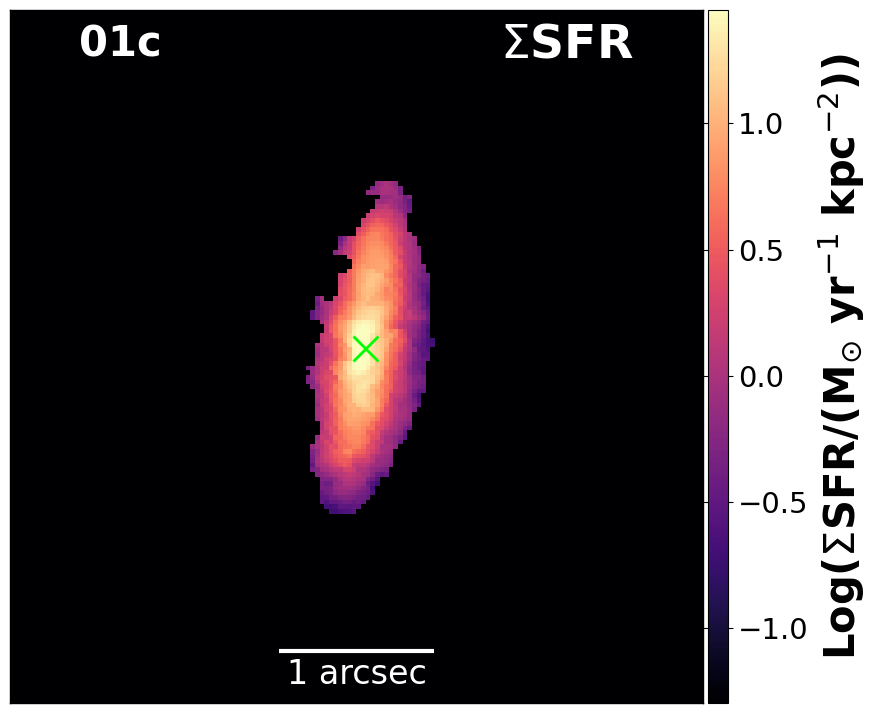}\\
\caption{\small Pixel-based maps of \texttt{CIGALE} bestfit parameters. Panels from top to bottom show \av, stellar mass surface density, and instantaneous SFR surface density\null.  From left to right, columns show source 01a, 01b, and 01c. In all maps north is up and east to the left, and a scale bar shows 1\arcsec\ or $\sim$8\,pkpc. The green cross marks the galaxy center.
Only pixels with parameters measured at $>2\sigma$ are shown in the \av\ and $\Sigma M_{\rm star}$ maps, and at $>1\sigma$ significance in the $\Sigma$SFR maps. This is because the SFRs are poorly constrained when the SEDs do not extend to long wavelengths (50\,$\mu$m--3\,mm), where dust emits. 
The white circles and ellipses in the left panel represent regions masking distinct galaxies from 01a.
}
\label{fig:resolved_maps} 
\end{figure*}

The selected targets have NIR sizes (based on the segmentation map described in Sect.~\ref{sec:jwst_obs}) ranging from 0\farcs8 to 3\arcsec\
corresponding to $\sim$7--25\,pkpc.  Their NIRCam images show asymmetric color gradients across them (Fig.~\ref{fig:targets}).  These color gradients can be driven by non uniform dust distributions or variations in relative stellar ages.  In order to analyze the origin of these color gradients, we built, for each source, a pixel-by-pixel SED using the eight NIRCam images and including only the pixels in the segmentation maps with 3$\sigma$ detections in at least four bands, but all pixels ended up having at least five band detections. The pixel flux density uncertainty was computed using eq.~\ref{eq:variance} and assuming $N_{\rm S}=1$. These pixel-based 0.9--4.4\,$\mu$m SEDs were then modeled with \texttt{CIGALE} using the same model components 
as for the integrated SEDs but without the radio component.  We thus obtained for each pixel with a sufficiently sampled SED a bestfit model and  associated physical parameters.  Fig.~\ref{fig:resolved_maps} shows resolved maps of the main physical parameters.
Only pixels with \av, and $M_{\rm star}$ measured at $>2\sigma$ ($M_{\rm star}>1.3\sigma$ for 01c), and with SFR at $>1\sigma$ are shown. This is because the SFRs are poorly constrained when the SEDs do not extend to long wavelengths (50\,$\mu$m--3\,mm). However, the smoothness of these maps indicate that the large uncertainties do not significantly affect the bestfit values, and that they can be used to qualitatively analyze the parameters' spatial variations.

\begin{figure*}[ht!]
\centering
\includegraphics[width=0.33\textwidth]{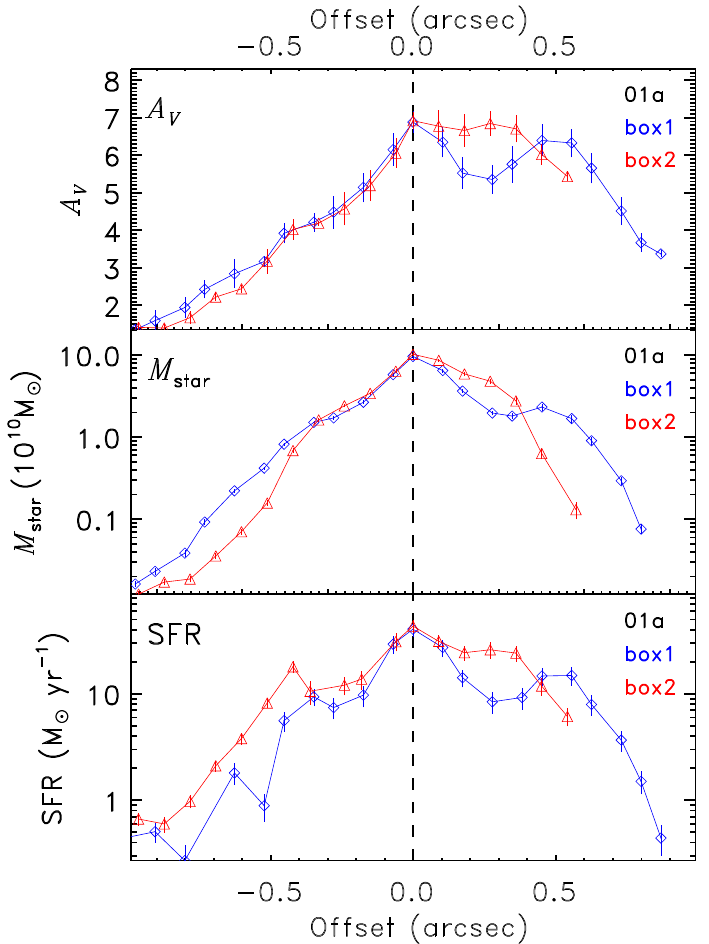}         \includegraphics[width=0.33\textwidth]{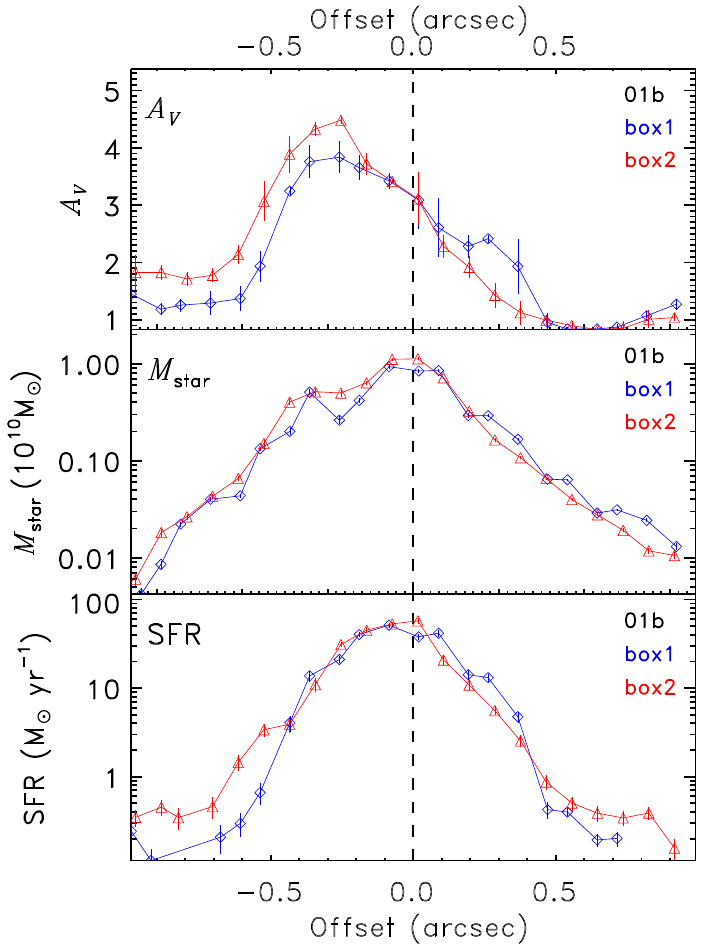}         \includegraphics[width=0.33\textwidth]{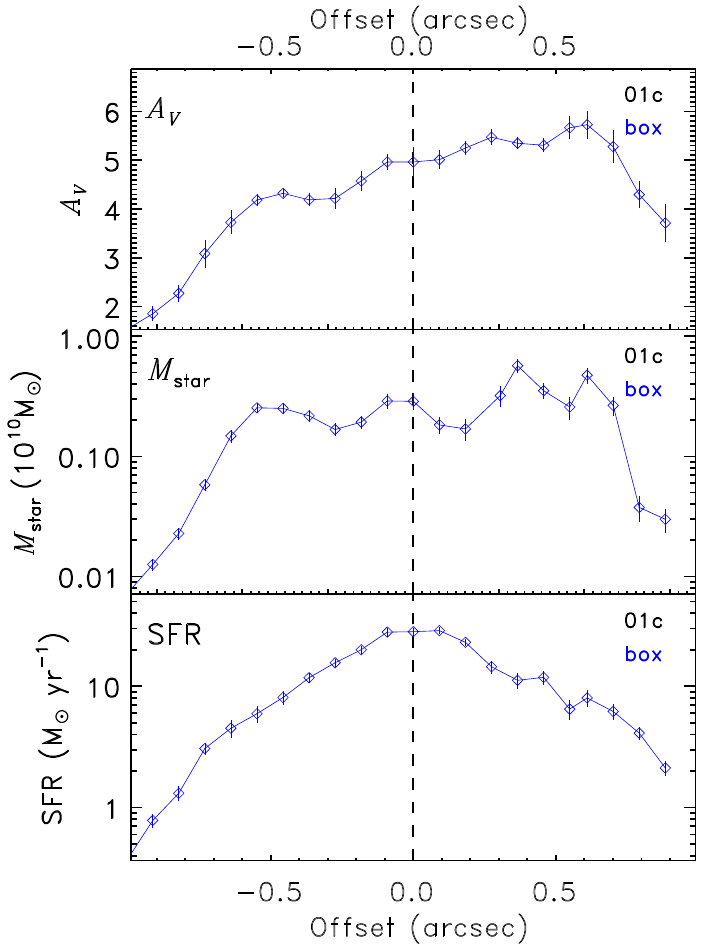}\\
\caption{\small Physical parameters in the boxes shown in Fig.~\ref{fig:targets}.  Columns show the three galaxies as labeled, and rows from top to bottom show mean \av,  $M_{\rm star}$, and instantaneous SFR.
The parameters were derived from the resolved maps in bins of 2--3 pixels within 0\farcs21 wide boxes spanning 2\arcsec, crossing the center and sampling the most dramatic color gradients across the whole galaxy. The abscissa  shows the distance of each bin from the galaxy center, indicated by the dashed vertical line, from  southeast to  northwest.  For 01a and 01b,  the two boxes with different orientations are indicated by blue diamonds and a blue line for box1 and red triangles and a red line for box2. (One box sufficed for 01c.) }
\label{fig:profiles} 
\end{figure*}

The resolved maps in Fig.~\ref{fig:resolved_maps} show that the derived physical parameters have an asymmetric distribution across each galaxy,  
and there are several substructures that are not symmetrically distributed with respect to the galaxy centers. 
To quantify these variations, we computed physical parameters in rectangular boxes, shown in Fig.~\ref{fig:targets}.
For 01a and 01b, two boxes are needed to capture variations at different azimuths. Fig.~\ref{fig:profiles} shows how parameters vary along the box. 
There are significant and asymmetric variations of dust extinction across all three galaxies, with the largest values measured away from the center in all of them. This is well illustrated in the top panels of Figs.~\ref{fig:resolved_maps} and~\ref{fig:profiles}. Extinction values \av$\gtrsim 7$ mag are measured in the central and western regions of 01a, \av$\sim 4$--5 mag to the southeast of the center in 01b, and \av\ gradually increases  to \av$\gtrsim 5$ mag from the south to the north in 01c. 
The $M_{\rm star}$ and SFR profiles peak instead at the center despite the asymmetric dust extinction distribution. The substructures visible in the NIRCam images show up as relatively small variations on top of quite smooth profiles.

\begin{figure}
\centering
\includegraphics[width=0.44\textwidth]{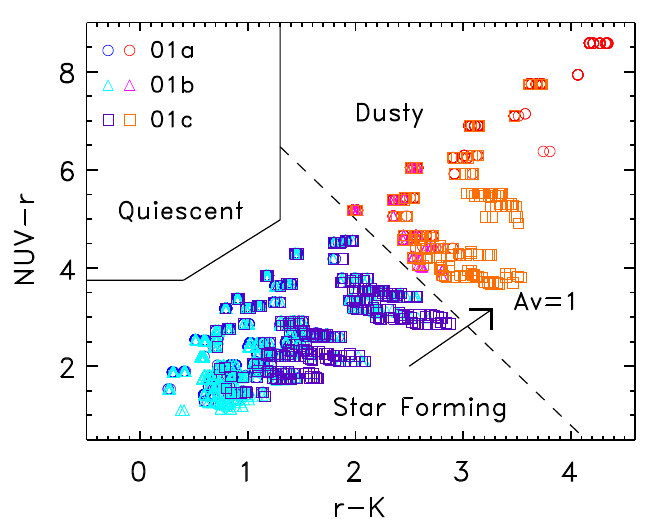}
\includegraphics[width=0.44\textwidth]{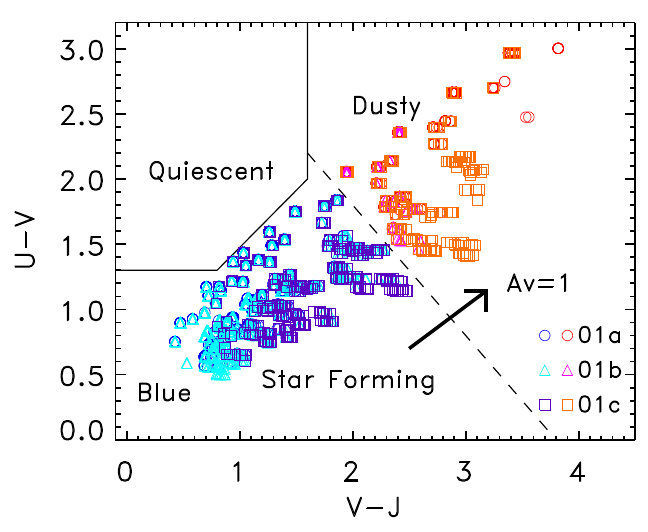}
\caption{\small Pixel-based rest-frame NUVrK, and UVJ diagrams \citep{williams09,arnouts13} of 01a (open circles), 01b (open triangles), and 01c (open squares). In each diagram the regions where quiescent and star-forming systems lie are outlined by a solid black line and labeled. The diagonal dashed line represents the criterion to identify the highly extincted pixels (red-shaded symbols), that is, NUV$-$r$>$9.2$-$2.1(r$-$K) or U$-$V$>$3.8$-$(V$-$J). The solid black arrow shows the change in colors in the case of \av=1 extinction.}
\label{fig:UVJ_NUVrK_diagrams} 
\end{figure}

\begin{figure*}
\centering
\includegraphics[width=0.3\textwidth]{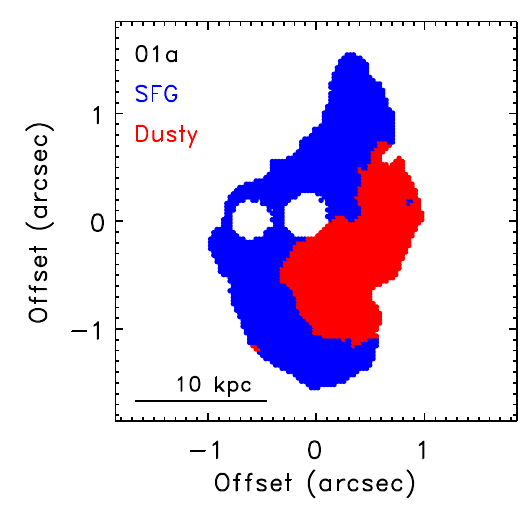}
\includegraphics[width=0.3\textwidth]{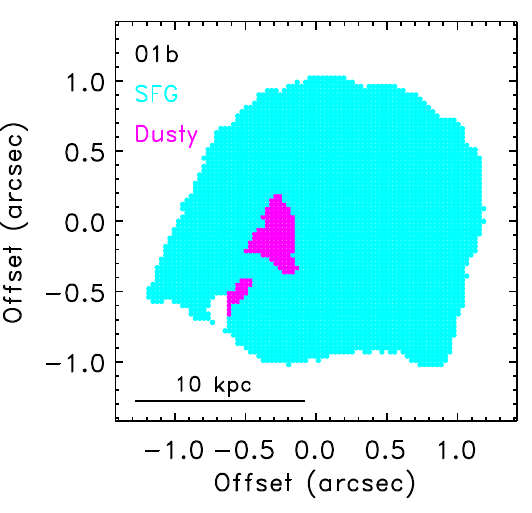}
\includegraphics[width=0.3\textwidth]{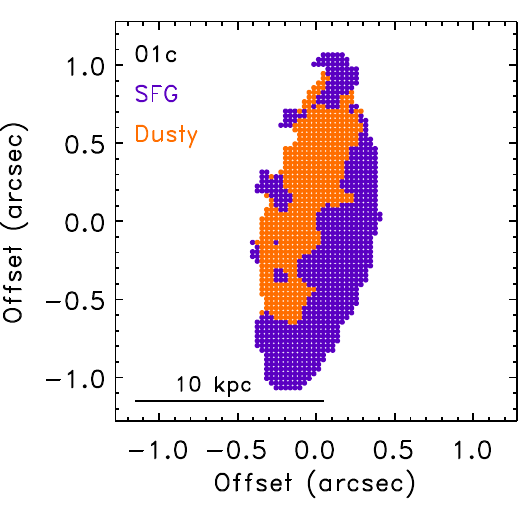}
\caption{\small Resolved maps of the three CO sources, 01a ({\it left panel}), 01b ({\it middle panel}), and 01c ({\it right panel}). Colors indicate the classification of each pixel based on the NUV$rK$ and $UVJ$ diagrams \citep{williams09,arnouts13}. There are no pixels classified as quiescent. Those classified as star-forming are shown in blue, cyan, or purple, and those classified as star-forming with high extinction---NUV$-r>9.2-2.1(r-K)$ or $U-V>3.8-(V-J)$---are in red, magenta, or orange.}
\label{fig:color_maps} 
\end{figure*}

Resolved color maps can identify regions within a galaxy experiencing different activity levels (e.g., quiescent, star-forming, or extremely dusty and star-forming). Two diagnostic diagrams,  rest-frame NUV$-r$ vs.\ $r-K$ and $U-V$ vs.\ $V-J$ \citep{williams09,
arnouts13},   are widely used to separate quiescent galaxies from SFGs up to $z{\lesssim}$5 \citep{shahidi20} and help distinguish dust reddening from age \citep{miller22,miller23}.  To derive the rest-frame magnitudes necessary to compute these colors, we convolved the filter transmission curves with the pixel-based bestfit models obtained using \texttt{CIGALE}\null.  All pixels in the three sources fall in the star-forming locus.  A subset of pixels---35\% in 01a, 5\% in 01b, and 38\% in 01c---have red colors, that is, $(NUV-r)>9.2-2.1(r-K)$ and $(U-V)>3.8-(V-J)$, consistent with being highly extincted. The position of each pixel in the three galaxies is shown in Fig.~\ref{fig:UVJ_NUVrK_diagrams}. Fig.~\ref{fig:color_maps} shows the resolved maps with the color-based classification. Based on this analysis, we do not identify any quiescent region within the three galaxies. As previously pointed out, the most obscured regions in our sources are not always in the center but are preferentially on one side of each galaxy \citep[see similar cases in ][]{smail23,kamieneski24b}. Furthermore, their distribution covers wide, contiguous regions extending over several kiloparsecs. 


Previous studies have reported that color gradients in star-forming galaxies at $0.5<z<2$ are mostly due to higher dust extinction, which is usually higher in their center, and in highly inclined galaxies \citep{liu16,liu17,miller22,miller23}.  This interpretation is supported by the  color--color diagrams, such as the $UVJ$ diagram \citep{wang17}, by the Balmer decrement \citep{nelson16}, and by different dust and stellar mass distributions \citep{tadaki20,gomez22}.  The color gradients in our galaxies can be also attributed to dust reddening, but the highest extinction values are not always in the center. 
Furthermore, based on the inclination of our sources, we can claim that the highest extinction values do not necessarily require a high inclination. For example, the extinction is higher in the face-on system 01a than in the edge-on galaxy 01c. Similar conclusions were drawn from other studies of optically faint sources \citep{gomez23,lorenz23}.  This implies that the extinction in some high-$z$ SFGs may be due to dust in dense star-forming regions rather than to dust in the diffuse interstellar medium (ISM). 

\section{Morphological analysis}\label{sec:morph}

To further investigate the origin of the color gradients in our galaxies and compare their properties with other SFGs and  \jwst\ red sources, we performed a detailed structural study of their stellar component \citep[e.g.,][]{gillman23,kokorev23,smail23,price23,nelson23}.
Fig.~\ref{fig:cutouts} shows the NIRCam single-band images of the three selected galaxies.  The galaxies' appearances  change significantly as a function of the wavelength.  In the rest-frame UV (F090W, F115W), all three galaxies are extremely faint. 
In the 01a rest-UV images, the light from three distinct sources (App.~\ref{app:01a_fgd_srcs}) in the vicinity dominate. The galaxy 01b is visible but much fainter than at longer wavelengths, and 01c is invisible in F090W and barely hinted at in F115W\null. 
In the rest-frame visible (F150W--F277W), all galaxies are much brighter and are clumpy. 
Source 01a shows multiple clumps, extended tails to the south and to the north (the latter is more visible in Fig.~\ref{fig:targets}), and two substructures that mimic spiral arms appear at $\lambda\ge2.7$~$\mu$m ($\lambda^{\rm rest}\ge0.78$~$\mu$m). 
Overall, 01a resembles a disturbed dusty spiral. The appearance of 01b is consistent with a face-on spiral galaxy, but at short wavelengths only the center and one spiral arm are visible.  Source 01c is an edge-on disk with several clumps along the disk major axis.  

Because of the dust extinction, a visible-light or single-band study of these galaxies would be affected by serious observational biases.  The best tracer of the stellar component and the least biased toward more massive stars is the F444W band (rest-frame NIR $\sim$1.3\,$\mu$m).  Nebular emission does not affect the analysis because the strongest emission lines at the sources' redshifts fall between the selected NIRCam filters.  Therefore the observed images are tracers of stellar radiation.

To characterize the galaxies' morphology in a quantitative way and compare them with other galaxies, we employed two codes that model surface brightness and return morphological parameters commonly used in the literature. The two codes are 
\texttt{statmorph}\footnote{https://statmorph.readthedocs.io/en/latest/}
 \citep{rodriguez19} and \textsf{Galfit} \citep{peng02,peng10}.  We used the WebbPSF
software \citep{perrin14} to generate F444W PSFs on the same
30\,mas pixel scale as our image mosaics.  Image cutouts with sizes 200$\times$200 pixel
(6\arcsec$\times$6\arcsec)  are 
large enough to contain all of the light from the galaxy of interest and to
characterize the noise.  We ran \texttt{statmorph} only on the F444W
cutouts using the segmentation map described in
Sect.~\ref{sec:jwst_obs}.  The code returns the half-light radius, the S\'ersic index ($n$), and the quantitative
morphological indicators concentration ($C$),  asymmetry ($A$), and  clumpiness ($S$) \citep[CAS;][]{abraham03,lotz08a,lotz08b,conselice14} plus Gini ($G$) and $M_{\rm 20}$. 
Higher CAS values indicate more concentrated,
asymmetric, and clumpier light profiles, respectively.
The Gini parameter defines the light distribution across the pixels.
$G = 1$ means that the light is concentrated in a single pixel, while a uniform surface brightness gives $G = 0$. The $M_{\rm 20}$ parameter measures the moment of a galaxy brightest regions containing 20\% of the total flux, normalised by the total light moment for all pixels. 
Negative $M_{\rm 20}$ values indicate a high concentration of light, although not necessarily at the center. More detailed descriptions of each morphological parameter were given by
\citet{lotz04} and \citet{snyder15}.

We ran \textsf{galfit} on all seven NIRCam wide-band images, fixing the center and the position angle (PA) of all objects to the F444W values.  In the case of 01a, we fit simultaneously 01a and the two bright nearby objects A1 and A3 to model their relative contributions in the overlapping regions. Object A2 is not included in the fit, because the code was not able to model it. Since A2 is fainter than all the other objects and located in the periphery of 01a, its presence should not affect the fit of 01a, and indeed it is present in the residual map.  The remaining galaxies, as well as those around 01b and 01c, were masked.  Both codes ran successfully and returned robust measurements (based on the quality
\textsf{flag} for \texttt{statmorph} and on the $\chi^2$ for
\textsf{galfit}). The uncertainty on the \texttt{statmorph} values were estimated by running \textsf{statmorph} 100 times, each time adding random noise to the cutout of magnitude equal to the original rms noise measured in the cutout. The final
properties (and uncertainty) for each galaxy are defined as the median (and standard deviation) of these 100 iterations.  The measured morphological parameters derived with \textsf{statmorph} are listed in Table~\ref{tab:statmorph_params}, and the reference images are shown in Fig.~\ref{fig:statmorph}. The \textsf{galfit} input images, bestfit model and residuals are shown in Fig.~\ref{fig:galfit}. 

\begin{table*}
\caption{\label{tab:statmorph_params}F444W \texttt{statmorph} morphological parameters.}
\centering 
\renewcommand{\arraystretch}{1.4}
\begin{tabular}{c c c c c c c c c} 
\hline\hline
Source & $R^{\rm half}$[kpc]&  $n$   &   $C$     &  $A$    &  $S$      & Gini ($G$)   &$M_{\rm 20}$&  $A_s$  \\
\hline
  01a  &   3.51$\pm$0.01             &  0.64$\pm$0.01  &   2.551$\pm$0.001  &  0.214$\pm$0.001 &   0.043$\pm$0.001 &   0.516$\pm$0.001 &   $-$1.60$\pm$0.006  &  0.108$\pm$0.004  \\
  01b  &   2.24$\pm$0.01             &  1.20$\pm$0.01  &   3.086$\pm$0.002  &  0.112$\pm$0.002 &   0.003$\pm$0.007 &   0.533$\pm$0.001 &   $-$1.88$\pm$0.003  &  0.011$\pm$0.0003  \\
  01c  &   4.46$\pm$0.01             &  0.32$\pm$0.01  &   2.674$\pm$0.004  &  0.113$\pm$0.006 &   0.032$\pm$0.004 &   0.460$\pm$0.002 &   $-$1.40$\pm$0.008  &  0.004$\pm$0.001  \\
\hline   
\end{tabular}\\
\tablefoot{Morphological parameters derived from the F444W image using
\texttt{statmorph}, $R^{\rm half}_{\rm circ}$: half-light radius that contains half
of the light emitted by the galaxy for a circular aperture, $n$: S\'ersic
index, $C$: concentration, $A$: asymmetry, $S$: clumpiness, Gini: Gini coefficient,
$M_{\rm 20}$: spatial moment of the brightest quintile of pixel flux values,
relative to the total moment, and $A_s$: shape asymmetry.
}
\end{table*}  

The \texttt{statmorph} parameters can be used to classify galaxies as mergers, ellipticals, or disk galaxies  \citep{lotz08b,conselice03,bershady00}. The comparison between the \texttt{statmorph} parameters and the classification regions defined by \citet{lotz08b}, \citet{conselice03}, and \citet{bershady00} is shown in Sect.~\ref{app:morph_analysis}. All in all, the morphological parameters suggest the three sources are late-type galaxies with no obvious signs of mergers ($A<0.35$). The parameter values are consistent with those obtained for a large SMG sample using \jwst\ images in the rest-frame NIR \citep{gillman23}. A recently introduced morphological indicator, sensitive to faint features in a galaxy's outer region, is the shape asymmetry $A_s$ \citep{pawlik16}.  Values of $A_s>0.2$ imply strongly asymmetric and disturbed systems, $0.1<A_s<0.2$ are expected in mildly asymmetric and disturbed objects, and $A_s<0.1$ indicates symmetric and undisturbed systems.  Sources 01b and 01c have $A_s$ values well below the 0.1 threshold, implying that they are undisturbed. Source 01a has $A_s=0.19\pm 0.01$ and would be considered mildly disturbed.

\begin{figure}[ht!]
\centering
\includegraphics[width=0.48\textwidth]{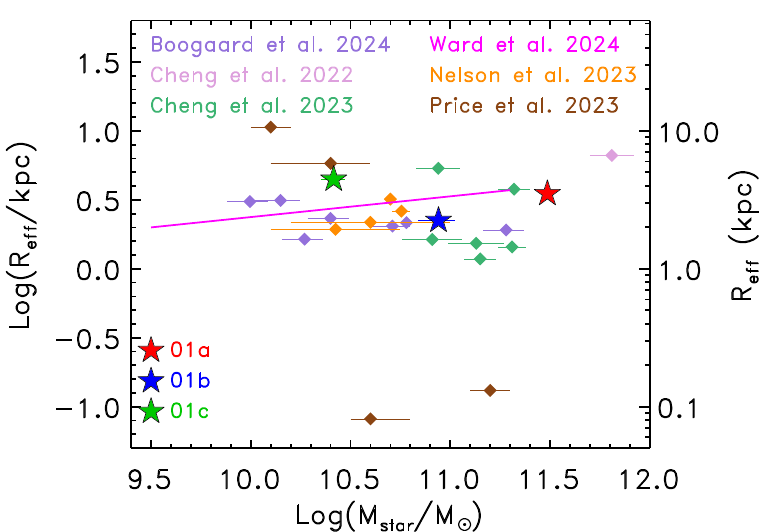}
\caption{\small Stellar effective (half-light) radius in F444W as a function of the stellar mass.  Filled stars show CO emitters identified by color as indicated in the legend.  Filled diamonds show a variety of sources at $2<z<3$  (lilac: ALMA-selected galaxies from \citealt{boogaard24}; pink: SMGs
in the SMACS J0723$-$7327 field from \citealt{cheng22}; sea green: some of the SMGs in PEARLS from \citealt{cheng23}; orange: HST-dark galaxies from \citealt{nelson23}; and brown: SMGs in the Abell\,2744 field from \citealt{price23}).  The magenta line shows the average relation derived for a sample of SFGs at $z=2.5$ by \citet{ward24}.}
\label{fig:size_mass} 
\end{figure}

\begin{figure}[h!]
\centering
\includegraphics[width=0.48\textwidth]{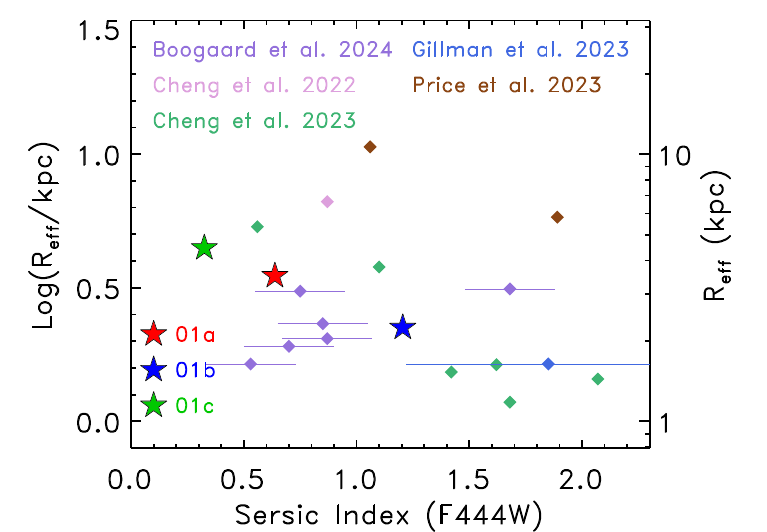}
\caption{\small Stellar effective (half-light) radius in F444W as a function of the S\'ersic index.  Filled stars show our three CO emitters  identified by color as indicated in the legend. Filled diamonds show sources at $2<z<3$ from the literature (lilac: ALMA-selected galaxies from \citealt{boogaard24}; pink: SMGs in the SMACS J0723$-$7327 field from \citealt{cheng22}; sea green: some of the SMGs in PEARLS from \citealt{cheng23}; royal blue: median values for the SMG sample from \citealt{gillman23} and brown: SMGs in the Abell 2744 field from \citealt{price23}).}
\label{fig:size_n} 
\end{figure}

Fig.~\ref{fig:size_mass} shows the effective radii for the three galaxies.  The CO emitters have effective radii ranging from 2.0 to 4.4\,kpc,  consistent with those of other sources at similar redshift  and with the mass--size relation of SFGs at $z=2.5$ \citep{ward24}. Fig.~\ref{fig:size_n} shows the comparison with S\'ersic index. All sources share similar stellar profiles with S\'ersic indices $n\sim1$, as expected for a disk galaxy, and relatively compact sizes, $R_{\rm eff}\sim1$--10\,kpc.
In summary, the stellar light of the CO emitters is well described by an exponential disk with size consistent with other SFGs at similar redshifts, and no signs of on-going major merging activity. Nonetheless, substructures such as clumps and arcs appear in the residual images. While these indicate disturbances, the substructures are not captured by the global morphological parameters, implying that their contribution to the light profile is small. The low values of the clumpiness parameter $S$ also imply that these substructures contain a small fraction of the total light \citep[][]{conselice11}. On the other hand, some studies find that clumpiness is low (consistent with local ellipticals) when derived from rest-frame NIR images \citep{baes20} or in high redshift galaxies \citep{gillman23} because of resolution effects. 

\begin{figure}[h!]
  \centering
 \includegraphics[width=\linewidth]{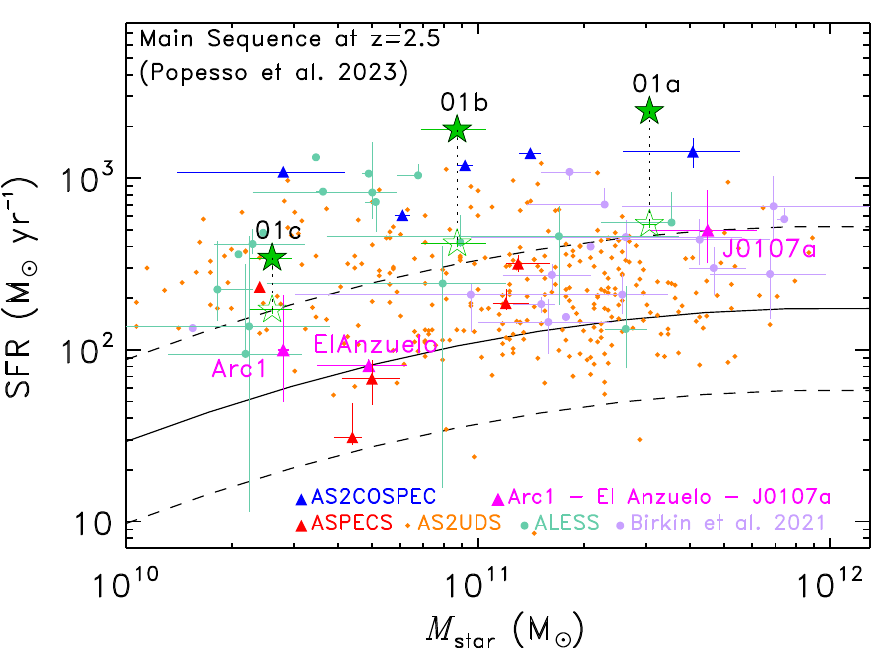}
 \caption{SFR as a function of the stellar mass. Filled green stars show the three selected targets. Comparison DSFGs from the literature include: the ASPECS sample of CO emitters from the Hubble Ultra Deep Field \citep[red triangles;][]{aravena19}, the AS2COSPEC sample of the brightest ($S_{\rm 870\,\mu m}>$12.4\,mJy) submm sources in the COSMOS field \citep[blue triangles;][]{liao24}; and the AS2UDS \citep[small orange diamonds;][]{as2uds20} and the ALESS \citep[small green circles;][]{danielson17} flux-limited SMG samples ($S_{\rm 850-870\,\mu m}>3.6$ and~2\,mJy, respectively) with follow-up ALMA observations.  CO observations are available for a subset \citep[small purple circles;][]{birkin21} of these samples. Labeled magenta triangles indicate three DSFGs studied in detail by \jwst\null: ALMA J010748.3$-$173028 (aka J0107a) at $z=2.467$ \citep{huang23}, Arc~1 in the PLCK\,G165.7$+$67.0 field at $z=2.24$ \citep{frye24} and El~Anzuelo at
$z=2.3$ \citep{kamieneski23}. The solid and dashed black lines represent the main sequence of star formation at
$z=2.5$ and its scatter \citep{popesso23}.}
   \label{fig:MS}
\end{figure}
\section{Star formation properties}\label{sec:SF_mode}

The SFRs of SFGs correlate with stellar mass according to the so-called main sequence (MS) relation. We refer to \citet{popesso23} for a recent compilation. This relation varies with redshift: for a given galaxy mass, the SFR is higher at higher redshift. Galaxies with SFRs that are $>$4 times higher than the MS relation are classified as starbursts \citep{rodighiero11}.
Fig.~\ref{fig:MS} shows the SFRs of the CO sources with respect to the MS relation. 

All three CO sources have SFR $>6$ times the MS value (Table~\ref{tab:cigale_params}),
implying that they are undergoing a starburst phase.  For two galaxies, the ratio is $>$4 even  considering only SFR$_{\rm 100\,Myr}$, and it is $\sim$3 in 01c. This suggests that the starburst phase might have lasted at least 100\,Myr, consistent with the model predicted  duration \citep[e.g.,][]{hopkins13}. The rates occupy the upper envelope of the SFR distribution at all masses and are among the highest known for DSFGs at $2<z<3$ (Fig.~\ref{fig:MS}).
The MS offsets are similar to those of the AS2COSPEC sources
\citep[median SFR $\sim$1400\,\msun\,yr$^{-1}$;][]{liao24} which by selection are the brightest ($S_{\rm 870\mu m}>$12.4\,mJy) submm sources in the COSMOS field. Interestingly, these sources are also characterized by large extinction values (\av=1.9--4.4) \citep{liao24}.  Finding four AS2COSPEC sources at $2<z<3$ with SFR$\sim$1000\,\msun\,yr$^{-1}$ in the ${\sim}1.4$\sqdeg\ COSMOS field implies a surface density of $\sim$3\,deg$^{-2}$. The AS2UDS survey \citep[flux-limited survey at $S_{\rm 850-870\mu m}>$3.6\,mJy;][]{as2uds20} found $\sim$10 DSFGs per square degree at $z=2$--3 \citep{as2uds20} at all SFRs, making the AS2COSPEC sources only $\sim$30\% of the total. Finding three DSFGs, of which two have SFR$>$1000\,\msun\,yr$^{-1}$, within 8\arcsec\ of each other suggests that our sources might reside in an overdensity (see Sect.~\ref{sec:environment}), or that such sources might be more common than previously thought but need deep NIR or mm observations to identify them.

\subsection{SFR assessment}\label{sec:sfr_assessment}

Because the \herschel\ beam size \citep[$FWHM=25$\arcsec\ at 350\,\micron;][]{griffin10} is larger than the source separations, the thermal dust emission from the CO emitters is poorly constrained. In addition to the CO sources, there might be other sources in the \herschel\ beam that  contribute to the total submm emission. The LOFAR 144\,MHz and NOEMA 3\,mm observations provide additional tracers of star formation activity, and hence of submm emission, in their $\sim$50\arcsec\ fields of view. No sources were detected in the NOEMA continuum map and in the LOFAR image except 01a and 01b, not even 01c, for which the bestfit model estimates $S_{\rm 250\,\mu m}\simeq 3$\,mJy and $S_{\rm 144\,MHz}\simeq 0.1$\,mJy. While other sources might be present and not detected, their contribution is likely  $\lesssim$3\% of the total submm and radio fluxes. Therefore most of the \herschel\ flux is due to 01a and 01b, and their summed SFRs might be overestimated by a few percent at most.

An obscuration-free SFR estimate independent of \herschel\ comes from the radio flux density, assuming it is due only to star formation.  The radio--SFR relations from the literature \citep{best23,gurkan18,smith21} yield SFR$_{\rm radio}$ estimates with up to 50\% differences (e.g., SFR$_{\rm radio}= 1507^{+264}_{-224}$\,\msun\,yr$^{-1}$ for 01a assuming the relation in \citet{best23}, and SFR$_{\rm radio}= 2327^{+232}_{-207}$\,\msun\,yr$^{-1}$ assuming that in \citet{gurkan18}). Furthermore, the SFR derived from the bestfit model depends on the timescale where it is computed.
To assess the agreement between the SFR derived from the bestfit model and the measured radio emission, we thus compare the observed (see Table~\ref{tab:lofar}) and the predicted radio flux densities. The bestfit models predict a 144\,MHz flux density of 1.0, 0.8, and 0.12\,mJy for 01a, 01b, and 01c, respectively. As illustrated in Fig.~\ref{fig:cigale_sed}, the predicted radio flux density agrees with the observed one for 01a, and 01c, while it is a factor of 2.5 smaller (8$\sigma$) than the observed one for 01b.

The excess of radio emission in 01b could be explained if the bestfit SED underestimated the SFR or if an AGN contributes to the radio flux.  The bright, red nucleus seen by NIRCam also suggests an AGN,
although there is no evidence of unresolved PSF like features in
its core as observed in other AGN candidates in \jwst/NIRCam images \citep{ortiz24}\null.  
The possibility that 01b hosts an AGN is of particular interest because of the frequent connection between AGN-driven radio activity and an overdense environment \citep{miley08} and several observational facts suggesting that 01b might reside in a proto-structure (see Sect.~\ref{sec:environment}). Despite the uncertainties in the radio-FIR relation, the comparison with the radio emission tends to confirm the large estimated SFRs and the reliability of the SED modeling.

\begin{figure}[h!]
\centering
 \includegraphics[width=0.475\textwidth]{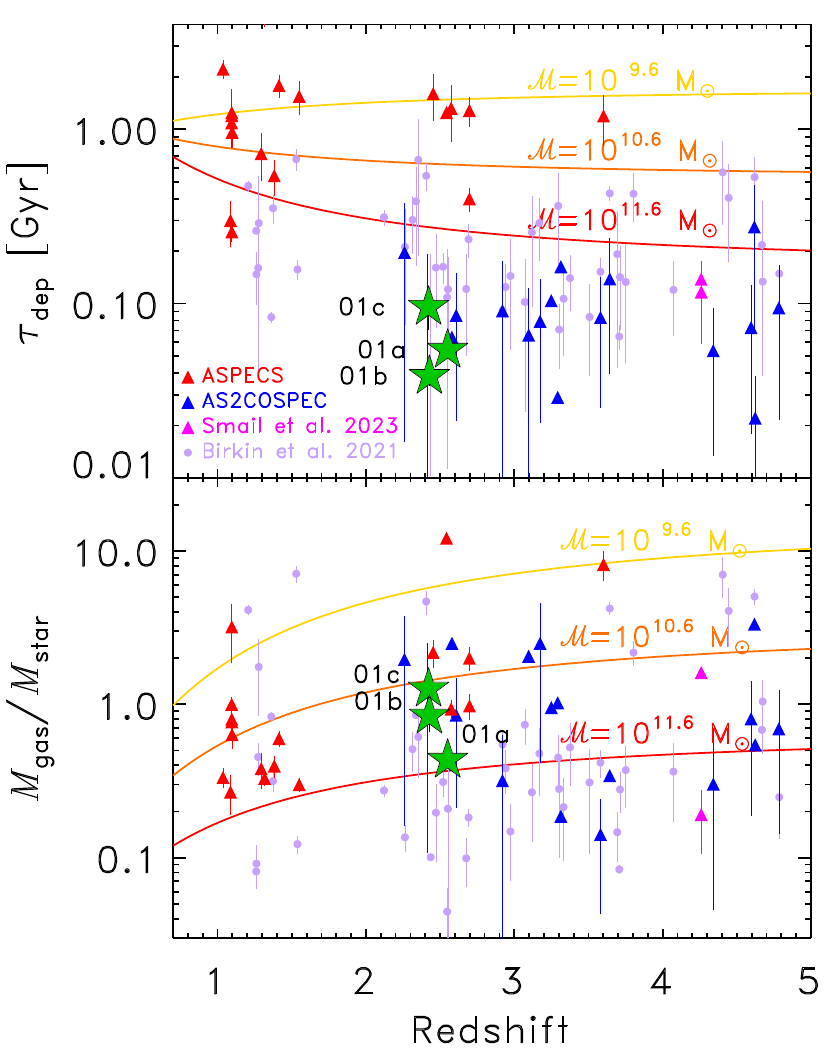}
 \caption{Molecular gas properties (gas depletion time in the top panel and gas-to-stellar mass ratio in the bottom panel) as a function of the redshift of the CO sources (filled green stars). Comparison samples are ASPECS \citep[red triangles:][]{aravena19}, AS2COSPEC \citep[blue triangles:][]{liao24}, PEARLS $z=4.6$ SMGs in Abell\,1489 \citep[magenta triangles:][]{smail23}, and other CO-detected SMGs \citep[light purple circles:][]{birkin21}. Solid lines show the relations derived by \citet{liu19} for SFGs with different stellar masses ($M_{\rm star}=10^{9.6}$, 10$^{10.6}$, and 10$^{11.6}$\msun\ in yellow, orange, and red, respectively).}   \label{fig:gas_properties}
\end{figure}

\subsection{Depletion times and gas-to-stellar mass ratios}
The NOEMA CO observations enable computations of the gas depletion time for the three sources. The depletion time is defined as the timescale required to exhaust the molecular gas mass with the current SFR (i.e., $\tau_{\rm depl}\equiv M_{\rm gas}$/SFR) and is equivalent to the inverse of the SFE.
The gas mass was estimated from the CO luminosity \LpCOone\ as 
$M_{\rm gas}=1.36 \alpha_{\rm CO}$\LpCOone, where $\alpha_{\rm CO}$ is the CO-to-H$_2$ factor in units of \msun\,pc$^{-2}$\,(K\,\kms)$^{-1}$, and the factor 1.36 takes into account the contribution of helium to the molecular gas mass \citep{bolatto13}.  The CO--H$_{\mathrm 2}$ conversion factor in galaxies above the MS is smaller than in normal SFGs \citep[e.g., eq.~2 of][]{castignani20}.  For starburst galaxies, a typically adopted value is  $\alpha_{\rm CO}\sim0.8$ \citep{casey14}. Adopting this value allows comparison with other SMGs in the literature. However, the correct $\alpha_{\rm CO}$ might differ from object to object because it depends on gas metallicity and molecular gas surface brightness \citep{narayanan12}. A different choice of $\alpha_{\rm CO}$ value could be that assumed for normal SFGs, that is, $\alpha_\mathrm{CO}$\,=\,3.5\,\msun\,pc$^{-2}$\,(K\,\kms)$^{-1}$ \citep{magdis17}. This choice would result in four times higher gas masses. The \LpCOone\ luminosity was derived from \LpCOthree\ assuming the brightness temperature ratio measured in SMGs: $r_{3,2}\equiv\LpCOthree/\LpCOone=0.63$ \citep{birkin21}. The estimated gas masses are listed in Table~\ref{tab:co_data}.
Fig.~\ref{fig:gas_properties} shows the gas-depletion times as a function of the redshift. 
Not surprisingly, the depletion times, such as those of the AS2COSPEC sample, are much shorter (implying higher SFEs) than those typically measured in SFGs of similar mass \citep{liu19}. Based on these values, our sources would exhaust all their gas and stop forming stars within $\Delta z\approx-0.1$ unless the gas is replenished.

Another important parameter for assessing whether an excess of cold gas leads to the large SFRs and SFEs of our sources is the molecular gas-to-stellar mass ratio ($M_{\rm gas}/M_{\rm star}$). This parameter is the fraction by which the stellar mass would grow if all the existing gas forms new stars. The estimated gas-to-stellar mass ratios are shown in the bottom panel of Fig.~\ref{fig:gas_properties}. The three sources have gas-to-stellar mass ratios consistent with the SFG scaling relation \citep{liu19}. Despite having gas-to-stellar mass ratios, stellar masses, and sizes (Fig.~\ref{fig:size_mass}) similar to those of normal SFGs, our sources are more efficient and powerful in forming stars. Consistent with other SMG studies \citep{liao24}, this suggests that what drives a galaxy to move above the MS must be greater SFE, not the amount of cold gas or stellar mass \citep[see also][for a similar study in galaxy mergers]{thorp22}. In Sect.~\ref{sec:discussion}, we discuss the possible mechanisms and conditions that might enhance the SFEs and SFRs in our sources.

\section{Dust properties}\label{sec:dust_properties}
\subsection{Dust extinction and morphology}

Our three CO sources are remarkable for their large extinction values---up to $A_V=7$ across kiloparsec-wide regions and $A_V=2.9$--3.8 on average for the whole systems (Table~\ref{tab:cigale_params}). 
An extinction $A_V=7$ corresponds to a molecular hydrogen column density $N_{\rm H_2}\simeq3\times 10^{21}$\,cm$^{-2}$ \citep{predehl95,nowak12,skalidis24}. Such large gas column densities are not typical of the diffuse ISM but of giant molecular clouds (GMCs) or of the molecular torus at the center of an AGN, both of which have sizes of hundreds of parsecs at most. Similarly high extinction values are usually confined to the central 1--2\,kpc of local, luminous IR galaxies \citep{mayya04,scoville15a}. Typical dusty SFGs at $2<z<3$ have  average extinction  \av$\sim$2.5 \citep{knudsen05,dacunha15,as2uds20}, although recently several systems with exceptionally high values have been found, e.g., the SMG 850.1 at $z=4.26$ with \av=5.1$\pm$0.2 \citep{smail23}, AS2COS0001.1 at $z=4.62$ with \av=5.83$\pm$0.14 \citep{liao24}, about 10\% of the SMGs studied by \citet{as2uds20} have \av$>$5 (and up to \av=7.7), and 5\% of \herschel-selected starburst galaxies at $0.5<z<0.9$ have \av=17--24 \citep{calabro19a}. When the extinction in SMGs is derived from the dust mass and a fixed gas-to-dust ratio, even higher values are found \citep[median \av$\simeq540^{+80}_{-40}$\,mag;][]{simpson17}, implying the presence of optically thick dust at submm wavelengths and no light escaping at NIR wavelengths \citep{papadopoulos10}. This is an intriguing possibility as it might explain the faint high-$J$ CO emission in 01a \citep{polletta22}. Optically thick dust would cause more absorption at higher frequencies (higher CO transitions) than at lower frequencies (lower CO transitions). The difference in absorption of the CO emission at different transitions might be further enhanced if the denser and more excited gas is mostly located in the optically thick regions, and the less excited gas is more extended \citep[see Fig.~9 in][]{papadopoulos10}. The presence of dust with such optical depths could have important repercussions on estimates of key properties such as stellar mass and SFR.

\begin{figure}[t]
\centering
\includegraphics[width=0.24\textwidth]{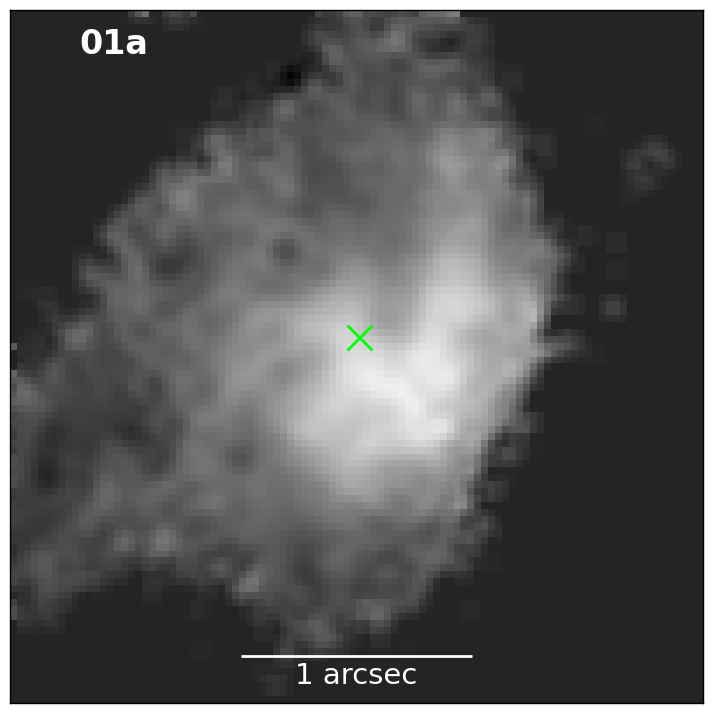}
\includegraphics[width=0.24\textwidth]{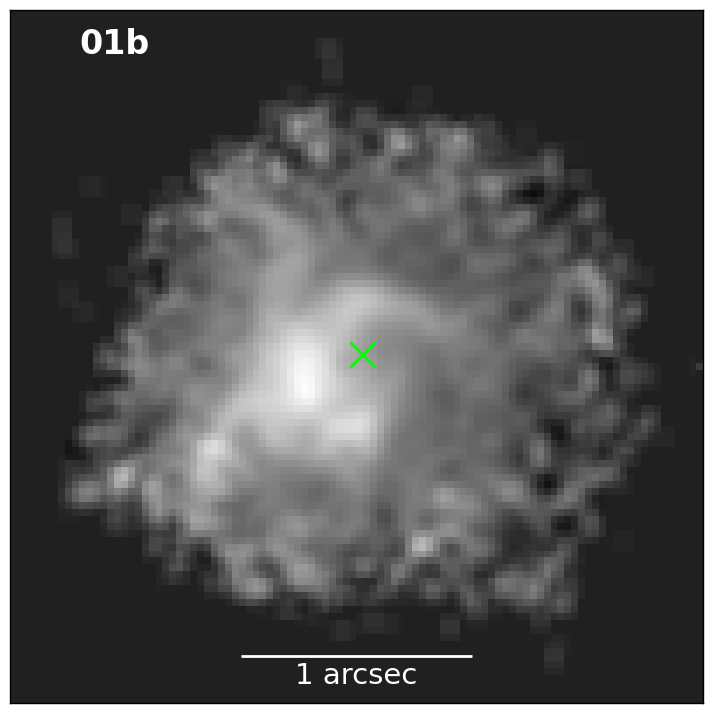}
\caption{\small 3\arcsec$\times$3\arcsec\ pseudo-extinction images of 01a (\textit{left}) and 01b (\textit{right}).  Lighter areas correspond to higher extinction, which was derived as $-2.5\times
\log(S_{\rm F277W}/S_{\rm F444W})$ and only for pixels with ${>}3\sigma$ detections in both bands. The green crosses mark the galaxy centers.}
\label{fig:ext_images} 
\end{figure}

Assuming that dust extinction is a good tracer of dust distribution (this requires the presence of an UV-visible light source behind or mixed in with the dust), we can infer the dust morphology from the extinction map. The CO sources' extinction maps (top panels of Fig.~\ref{fig:resolved_maps}) show the presence of substructures. 
These substructures are also perceptible in the SFR maps (bottom panels of Fig.~\ref{fig:resolved_maps}), in spite of those maps' irregular coverage. These similarities support the dust related origin of these structures and rule out the possibility that they might be caused by older stellar ages. As further validation of this interpretation, the stellar age maps of the three sources show only small variations that do not match those  of the extinction maps. The negligible variations in the stellar ages across each galaxy imply that the stellar populations have uniform intrinsic colors. Given constant intrinsic colors,  a more detailed image of dust attenuation can be obtained simply by dividing any two images. The most leverage is to use the longest available wavelength, F444W, and a shorter wavelengths  with sufficient signal across most of the galaxy. Fig.~\ref{fig:ext_images} shows the results using F277W\null.  (We do not show the results for 01c because, with the exception of some clumps along the major axis, its edge-on orientation makes the substructure details indistinguishable.) The substructures revealed in the extinction maps are clearly visible. They resemble spiral arms with significant bends, depicted by changes in their pitch angle, and with widths of 0\farcs12--0\farcs17 (equivalent to 1.0--1.4\,pkpc), consistent with the width of the spiral arms in nearby grand-design spiral galaxies such as the Whirlpool Galaxy \citep{marchuk24}. The arms contain clumps (with radii $\sim$0\farcs05--0\farcs09, equivalent to 0.40--0.75\,pkpc) and also straight sections (up to 0\farcs3--0\farcs4 or 2.5--3.3\,pkpc in length) with notable deflections ($\sim$50\deg--70\deg).  The sizes and shapes of these substructures are consistent with spiral arms, but their bends are indicative of disturbances, likely due to interactions.

\subsection{Dust attenuation curve}\label{sec:dust_att}

Attenuation corrections are critical for inferring key physical properties such as SFR, stellar mass, and dust content \citep{kriek13,salim20,robertson24,lofaro17,hamed21,boquien22,markov23}. Meaningful corrections require distinguishing between a galaxy's attenuation curve and the dust's extinction curve \citep{gordon97}.  The latter is directly related to the physical properties of the dust grains, such as their composition and size distribution  \citep{fitzpatrick99,nozawa16,hsu23}. The attenuation curve, defined as the change in the galaxy's flux due to the presence of dust, depends on the extinction curve but also on the spatial distributions of the stars and dust \citep{witt92}. At high redshifts, we can expect different extinction curves than observed in the local Universe because the processes affecting the formation, growth, and disruption of dust grains might differ. Likewise, extremely dusty galaxies might be characterized by ISM conditions resulting in atypical attenuation laws. \jwst\ is already providing new insights on the variety of attenuation curves by investigating systems at high redshift \citep{witstok23,markov23} and along different sightlines through resolved studies of nearby galaxies \citep{keel23,robertson24}. Here, we take advantage of the asymmetric dust extinction distribution of 01a and 01b to measure their dust attenuation curves.

\begin{figure}[t]
\centering
\includegraphics[width=0.24\textwidth]{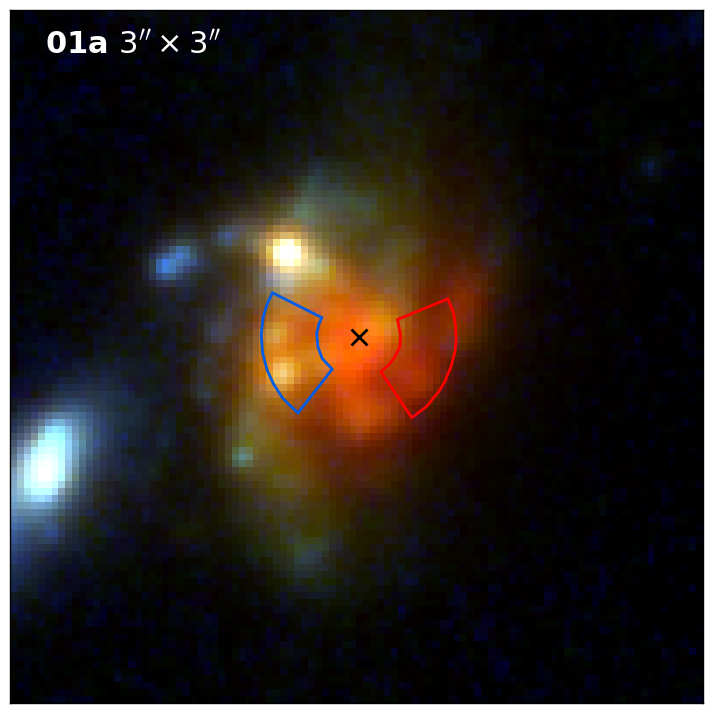}
\includegraphics[width=0.24\textwidth]{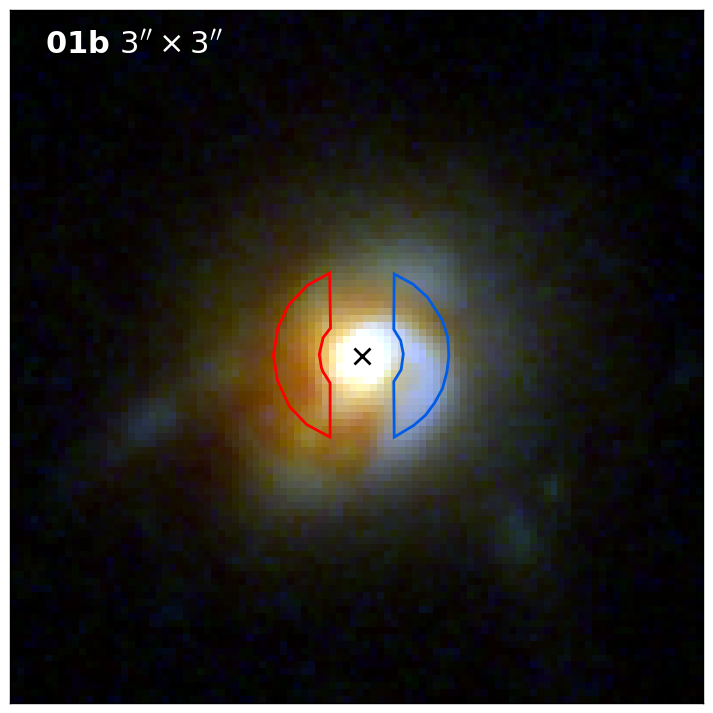}
\caption{\small Multicolor 3\arcsec$\times$3\arcsec\ (corresponding to 25\,pkpc$\times$25\,pkpc) \jwst/NIRCam images of 01a (\textit{left}) and 01b (\textit{right}). Colors are as in Fig.~3. The areas used to compute the dust-attenuation curve are outlined in blue for the less attenuated regions  and in red for the more attenuated.  In 01a, the ring has minimum radius of 0\farcs18 and maximum of 0\farcs42, and in 01b, it has minimum radius of 0\farcs18 and maximum of 0\farcs38. The center of each galaxy is indicated by a black cross. North is up and east to the left.}
\label{fig:ext_regions} 
\end{figure}

\begin{figure}[h!]
\centering
\includegraphics[width=0.48\textwidth]{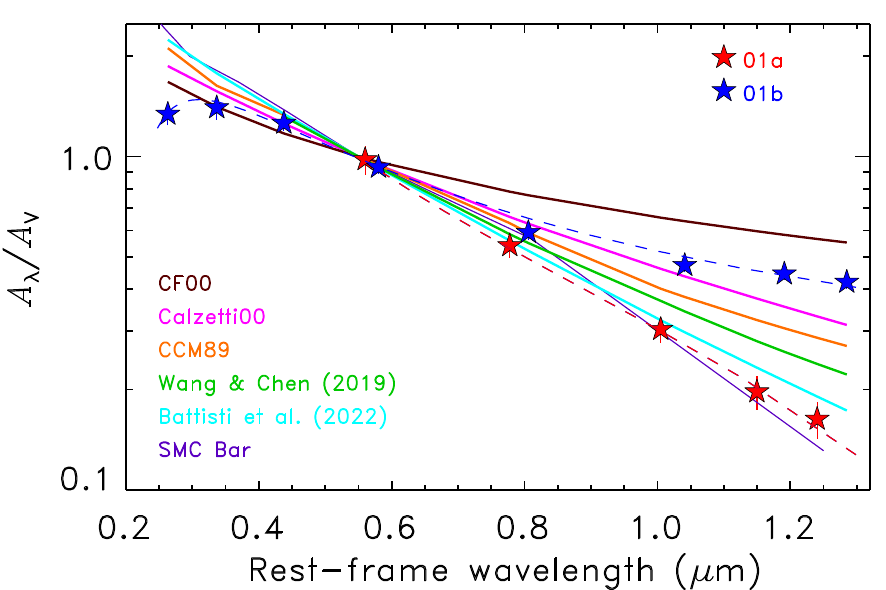}
\caption{\small Normalized dust attenuation values ${A}_{\lambda}/{A}_{V}$ for 01a (filled red stars) and for 01b (filled blue stars). Solid lines show various attenuation laws from the literature: the CF00 attenuation law in brown \citep{charlot00}; the law measured in the inner regions of starburst galaxies in magenta \citep[][Calzetti00]{calzetti00}; a Milky Way extinction law in orange \citep[][CCM89]{cardelli89}; an extinction law based on red clump stars in green \citep{wang19}; an attenuation law from a sample of galaxies at $0.75<z<1.5$ in cyan \citep{battisti22} and the extinction law measured in the Small Magellanic Cloud bar in purple \citep{gordon03}. The dashed red and blue lines represent the bestfit cubic polynomial to the 01a and 01b values, respectively.}
\label{fig:ext_curve} 
\end{figure}

To measure the attenuation curve, we adopted a method similar to the pair method \citep{stecher65} but using two regions of the same galaxy instead of two stars of the same spectral type. The attenuation curve can be directly derived from the ratio between more and less attenuated SEDs of two regions with the same intrinsic spectrum, provided the attenuation law is the same for the chosen pair of regions. If the underlying stellar populations, the dust mixtures, and the radiation fields are the same in the two paired regions, any SED difference between the two regions must come from different dust extinctions. In both 01a and 01b, the western and eastern sides exhibit exhibit different apparent colors (Fig.~\ref{fig:targets}) that resulted in different extinction values from the resolved SED fitting procedure (see top-left and top-middle panels of Fig.~\ref{fig:profiles} and top panels in Fig.~\ref{fig:resolved_maps}). In each galaxy, we selected two regions, one per side, as shown in Fig.~\ref{fig:ext_regions}. The two regions were selected at the same distance from the center of each galaxy and of approximately the same size. The attenuation values were derived as
\begin{equation}
A_{\lambda} = -2.5\times \log(S^{\rm red}/S^{\rm blue})\quad,
\end{equation}
where $S^{\rm red}$ is the flux density of the more reddened SED, and $S^{\rm blue}$ that of the less reddened SED \citep[following][]{battisti22} normalized by the region size.
Fig.~\ref{fig:ext_curve} shows the normalized attenuation curves of 01a and 01b derived for all NIRCam bands in all locations where there is ${>}5\sigma$ signal. The 01a curve is not constrained at rest-frame $\lambda<0.5\,\mu$m because the western region is undetected in the F090W, F115W, and F150W bands. 
In the visible--NIR ($\lambda_{\rm rest}>0.3~\mu$m), the dust attenuation curve can be expressed by a cubic polynomial \citep{battisti22}. The bestfit models for 01a and 01b are also shown in the figure.

The attenuation curves measured in 01a and in 01b differ significantly from each other. The 01b curve is flatter than most curves, while that of 01a is as steep as the extinction law measured in the bar of the Small Magellanic Cloud \citep{gordon03}. The observed diversity could be due to differences in the dust-to-star geometry, a varying grain size distribution, or a combination of both \citep{cardelli89,fitzpatrick99,kriek13,nozawa16,hsu23}.  Shallow attenuation curves, as derived for 01b, have been also measured in other highly extincted SFGs at $z\sim 1.5$--3.0 \citep{salmon16,lofaro17,hamed23}. The slope of the attenuation curve slope in these SFGs correlates with total extinction (with shallower curves in more-extincted SFGs), a result attributed to the effects of scattering, star--dust geometry, and dust grain size \citep{salmon16}. A shallow attenuation curve can be the result of a patchy dust distribution because the emerging visible--NIR light of a galaxy is dominated by the least attenuated stars \citep{witt92,witt00}. With 01a more extincted than 01b, it is surprising that 01a exhibits a steeper attenuation curve. This behavior might indicate that the assumptions made to derive the attenuation law are not valid or that the correlation between extinction and the slope of the attenuation law breaks in some specific cases.

As stated above, the attenuation curve depends on both the physical properties of the dust grains (i.e., composition and grain size) and the grains' spatial distribution with respect to the stars and the clumpiness of the ISM\null. Our sources exhibit extended regions with extremely high extinction values. It is therefore plausible that these regions might be characterized by dust properties and geometries that are different from those measured along the sightlines used for the ``standard'' laws. Notably, the semi-analytic galaxy-formation model SHARK \citep{lagos18} is capable of reproducing several observed properties of SMGs (e.g., SEDs, number counts, and redshift distributions) thanks to the adoption of attenuation curves that scale with the dust surface density  \citep{lagos19}. Furthermore, \citet{donnan24} showed that differential extinction is necessary to reproduce the spectra of highly obscured galaxies, where neither simple screens nor uniformly mixed dust distributions fit the data. To interpret our attenuation curves, it would be useful to have predictions from simulations for a variety of sightlines, dust grain distributions, and dust formation and evolution models \citep[see e.g.,][]{cochrane24}.

Adopting a uniform ``standard'' law in retrieving a galaxy's stellar properties might bias the estimates of stellar mass and SFR\null. A steeper attenuation curve at longer wavelengths translates into smaller stellar masses compared to standard laws and vice versa. The attenuation curve might vary within the same galaxy if it depends on the extinction, for example, in the diffuse ISM versus molecular clouds. High redshift DSFGs, with their heavily obscured sightlines,  offer the opportunity to explore such a relation.  DSFGs are important contributors to the cosmic SFR and are thought to be the precursors of the most massive local galaxies, and therefore properly quantifying their dust attenuation and overall properties is fundamental to understand galaxy evolution. 

\subsection{Dust-to-stellar mass ratio}\label{sec:dsr}
The dust-to-stellar mass ratio (DSR) might provide indications on a galaxy's evolutionary stage and on  dust-destruction processes \citep{calura17,popping17,donevski20}.  
\begin{figure}[t]
  \centering
 \includegraphics[width=0.5\textwidth]{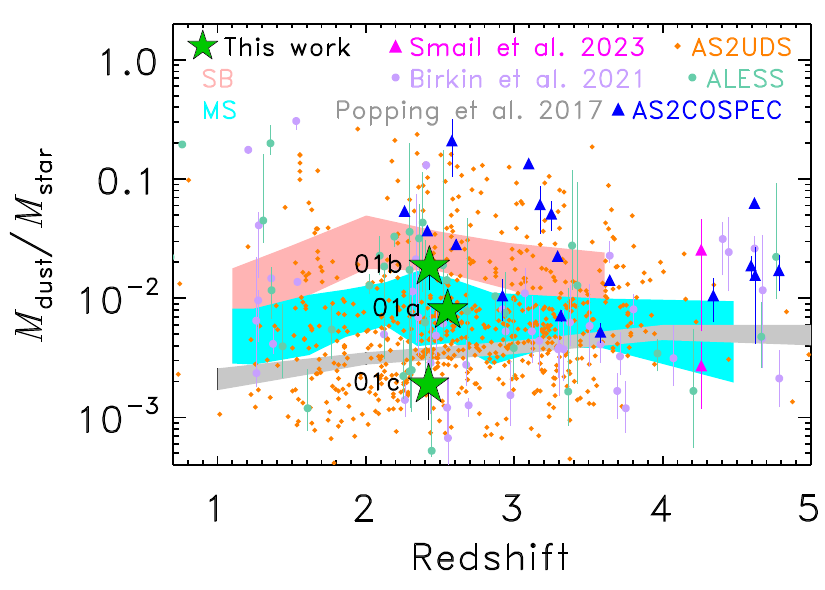}
 \caption{$M_{\rm dust}/M_{\rm star}$ versus redshift  of our sources (green stars), and of various SMG samples from the literature (AS2COSPEC \citep[blue triangles;][]{liao24}, PEARLS SMGs in Abell\,1489 \citep[magenta triangles;][]{smail23}, AS2UDS \citep[orange diamonds;][]{as2uds20}, ALESS \citep[teal circles;][]{danielson17}, and the ALMA SMGs from \citet{birkin21} (light purple circles). The average values and 68\% probability ranges for main sequence (MS) and starburst (SB) dusty galaxies from the sample of DSFGs in \citet{donevski20} are shown as shaded cyan and salmon areas, respectively. The gray region represents the predicted DSRs for galaxies with $\log(M_{\rm star}/M_{\odot})=10.5-11.5$ from the fiducial model of \citet{popping17}.}
   \label{fig:DSR}
\end{figure}
To compute the DSRs of our sources, we considered the dust and  stellar masses derived from fitting the whole SED with \texttt{CIGALE} (Table~\ref{tab:cigale_params} and Sect.~\ref{sec:cigale}).
SMGs show a wide range of DSRs, as shown in Fig.~\ref{fig:DSR}. 
The two dex range is much larger than the range in dusty MS and SB galaxies \citep{donevski20}, and the three CO emitters cover more than half of the large SMG range. 
The vast majority of SMGs exhibit higher DSRs than those predicted for galaxies with $\log(M_{\rm star}/M_{\odot})=10.5-11.5$ by the fiducial model of \citet{popping17}. 
High DSRs are usually associated with efficient and rapid dust grain formation \citep{donevski20}.
The wide range for SMGs might be due to differences in baryonic and stellar masses, dust extent, or evolutionary phase. DSRs depend on local conditions and can vary by large factors within a galaxy, but these variations are averaged out when considering integrated quantities. The large spread of DSRs in  SMGs likely reflects ``local conditions'' occupying a greater fraction of the galaxies. Additional disparities might be introduced by  the methods employed to estimate the dust masses \citep[e.g., from visible--NIR SED fitting, from the mm continuum flux density, from fitting the far-IR SED with a gray body model;][]{berta16}, and by uncertainties associated with the stellar masses. Our dust-mass estimates are poorly constrained because of the large uncertainties associated with the submm flux densities (Table~\ref{tab:noema}). New ALMA observations of 01a will enable a more accurate estimate of the dust extent and mass and a study of the resolved DSR\null. Such a study might reveal variations associated with the level of star formation activity and dust grain formation processes.

\section{Comparison with SMGs and  {\bfseries\slshape JWST} red sources}\label{sec:comparison}

\subsection{The nature of the  \jwst\ red sources}
\jwst\ has revealed an abundance of red galaxies, some invisible at shorter wavelengths even in HST images.  In the literature, they are often referred to as little red dots \citep[LRD;][]{kokorev24,perez24,matthee24,wang24_rubies,kocevski24}, HST-dark or HST-faint galaxies \citep{perez23,barrufet23,frye24,kamieneski24a}, optically dark or optically faint galaxies \citep[OFGs;][]{gomez23}, or extremely red objects \citep[EROs;][]{barro24}. These objects include DSFGs, obscured AGN, and $z\ga 5$ galaxies.  According to \citet{rodighiero23}, these red NIRCam sources are either massive DSFGs at $3<z<7$, quiescent galaxies at $3<z<5$, or galaxies with dwarf masses and high extinction \citep[e.g., $A_V\sim5.5$\,mag,][]{xiao23,bisigello23}. \citet{kocevski24} found instead that they are heavily reddened AGN, whereas \citet{perez24} claimed that they are mostly extremely intense and compact starburst galaxies with some contribution from an obscured AGN\null. Followup studies of individual sources  \citep{fujimoto23,killi23,furtak23,kokorev23,akins23,killi23,kokorev24,zavala23,meyer24} have found that the objects' redness is mostly due to heavy obscuration and to high redshifts. This claim is supported by simulations that reproduce these objects' properties by assuming that they are at $z=4$--7 and heavily obscured \citep[\av\,=\,2--4;][]{cochrane24}.

\begin{figure}[h!]
\centering
\vspace{-4pt}
\includegraphics[width=0.44\textwidth]{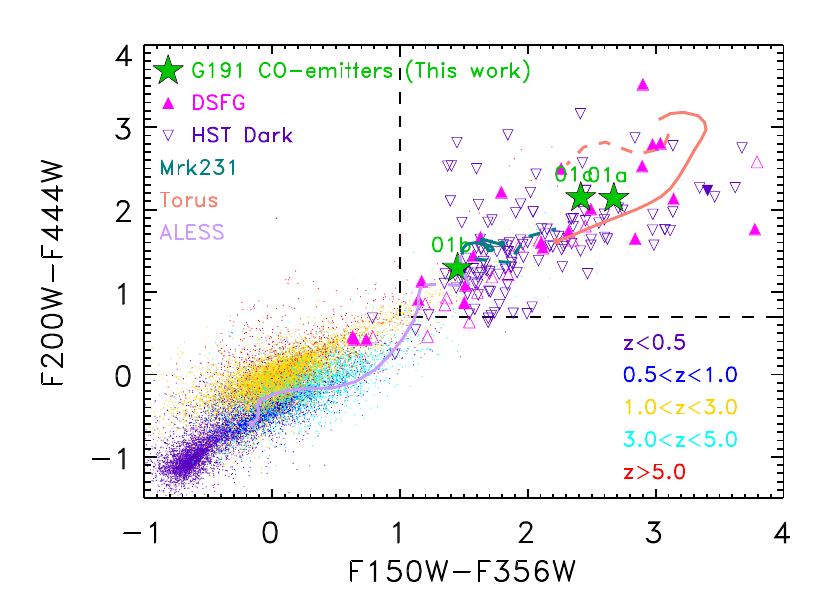}
\null\vspace*{-10pt}
\includegraphics[width=0.44\textwidth]{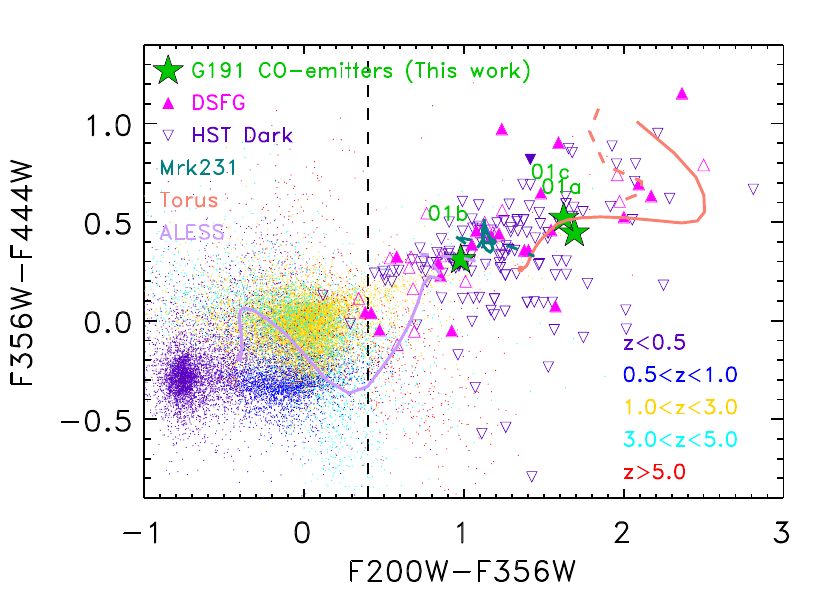}
\null\vspace*{-10pt}
\includegraphics[width=0.44\textwidth]{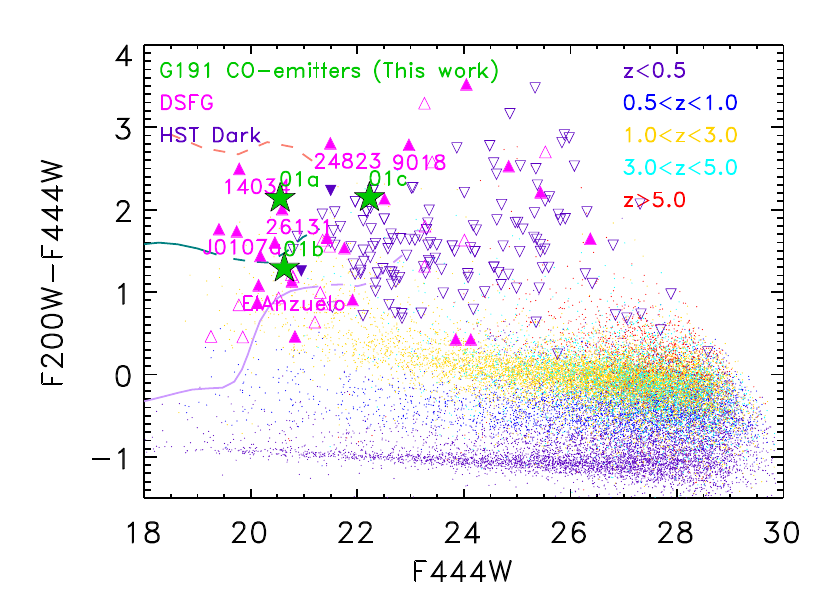}
\null\vspace*{-10pt}
\caption{\small  \jwst\ color--color and color--magnitude diagrams for a variety of galaxies. Filled green stars represent the three CO targets. Small dots represent galaxies from the UNCOVER program \citep{wang24} with their colors indicating the photometric redshift: purple for $z<0.5$, blue for $0.5<z<1.0$, yellow for $1.0<z<3.0$, cyan for $3.0<z<5.0$, and red for $z>5$.   Other symbols represent HST-dark galaxies \citep[purple upside-down triangles;][]{barro24,nelson23,barrufet23,perez23,frye2023,wang24_rubies} and DSFGs \citep[magenta triangles;][]{peng23,zavala23,huang23,cheng23,mckay23,smail23,kamieneski23,sun24,meyer24}. Filled symbols indicate sources with spectroscopic redshifts. Dashed lines mark the typical colors of \jwst\ red sources.  DSFGs with spectroscopic redshifts $\sim$2.5 are labeled in the bottom panel (i.e., El Anzuelo at $z=2.3$ \citep{kamieneski23}, J0107a at $z=2.467$ \citep{huang23}, and four galaxies at $2.1<z<2.7$ from \citet{price23}. The lilac, salmon, and teal curves show tracks for the average SED from the ALESS SMG sample \citep{danielson17}, the Torus template \citep{Polletta2007}, and Mrk~231, respectively, with the solid curve showing  $0.1<z< 2.5$ and the dashed curve  $2.5<z<5$.
\vspace{-6pt}}
\label{fig:color_diagrams} 
\end{figure}

The three CO emitters are bright and red at near- and mid-IR wavelengths and extremely faint at visible wavelengths, putting them in the category of EROs. Fig.~\ref{fig:color_diagrams} compares our targets' colors and magnitudes with those of various HST-dark galaxies, submm/mm-selected  SFGs, and the full galaxy population from the \jwst\ UNCOVER program \citep[PIs: Labb\'e and  Bezanson;][]{wang24}. The CO emitters have colors similar to those of the HST-dark galaxies and DSFGs,  which span about 3\,mag. These galaxies' colors are distinct from the bulk of  \jwst\ sources, having $\rm F200W-F444W>0.7$, $\rm F150W-F356W>1$, and $\rm F200W-F356W>0.5$.  Only 1.5\% of the full UNCOVER sample (with ${>}5\sigma$ detections) have such red colors. The CO emitters are  among the sources with the brightest F444W magnitudes, $\sim$20.6--22.2 versus 22--26 for the DSFGs with missing or $>2.5$ redshift and the majority of HST-dark galaxies. The DSFG El~Anzuelo is intrinsically $\sim$2.5\,mag fainter than shown in Fig.~\ref{fig:color_diagrams} once the magnification from strong gravitational lensing is taken into account \citep{kamieneski23}.

\citet{gottumukkala23} and \citet{cochrane24} suggested that the \jwst\ red sources constitute a new population. A few sources with similar red colors had already been discovered prior to \jwst\ but at lower redshifts and higher luminosities than the bulk of the  \jwst\ red population.  Some examples are the BAL QSO Mrk~231 at $z=0.042$ \citep{berta05,Polletta2007} and the heavily obscured QSOs SWIRE J104409.95+585224.8 at $z=2.54$ \citep{polletta06}. The predicted colors of these sources for $0.1<z<5$ overlap with those of the red  \jwst\ sources, as shown in Fig.~\ref{fig:color_diagrams}, which also shows the expected colors derived from the average SED of the ALESS SMG sample \citep{danielson17}. The ALESS colors overlap with the red  \jwst\ sources only for $z> 2.5$. The visible--NIR colors of the red  \jwst\ sources are thus consistent with those of SMGs at $z>2.5$ or with heavily obscured AGNs that were known before \jwst.
In the color--magnitude diagram (Fig.~\ref{fig:color_diagrams} bottom),  the three CO sources fall among the mm-selected  DSFGs.  Most of these have F444W magnitudes $<$22, in contrast to the optically selected, HST-dark red sources, most of which have F444W magnitudes between 22 and 26.  Both groups have similar F200W$-$F444W colors, much redder than the UNCOVER galaxies. One possibility is that the HST-dark galaxies and high-$z$ DSFGs represent the same population with the latter being typically more luminous. 

In spite of the success of obscured-AGN templates in reproducing the 2--5\,\micron\ colors of the red  \jwst\ sources and recent spectroscopic evidence of AGN activity in a subset characterized by blue UV spectra \citep{greene24},  AGN emission  might not be the dominant luminosity source.  \citet{williams24} found that the SEDs of a sample of LRDs differed from the Mrk~231 template at observed wavelengths $\lambda>7\,\mu$m. These authors claimed that the LRDs are moderately dusty and old galaxies rather than obscured AGN \citep[see also][]{perez24}. This claim is supported by our sources because they do not show any obvious signs of AGN activity.

\subsection{The CO sources and the SMG population}

Based on the estimated 850\,$\mu$m fluxes from our bestfit SED model (i.e., $S_{\rm 850\,\mu m}$ = 11 and 7\,mJy for 01a and 01b, respectively), as well as on the typical 870\,$\mu$m/3\,mm flux ratio for SMGs \citep[80$\pm$40;][]{birkin21}, sources 01a and 01b can be considered SMGs. The comparison between their main properties and those of various SMG samples from the literature  (Sect.~\ref{sec:SF_mode}) indicates that 01a and 01b are among the SMGs with the highest SFRs and shortest depletion times at $2<z<3$. However, they do not stand out in other properties such as stellar mass and gas-to-stellar mass ratio. Their morphologies (Sect.~\ref{sec:morph}) are also consistent with those of other SMG samples from the literature \citep{gillman23,cheng23,price23,boogaard24}. Interestingly, there is some evidence of an overdense environment around source 01b because of its proximity to 01c, in line with the SMG expected environmental properties. Sect.~\ref{sec:resolved_seds} showed that the extinction varies significantly across the galaxies. This result emphasizes the importance of resolved studies of the stellar, dust, and gaseous components of this kind of sources in order to understand their formation and evolution \citep[e.g.,][]{cheng22,smail23,kamieneski23,zavala23,kokorev23} and of establishing a common framework for meaningful comparisons. This study adds one additional piece of information to the SMG picture and lays the basis for followup mm observations.

\section{Discussion}\label{sec:discussion}

\subsection{The powering mechanism of the estimated high SFRs}

The sources examined in this work are starburst galaxies with some of the highest estimated SFRs in the galaxy population at cosmic noon. They are also quite peculiar because of their heavy off-center obscuration and large extinction gradients (Sect.~\ref{sec:resolved_seds}).  At the same time, they have relatively smooth and symmetric stellar-mass distributions, well described by a disk like profile with similar sizes and S\'ersic indices as SMGs and normal SFGs at $z\sim2.5$ (Sect.~\ref{sec:morph}). The residual maps obtained after subtracting an exponential disk show substructures, including arcs mimicking spiral arms and clumps, although the substructures are not dominant. Indeed  modeling the observed light with a S\'ersic  $n\sim1$ profile  yields a good fit, and the galaxies have a low clumpiness parameter. Notably, many highly SFGs at $1<z<3$ and SMGs display disks with substructures similar to our sources \citep[e.g.,][]{glazebrook95,abraham96,vandenbergh96,cowie95,elmegreen07a,elmegreen09,forster09,cheng22,wu23,liu24}. In our sources, these substructures seem to be associated with variations in the obscuration level. Asymmetric dust distributions have also been reported in other SMGs such as  COSBO7 \citep{ling24}, COSMOS 1648673 \citep[also known as PACS-819;][]{liu24}, El Anzuelo \citep{kamieneski23}, and Arc~1 in the PLCK~G165.7$+$67.0 field \citep{frye24}. In summary, the properties of our sources, their large SFRs, stellar disk profiles, substructures, and asymmetric obscuration are common in the $1<z<3$ SMG population.

Starburst galaxies in the local Universe are the result of a major merger \citep[mass ratio $\geq$1:4;][]{armus87,sanders96,kartaltepe10,ellison13b,engel10,casey16}. However, this does not seem to be always the case at high redshifts, where different conditions, such as the availability of more gas might result in higher gas accretion rates and a different mode of star formation \citep{scoville16,liu19}. Starburst activity may thus be driven secularly by compressive, gravitationally unstable, gas-rich disks \citep[e.g.,][]{ceverino10,tadaki18}. The transition between different growth regimes is expected to occur at $2\lesssim z\lesssim3$ \citep{oser10}.

Our CO emitters, with their exceptional SFRs, are ideal laboratories to test the predictions from the various proposed star formation driving mechanisms. If a major merger were responsible for their vigorous star formation, we would expect to see double nuclei or a galaxy pair.  The lack of those features as well the CAS-based morphological diagnostics (Sect.~\ref{sec:morph}) do not support the major merger scenario. However, identifying a major merger depends on the source separation and the merger stage. Morphological indicators can be quite incomplete in this regard, especially in the post-coalescence phase \citep{bignone17}.  Similar to our sources, \citet{engel10} found that most SMGs are compact, single sources, but those authors argued that this is consistent with end-stage, coalesced major merger products. This interpretation is supported by merger simulations that predict that the SMG phase occurs nearly a Gyr after the last major merger \citep{mihos96,springel05}, and by observations of late stage mergers with low clumpiness \citep{calabro19a}. According to \citet{calabro19a}, low clumpiness can be observed in late stage mergers after coalescence because of the clumps' rapid destruction by strong stellar radiation or AGN feedback, or of their migration toward the center.

During a major interaction, orbits of stars are violently perturbed, gas is torqued to the center, shocked, and compressed by tidal effects boosting the SFR \citep{sparre22}. The highly dissipative gas collapse (due to tidal torques) causes a contraction of the stellar component and a severe reduction of the disk mass fraction \citep{barnes96,bournaud11}. A major merger would then destroy the disk in favor of the buildup of a spheroid. The remnant galaxy would be an early type galaxy with a small radius and a high S\'ersic index \citep{bournaud11,wuyts10}. This scenario is not consistent with the properties of our sources. However, \citet{moreno19} have shown, through simulations at high resolution, that the turbulence created by a merger makes the ISM stable against collapse and leads to enhanced fragmentation into cold clouds and to SFR enhancement that is radially extended. Also \citet{springel05} demonstrated that the disk is not destroyed in a major merger between disk galaxies with large gas fractions.  Cooling can quickly reform a disk, yielding a remnant that, structurally and kinematically, more closely resembles a spiral galaxy than an elliptical, but tidal torques form a central, rotating bulge.  Hence the remnant is a galaxy with a bulge and an extended star-forming disk. The low S\'ersic index, the centrally peaked SFR profile, and the negligible variations in stellar ages across our galaxies disfavor this possibility. Another prediction from simulations is the onset of AGN activity following a major merger  \citep{hopkins08}. Because major mergers favor the transport of gas into the galaxy central region, the central super massive black hole (SMBH) can grow fed by this gas and eventually reach a sufficiently high luminosity to affect the galaxy gas and star formation activity through feedback. The lack of AGN activity in our sources might thus be explained by the limited gas supply to the galaxy center predicted by this picture. Although AGN activity can be present in massive disk galaxies with no sign of interaction or mergers \citep[e.g., Rubin's galaxy;][]{holwerda21}. Based on these considerations the major merger scenario as driving mechanism in our sources seems disfavored. 

At the same time, we cannot rule out that a major merger might have occurred hundreds of Myrs in the past. To support this claim, we refer to a beautiful example of a late-stage merger ($\sim$500\,Myrs in the past), the IR-luminous galaxy NGC~3256 at $z=0.01$. The galaxy hosts two nuclei separated by $\sim0.8$\,kpc \citep{sakamoto14}.  The secondary nucleus is visible only in high resolution radio images because it is heavily obscured \citep[\av=16;][]{lira08}.  The galaxy exhibits asymmetric dust extinction that gives it an appearance at visible wavelengths that is strikingly similar to 01a and 01b.\footnote{https://webbtelescope.org/contents/media/images/2020/49/4742-Image, credits: NASA, ESA, Aaron S. Evans (UVA, NRAO, State University of New York at Stony Brook), Hubble Heritage–ESA/Hubble Collaboration.} It was only through the extensive multiwavelength spectroscopic and photometric analysis carried out at high resolution that its merging nature was uncovered. 

An alternative mechanism that can drive star formation and produce high SFRs is gas accretion. \citet{ho19} used EAGLE simulations to examine the effects of gas accretion on star formation on galaxies' disks. Because the accreted gas lacks angular momentum to maintain circular orbits, it flows radially inward and ends up with an anisotropic distribution and forms streams and disk like structures. Most of the gas that will be converted into stars is expected to reside first in the spiral arms and then concentrate near the galaxy center. In the case of gas accretion, we thus expect to see star formation throughout the disk, detecting it to large radii \citep[e.g.,][]{dannerbauer17} and along spiral arms, and also a dusty star-forming core. This picture is qualitatively consistent with the appearance of our sources. Inflowing gas might stem from the circumgalactic medium (CGM), the intergalactic medium (IGM), galactic halos and groups of galaxies \citep[e.g.,][]{westmeier05}, may be stripped from infalling satellites, or may infall directly via a smooth cold, flow along filamentary structures. Interactions (flybys and minor mergers) can also supply gas and induce extended star formation  \citep{pan19,thorp19}. Source 01a might be currently interacting with the neighbor (at 1\farcs5) galaxy A3 and exhibits a tail of stars to the north consistent with being stripped from the disk by the forces at play during a strong interaction. Sources 01b and 01c are close in projection (5\farcs7 apart, or $\lesssim$50\,pkpc) and at similar redshifts and they have three nearby galaxies ($<$5\arcsec) with consistent photometric redshifts, also supporting on-going interactions (Sect.~\ref{sec:environment}). Also the width of the observed \COthree\ lines suggests a complex kinematics consistent with disturbances from an interaction or minor mergers. 

Observational evidence of inflowing gas is often elusive as the gas is expected to be relatively cold (at $T\sim10^4$\,K) and faint. On large scales, such gas can be traced by \lya\ filaments \citep{daddi22}, and it is often associated with massive halos and overdensities at the core of galaxy protoclusters \citep{valentino16,daddi22,pensabene24}. Accreting gas streams might be also traced by inflowing and extended cold molecular gas \citep[e.g.,][]{herrera_camus20,ginolfi17,berta21,vidal_garcia21}.  Indeed, cooling and gravitational collapse in gas streams may lead to clump condensation out of the CGM and to star formation
\citep{dekel09a,bouche13,pallottini14,ceverino16,nelson16}. However, signatures of infalling molecular gas are notoriously difficult to observe \citep{berta21}.  Moreover, even when observed, infalling molecular gas can have many interpretations such as unresolved satellites and merging companions or gas that was previously expelled and is falling back into the CGM, where it is able to cool \citep{vidal_garcia21}. To probe the gas-accretion scenario and definitively untangle the mechanisms that triggered and sustain star formation in our sources, we would need to measure the resolved star formation and the molecular gas kinematics \citep[e.g.,][]{liu24}.

\subsection{The environment of the CO emitters}\label{sec:environment}

The large stellar masses of the three CO emitters and the stellar-to-halo mass relation \citep{girelli20} suggest that they live in massive halos ($M_{\rm halo}=(2.9\pm 0.6)\times 10^{12}$\,\msun\ for 01b and $(2.4\pm 0.4)\times10^{13}$\,\msun\ for 01a). Other models in the literature \citep[e.g.,][]{behroozi19} would imply halo masses up to four times larger. According to \citet{chiang13}, a halo with mass above 10$^{13}$\,\msun\ at $z\simeq2.5$ will evolve into a cluster with mass above 10$^{14}$\,\msun\ by $z=0$. Assuming the halo masses estimated above, 01a might reside in a galaxy protocluster, while 01b might be in a proto-group. 

Quite a bit is already known about the environment of sources 01b and 01c. They are at similar redshifts ($\delta v<500$\,\kms) and at a projected distance of only $\sim$5\farcs7, equivalent to ${\lesssim}50$\,pkpc. The apparent proximity supports the proto-group or protocluster hypothesis. 
As stated earlier, the G191 field was selected as a protocluster candidate because of its bright \planck\ signal and the presence of an overdensity of \herschel\ sources and of red \spitzer\ sources. Our sources have the red mid-IR colors of members of such overdensities and are associated with a \herschel\ source. Furthermore, SMGs are  considered tracers of overdensities \citep{calvi23}. Several other examples of 
hyperluminous SMGs with no obvious signs of a major merger but located in large galaxy overdensities are known. Examples include SMM~J084933 at $z = 2.410$ \citep{ivison13} and GN20 at $z=4.055$ \citep{hodge12}, although recent \jwst\ observations at mid-IR wavelengths of the latter reveal a double nucleus consistent with a late-stage merger \citep{colina23}. Our sources and other luminous, non binary SMGs such as those mentioned may live in overdense environments, where related conditions might provide support to their star-forming disks and large SFRs. Environment might be the common trait of all these sources that explains their extreme star formation activity.

\begin{figure}[t]
  \centering
 \includegraphics[width=0.24\textwidth]{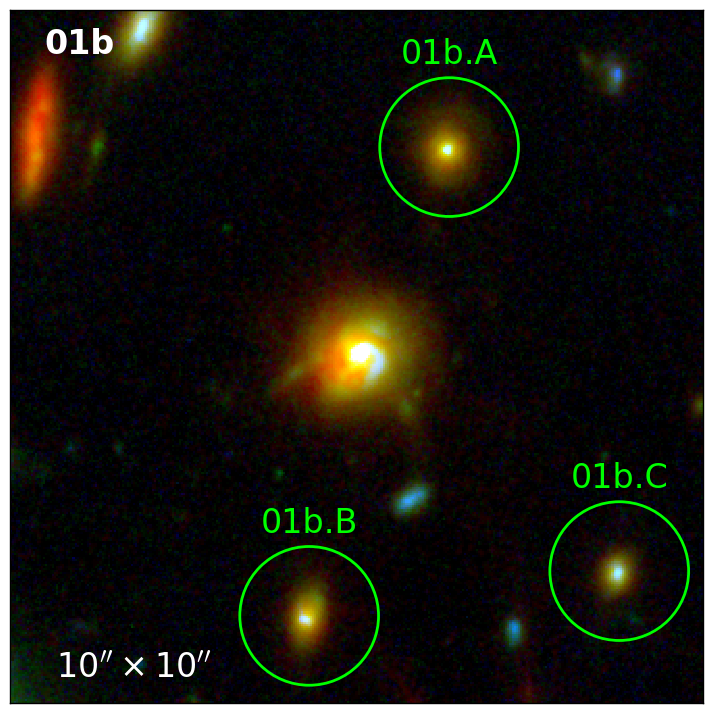}
 \includegraphics[width=0.24\textwidth]{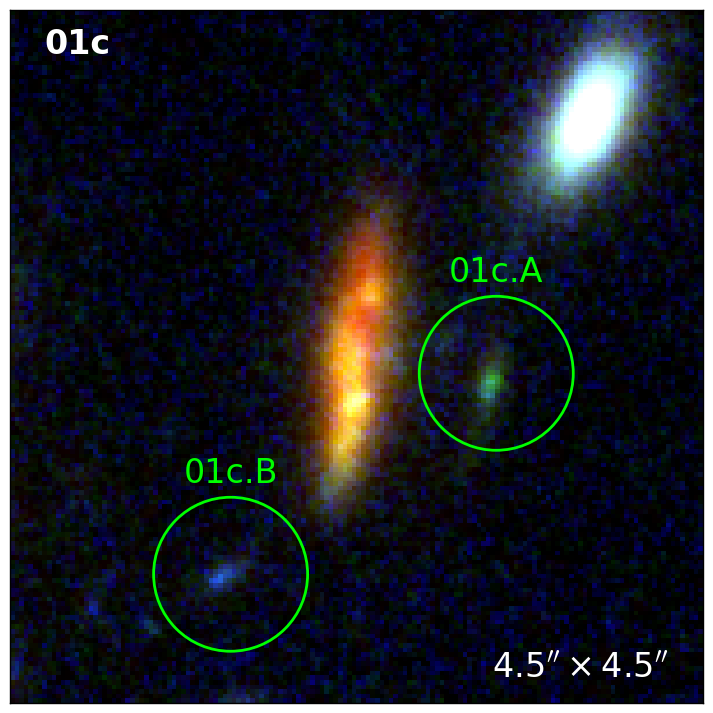}
   \caption{Multicolor \jwst/NIRCam images (with filter assignments as in Fig.~3) centered on 01b ({\it left}) and on 01c ({\it right}). Images sizes are 10\arcsec$\times$10\arcsec\ and 4\farcs5$\times$4\farcs5, respectively. Photometric redshifts were derived for the nearby sources marked with green circles and labeled to search for potential members of an overdensity at $z\simeq 2.42$. The galaxies designated as 01b.A, 01b.B and 01b.C in the left panel have all photometric redshifts consistent with the redshifts of 01b and 01c, while 01c.A and 01c.B in the right panel have lower redshifts.  }
   \label{fig:nrb_gals}
\end{figure}

To further investigate the  environment of 01b and 01c and assess the possibility that they might be members of a large-scale structure, we searched for nearby galaxies that might be at $z\simeq2.42$.
There are five galaxies that are either $<$2\arcsec\ away or up to 5\arcsec\ away but with similarly red colors: three around 01b and two near 01c (Fig.~\ref{fig:nrb_gals}). Table~\ref{tab:nrb_gals} gives the galaxies' designated source names, coordinates, and photometric redshifts, the latter measured with with \texttt{CIGALE} based on NIRCam flux densities measured using the method described in Sect.~\ref{sec:jwst_obs}.
 
\begin{table}
\caption{\label{tab:nrb_gals} Selected galaxies around 01b and 01c.}
\centering 
\begin{tabular}{c c c c c c c c c} 
\hline\hline
Source & $\alpha_{JWST}$ & $\delta_{JWST}$ & $z_{\rm phot}$ \\
ID     & (h:m:s)       & (\deg:\arcmin:\arcsec) &         \\
\hline
 01b.A  & 10:44:38.94  &  33: 51:10.59 & 2.27$\pm$0.28 \\
 01b.B  & 10:44:39.11  &  33: 51:03.83 & 2.43$\pm$0.17 \\
 01b.C  & 10:44:38.75  &  33: 51: 4.48 & 2.44$\pm$0.13 \\
 01c.A  & 10:44:39.35  &  33: 51:10.66 & 1.90$\pm$0.24 \\
 01c.B  & 10:44:39.49  &  33: 51:09.35 & 2.12$\pm$0.34 \\
\hline   
\end{tabular}\\
\end{table} 

The three galaxies close to 01b have similar appearance and colors. Their photometric redshifts are all consistent with that of 01b, as expected if they were all members of a proto-group or protocluster. The 01c neighbors have bluer SEDs and lower photometric redshifts than 01b and 01c, implying that they are distinct sources.
This analysis suggests that 01b and 01c might be members of a proto-structure with five members in a 10\arcsec$\times$10\arcsec\ ($\sim$200~ckpc) region. 
Only the densest protocluster cores contain a similar number of members in such a small region: four members in the Distant Red Core at $z=4.0$ \citep{oteo18}, five in SPT-CL J2349$-$5638 at $z=4.3$ \citep{miller18}, and seven in ClG J1001+0220 at $z=2.51$ \citep{xu23}.  The densest regions of most  $z\sim2.5$ protoclusters contain from one to three members across 10\arcsec\ 
\citep{darvish20,polletta21,perez_martinez23,perez_martinez24}.  If confirmed, the proto-structure with 01b and 01c may have an exceptionally dense core. 

The G191 field also contains one of the most significant overdensities of red \herschel\ sources in the PHz sample.  With 01b being associated with \herschel\ 01, it is possible that an overdense structure might extend to distances $\sim$5\arcmin, where the other \herschel\ sources are located \citep[][]{planck15,polletta22}. Spectroscopic observations of the potential members around 01b and 01c and of the galaxies in the two overdensities (the red MIR sources and the red \herschel\ sources) in the G191 field are necessary to confirm and characterize such a structure.

\section{Summary and conclusions}\label{sec:summary}

This work has analyzed the stellar, dust, and cold gas components of three starburst galaxies at $z\sim2.5$ using  \jwst/NIRCam imaging from the PEARLS GTO program and NOEMA observations. Their visible--radio SEDs imply SFRs of hundreds to thousands of \msun\,yr$^{-1}$,  stellar masses $M_{\rm star}\simeq10^{11}$\,\msun, and average extinction $A_V\sim3$--4 mag. Compared to the star-forming main sequence (MS) at their redshifts, their SFRs are from 6 to 17 times higher, placing these galaxies among the starburst population and among the most powerful SMGs at $2<z<3$. Our main results are summarized below.

\begin{itemize} 
\item[$\bullet$] {\bf Morphology and structural components:} The selected sources are extremely red with faint rest-frame UV fluxes and bright rest-frame visible emission. Their red colors are attributed to dust obscuration. Their stellar emission has surface-brightness profiles consistent with an exponential disk (S\'ersic index $n\sim1$) with effective radii of a few kiloparsecs. These stellar sizes are in good agreement with the effective sizes of mass-matched SFGs at $z = 2.5$. Two galaxies are almost face-on disks, while one is an edge-on disk. All three systems have however a disturbed appearance with several substructures such as clumps and spiral arms across them. These substructures exhibit extinction values up to $A_V\simeq5$--7, extend over several kiloparsecs and have shapes consistent with spiral arms that have undergone some major disturbances causing them to bend. The morphology and the physical conditions derived for these substructures suggest that interactions have formed them and that they contribute to the enhanced star formation activity of these galaxies.
\item [$\bullet$] {\bf Star formation properties:} All three galaxies have short depletion timescales ($\leq 100$\,Myr), consistent with the most powerful SFGs at their redshift. Despite their large gas reservoirs ($M_{\rm gas}\gtrsim10^{11}$\,\msun), their gas-to-stellar mass ratios ($M_{\rm gas}/M_{\rm star}\simeq 0.4$--1.3) are consistent with those found in the SFG population with similar stellar masses and redshifts.  This result suggests that what has driven these galaxies above the MS must be linked to higher efficiency and not to the amount of cold gas or stellar mass, in agreement with other SMG studies \citep{liao24}. Because of their large SFEs, the galaxies are expected to deplete their large gas reservoir and quench in $\lesssim$100\,Myr unless the gas is replenished. 
\item [$\bullet$] {\bf Dust properties:} The analyzed sources are characterized by elevated extinction levels that can reach \av=7 even in regions several kiloparsecs in size. Interestingly, these high extinction values are observed away from the galaxy centers and only on one side of each galaxy, suggesting that extinction in some DSFGs may be due to dust in dense star-forming regions rather than to dust in the diffuse ISM. We computed dust attenuation curves in the rest-frame wavelength range of 0.4--1.2\,$\mu$m for two of the systems. They differ from most of the standard attenuation curves \citep{cardelli89,calzetti00,wang19,battisti22}, possibly
implying that the dust-star geometry or dust properties (e.g., size distribution, composition) in these galaxies or along heavily extincted sightlines might be different than what commonly observed. It is beyond the scope of this work to determine new attenuation laws, but we have shown that \jwst\ observations may be used for this purpose and for exploring the dust properties.
Our analysis implies that in galaxies with extreme dust extinction 
a differential attenuation curve with an extinction-dependent shape might be necessary to correctly estimate unattenuated fluxes and ensuing physical properties, such as the stellar mass. 
\item [$\bullet$] {\bf Comparison with other red \jwst\ sources:} Our three sources belong to the population of  \jwst\ red galaxies that usually includes DSFGs, and HST-dark objects. Their red colors are attributed to their high redshift, $z\gtrsim2.5$, and  heavy extinction. 
Sources with similarly red colors represent $\sim$1.5\% of the whole  \jwst\ population. Based on the analysis of galaxy templates of various types, we find that similarly red colors are reproduced by heavily obscured AGN or by DSFGs at $z>2.5$, and that there is no need for a new population.
\item [$\bullet$] {\bf Star formation powering mechanism:}
The galaxy-scale star formation activity, the presence of a single nucleus, the lack of a prominent bulge, and the classification based on the non parametric morphological parameters disfavor the major merger scenario as powering mechanism in our sources. Their disk appearance, extended SF, signs of disturbance such as tidal tails, spiral arms with significant changes of pitch angle, clumps and an asymmetric dust extinction distribution favor instead gas inflows at the origin of their powerful SFRs and enhanced SFEs. The presence of galaxies near in projection to sources 01a and 01b and with consistent photometric redshifts (see App.~\ref{app:01a_fgd_srcs} and Table~\ref{tab:nrb_gals}) indicates that minor mergers and flybys might provide the necessary gas to sustain their star formation activity. This picture might also explain the lack of AGN activity because of the limited supply of gas that reaches the galaxy center, contrary to what would be expected in case of a major merger.
\item [$\bullet$] {\bf Environment: }
The CO sources are associated with two overdensities (of red \herschel\ sources and of red \spitzer\ sources) spread over several arcmin. In addition, their stellar masses imply that they are hosted by massive dark matter haloes. All sources have neighbors with similar (spectroscopic or photometric) redshifts. In summary, there are multiple indications that these CO sources might be members of a galaxy protocluster or proto-group, as first suggested by their association with a \planck-selected high redshift source.
\end{itemize}

In conclusion, these three powerful SFGs at $z\sim2.5$ stand out with respect to the average SFG population at similar stellar masses and redshifts because of their highly efficient and powerful star formation activity. Their exceptional properties are attributed to accretion of gas most likely through minor mergers and flybys that sustain their star formation and supports the disk. Instabilities of the gas can explain the spatially widespread star formation and the formation of substructures with diverse dust obscuration. These properties, also observed in other luminous non binary SMGs, might be associated with an overdense environment in a massive halo. For the future, the positions of these sources relative to any underlying protocluster will be investigated through the analysis of the galaxy population over a wide area ($\sim5$\arcmin) where overdensities of \herschel\ and red \spitzer\ sources are located. In addition, sources at similar redshifts will be searched through ground-based NIR spectroscopic observations, and CO observations at mm wavelengths. Finally, recently obtained ALMA observations of source 01a at high spatial resolution will be able to test the proposed fueling scenario. This can be achieved by measuring the gas turbulence in the disk and through the search of nonrotating gas components with a resolved study of the gas kinematics.


\begin{acknowledgements}
It is a pleasure to thank the anonymous referee for a timely and positive response that helped to improve the presentation of this work.
We kindly thank Professors Ian Smail and Adolf N. Witt for valuable comments and discussions. We thank Zhaoxuan Liu for sharing his work and useful comments.
M.P. acknowledges financial support from INAF mini-grant 2023 "Galaxy growth and fueling in high-$z$
structures."
B.L.F.~obtained student support through a Faculty Challenge Grant for Increasing Access to Undergraduate Research, and the Arthur L. and Lee G. Herbst Endowment for Innovation and the Science Dean’s Innovation and Education Fund, both obtained at the University of Arizona.
R.A.W., S.H.C. and R.A.J. acknowledge support from NASA JWST Interdisciplinary Scientist grants NAG5-12460, NNX14AN10G and 80NSSC18K0200 from GSFC.
%
%
We thank the  \jwst\ Project at NASA GSFC and  \jwst\ Program at NASA HQ for their many-decades long dedication to make the  \jwst\ mission a success.  We especially thank Tony Roman, the  \jwst\ scheduling group and Mission
Operations Center staff at STScI for their continued dedicated support to
get the  \jwst\ observations scheduled.  This work is based on observations
made with the NASA/ESA/CSA {\it James Webb} Space Telescope.  The data were
obtained from the Mikulski Archive for Space Telescopes (MAST) at the Space
Telescope Science Institute, which is operated by the Association of
Universities for Research in Astronomy, Inc., under NASA contract NAS
5-03127 for  \jwst.  These observations are associated with  \jwst\ programs
1176. 
We are very grateful to the IRAM staff for preparing and carrying out the
observations, and for their support and help in reducing the data.
Based on observations carried out with the IRAM Interferometer NOEMA under
project W21DA.  IRAM is supported by INSU/CNRS (France), MPG (Germany) and
IGN (Spain).
%
The \herschel\ spacecraft was designed, built, tested, and launched under a
contract to ESA managed by the \herschel/\planck\ Project team by an industrial
consortium under the overall responsibility of the prime contractor Thales
Alenia Space (Cannes), and including Astrium (Friedrichshafen) responsible
for the payload module and for system testing at spacecraft level, Thales
Alenia Space (Turin) responsible for the service module, and Astrium
(Toulouse) responsible for the telescope, with in excess of a hundred
subcontractors.  SPIRE has been developed by a consortium of institutes led
by Cardiff University (UK) and including Univ.  Lethbridge (Canada); NAOC
(China); CEA, LAM (France); IFSI, Univ.  Padua (Italy); IAC (Spain);
Stockholm Observatory (Sweden); Imperial College London, RAL, UCL-MSSL,
UKATC, Univ.  Sussex (UK); and Caltech, JPL, NHSC, Univ.  Colorado (USA). 
This development has been supported by national funding agencies: CSA
(Canada); NAOC (China); CEA, CNES, CNRS (France); ASI (Italy); MCINN
(Spain); SNSB (Sweden); STFC, UKSA (UK); and NASA (USA).
LOFAR data products were provided by the LOFAR Surveys Key Science project (LSKSP; \url{https://lofar-surveys.org/}) and were derived from observations with the International LOFAR Telescope (ILT). LOFAR \citep{vanhaarlem13} is the Low Frequency Array designed and constructed by ASTRON. It has observing, data processing, and data storage facilities in several countries, which are owned by various parties (each with their own funding sources), and which are collectively operated by the ILT foundation under a joint scientific policy. The efforts of the LSKSP have benefited from funding from the European Research Council, NOVA, NWO, CNRS-INSU, the SURF Co-operative, the UK Science and Technology Funding Council and the Jülich Supercomputing Centre.
%
This work has made use of data from the European Space Agency (ESA) mission
{\it Gaia} (\url{https://www.cosmos.esa.int/gaia}), processed by the {\it Gaia}
Data Processing and Analysis Consortium (DPAC,
\url{https://www.cosmos.esa.int/web/gaia/dpac/consortium}). Funding for the DPAC
has been provided by national institutions, in particular the institutions
participating in the {\it Gaia} Multilateral Agreement.
{\it Software:} This research made use of astropy, a community developed core Python package
for astronomy \citep{astropy}, APLpy, an open-source plotting package for
Python \citep{aplpy}, the IDL Astronomy Library \citep{Landsman1993}, \texttt{CIGALE} \citep{boquien19}, \texttt{starmorph} \citep{rodriguez19}, \texttt{Galfit} \citep{peng02}, GILDAS (https://www.iram.fr/IRAMFR/GILDAS), CARTA \citep[https://cartavis.org/;][]{comrie21}, and TOPCAT (http://www.starlink.ac.uk/topcat/).
This research has made use of the VizieR catalogue access tool, CDS,
Strasbourg, France \citep{10.26093/cds/vizier}. The original description 
of the VizieR service was published in \citet{vizier2000}.

\end{acknowledgements}

\begin{appendix}
\section{Neighbors' contamination in 01a}\label{app:01a_fgd_srcs}

Three sources, designated A1, A2, and A3, lie within 0\farcs5--1\farcs5 of the center of 01a. These sources seem to be located (Fig.~\ref{fig:targets}) along an arc or spiral arm extending from the major axis of A3 to the center of 01a, suggesting that the A sources might be the result of an on-going interaction with 01a and among themselves. If so, they would be at the same redshift, and at least A1 and A2 would be part of 01a, but their bluer colors suggest that they are distinct sources. To assess this hypothesis, we built their SEDs by measuring all the flux within specific apertures and used \texttt{CIGALE} to estimate their photometric redshifts.  The centers and sizes of the three photometric apertures are listed in Table~\ref{tab:fgd_srcs} and shown in the left panel of Fig.~\ref{fig:targets}. 

\begin{figure}[h!]
\centering
\includegraphics[width=0.45\textwidth]{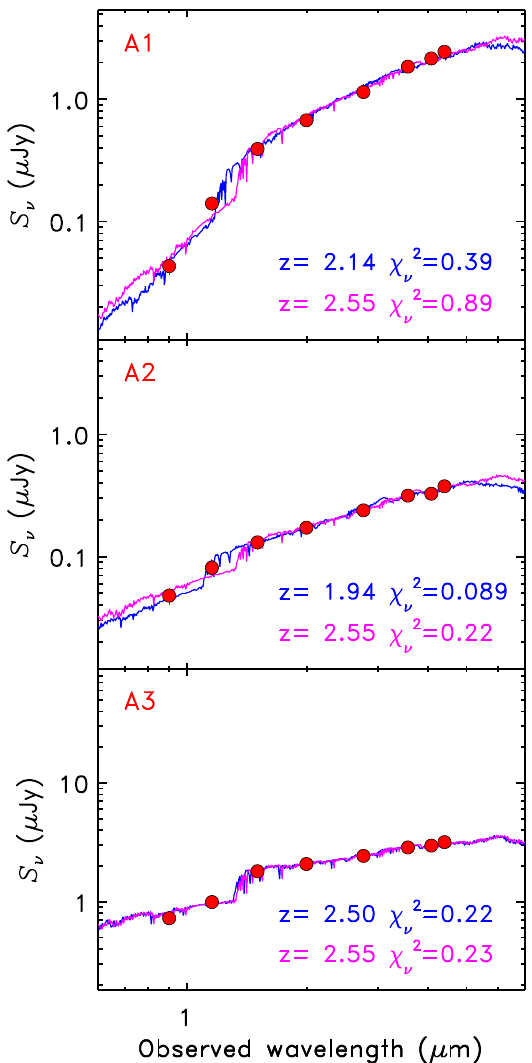}
\caption{\small SED of the sources within 1.5\arcsec\ of source 01a and considered distinct sources (filled red circles; A1 in the top panel, A2 in the middle panel, and A3 in the bottom panel).  In each panel, the best fit model, associated photometric redshift, and reduced-$\chi^2$ are shown in blue, and the model at the same redshift of 01a and corresponding reduced-$\chi^2$ are shown in magenta.}
\label{fig:seds_fgd_srcs} 
\end{figure}

We  fit the galaxies' SEDs with the redshifts unconstrained (Table~\ref{tab:fgd_srcs}) and also fixed to the spectroscopic redshift of 01a.  The two bestfit models are shown in Fig.~\ref{fig:seds_fgd_srcs}. The reduced $\chi^2$ of A1 and A2 are 2.3--2.5 times smaller at the best photometric redshifts than at $z=z_{\rm 01a}$, but $z=z_{\rm 01a}$ yields also a good fit for A2 and A3 ($\chi_{\nu}^2\leq0.23$). The $\chi^2_\nu$ for A1 at $z=z_{\rm 01a}$ makes it a foreground source. Source A3 is 1\farcs5 from 01a, and if it were at the same redshift of 01a, it would be a companion galaxy with a mass ratio of $\sim$1:22. The association between A1 and 01a is the most complex because A1 could be a super star cluster (SSC) or a star-forming region within 01a. 
The F444W flux density of A1 is $\sim$11\% of the whole 01a galaxy. At $z=z_{\rm 01a}$, A1's stellar mass would be ${\lesssim}3\times10^{10}$\,\msun, and its SFR would be $\sim$30\,\msun\,yr$^{-1}$. A1's size, brightness, SFR, and $M_{\rm star}$ are extremely large for a SSC or a star-forming region \citep[see e.g.,][]{mok20}, implying that it is more likely a distinct galaxy rather than part of 01a. We have assumed that A1 and A2 are in the foreground of 01a, and we have masked them in the analysis of 01a, but we cannot confidently rule out that these sources might be at 01a's redshift and undergoing an interaction. Source A3 was also masked because it is a distinct galaxy, although likely interacting with 01a. To fully capture the properties of 01a, it is important to establish whether A1 and A2 are part of it or distinct galaxies along the line of sight or interacting with it, although this information would not change the results reported in this work. It is nonetheless important to estimate a potential contribution to the submm and CO emission that we might have wrongly assigned to 01a from these distinct sources. The bestfit SED models and the recently obtained ALMA data rule out any significant contribution to 01a' submm and CO emission from these two sources.

\begin{table}
\caption{\label{tab:fgd_srcs}Masked sources in 01a.}
\centering 
\setlength{\tabcolsep}{2.0pt}
\begin{tabular}{l cc c c c} 
\hline\hline
  Source    & $\alpha$    & $\delta$   & Radius\tablefootmark{a} & $z_{\rm phot}$  & $\chi_{\nu}^2$ \\
  name      &   (h:m:s)     &  (\deg:\arcmin:\arcsec)     & Semi-axis    &                 &        \\
\hline
  A1  & 10:44:38.51 & 33:51:02.63 & 0.202            &  2.14$\pm$0.22  &  0.39  \\
  A2  & 10:44:38.54 & 33:51:02.63 & 0.183            &  1.94$\pm$0.18  &  0.09  \\
  A3  & 10:44:38.59 & 33:51:01.69 & 0.31$\times$0.55 &  2.50$\pm$0.21  &  0.19 \\
\hline                                                                                            
\end{tabular}\\
\tablefoot{
\tablefoottext{a}{Radius of circular aperture or major and minor semi-axis
of ellipse in arcsec centered on coordinates $\alpha$ and $\delta$.}
}

\end{table}                                                                                     

\section{Comparison between total and pixel-based parameters}

\begin{figure}[t]
\centering
\includegraphics[width=0.46\textwidth]{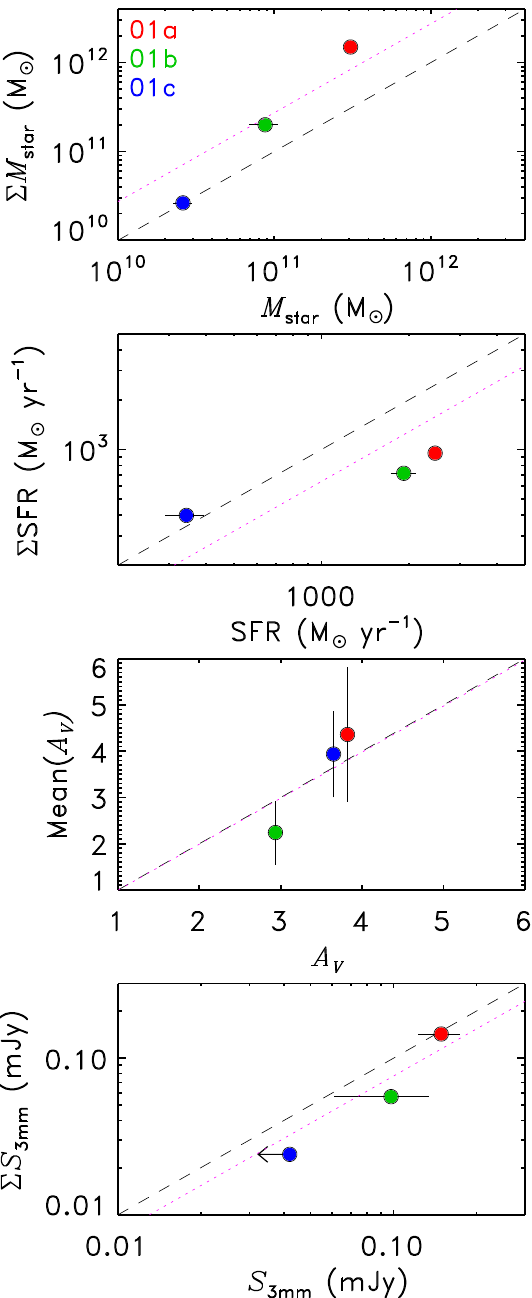}
\caption{\small Comparison between the sum of the resolved parameters and the parameters derived from fitting the integrated SEDs. Filled red, green, and blue points represent, respectively, sources 01a, 01b, and 01c. The panels show, from top to bottom, the stellar mass, the instantaneous SFR, the extinction, and the 3\,mm flux density. The parameter mean(\av) is the weighted mean of the pixel-derived values. The left-pointing arrows represent 5$\sigma$ upper limits (Table~\ref{tab:noema}). The dashed black line shows the 1:1 relation, and the dotted magenta line the average ratio for the three galaxies (2.7, 0.6, 1.0, and 0.8 from top to bottom). The resolved sums include only pixels with a value above 1--2$\sigma$ as was done in building the resolved maps shown in Fig.~\ref{fig:resolved_maps}.}
\label{fig:resolved_vs_whole} 
\end{figure}

Physical parameters derived from resolved SED studies do not necessarily recover the values obtained from integrated SED fitting \citep[e.g.,][]{sorba18}.  This is a concern here because we could use only the eight NIRCam data points in the resolved SED, while we included constraints from the submm, mm, and radio data when fitting the integrated SED\null. To limit this potential problem, we adopted the same components and model parameters, with the exception of the radio component, in fitting both the resolved and the integrated SEDs. Fig.~\ref{fig:resolved_vs_whole} compares the main parameters analyzed through resolved SED fitting obtained from the two approaches. The resolved SED fitting yields stellar masses that are, on average, a factor of 2.7 (0.4 dex) higher than those derived from the integrated SED. Such a bias had been previously reported and estimated to increase with the specific SFRs \citep[sSFRs;][]{sorba18}.
Because of the large sSFRs of our sources, they might thus suffer by a large bias in the global stellar mass estimates. Interestingly, the largest discrepancy is observed in 01a, the source with the highest sSFR of the three, whereas the agreement between the two mass estimates is excellent in 01c that has a much lower sSFR. This comparison suggests that the global mass estimates of 01a and 01b might need to be multiplied by a factor of 4.9 and 2.3, respectively. These corrections would not change the overall results obtained for these sources.

Contrary to the stellar mass, the sum of the SFRs from the resolved SED fitting is on average 0.6 times (0.2 dex) lower than the SFR from the unresolved SED. The two estimates agree for 01c, and similar offsets are measured for 01a and 01b. As previously stated, SFRs derived from the resolved SED fitting are highly uncertain due to the lack of constraints at submm/mm/radio wavelengths that are instead available in the unresolved SED. The resolved maps of SFR are thus to be considered with caution, but it should be safe to analyze the SFR distribution across each galaxy assuming that the offset between resolved and unresolved SFRs in each galaxy is systematic.

As a final check we also compared the extinctions and the mm flux densities from the two approaches. These quantities are both consistent implying that the distribution of extinction in the resolved maps can be safely analyzed. 

\section{Morphological analysis}\label{app:morph_analysis}

\begin{figure*}[h!]
\centering
\includegraphics[width=0.75\textwidth]{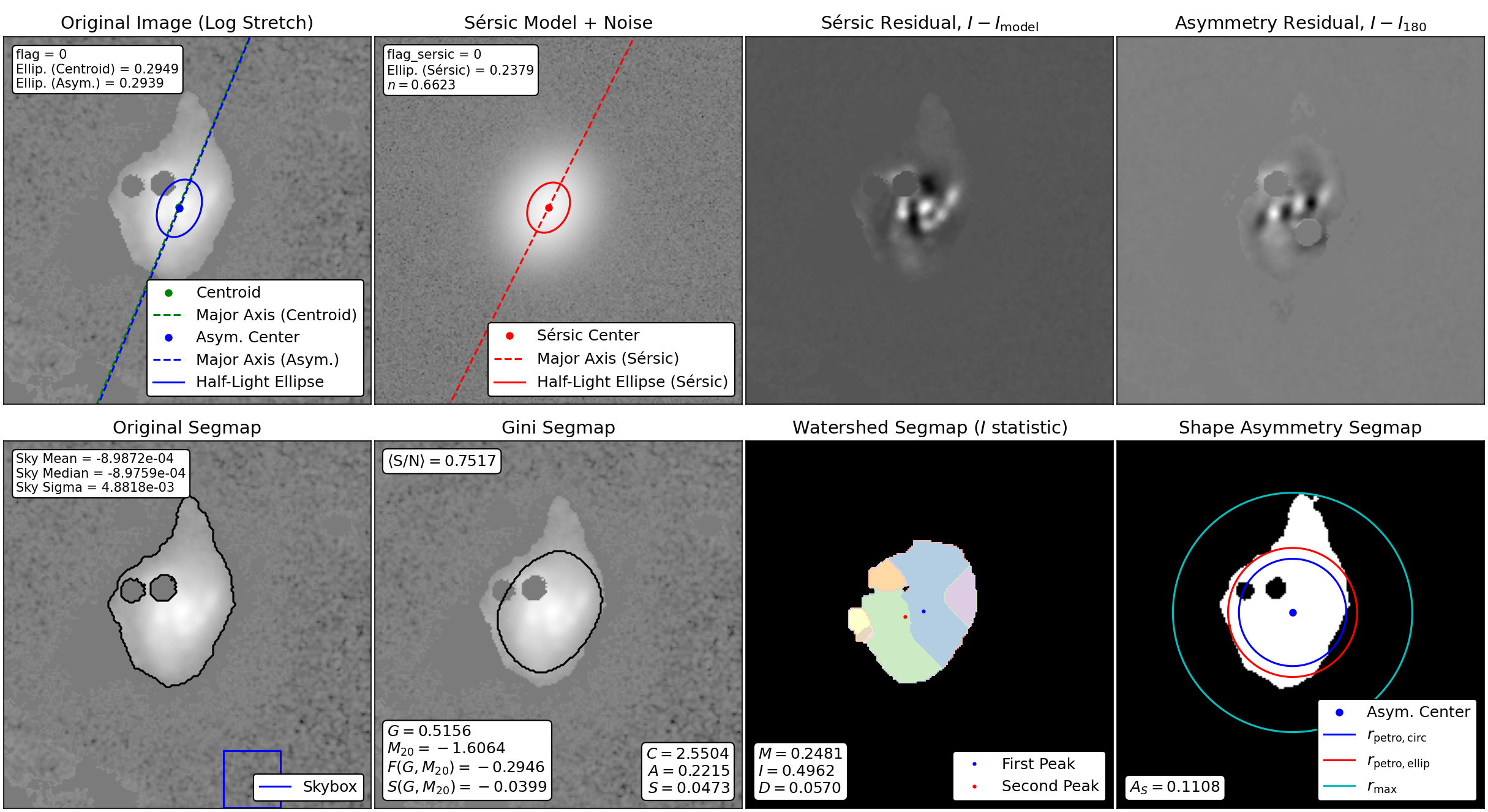}
\includegraphics[width=0.75\textwidth]{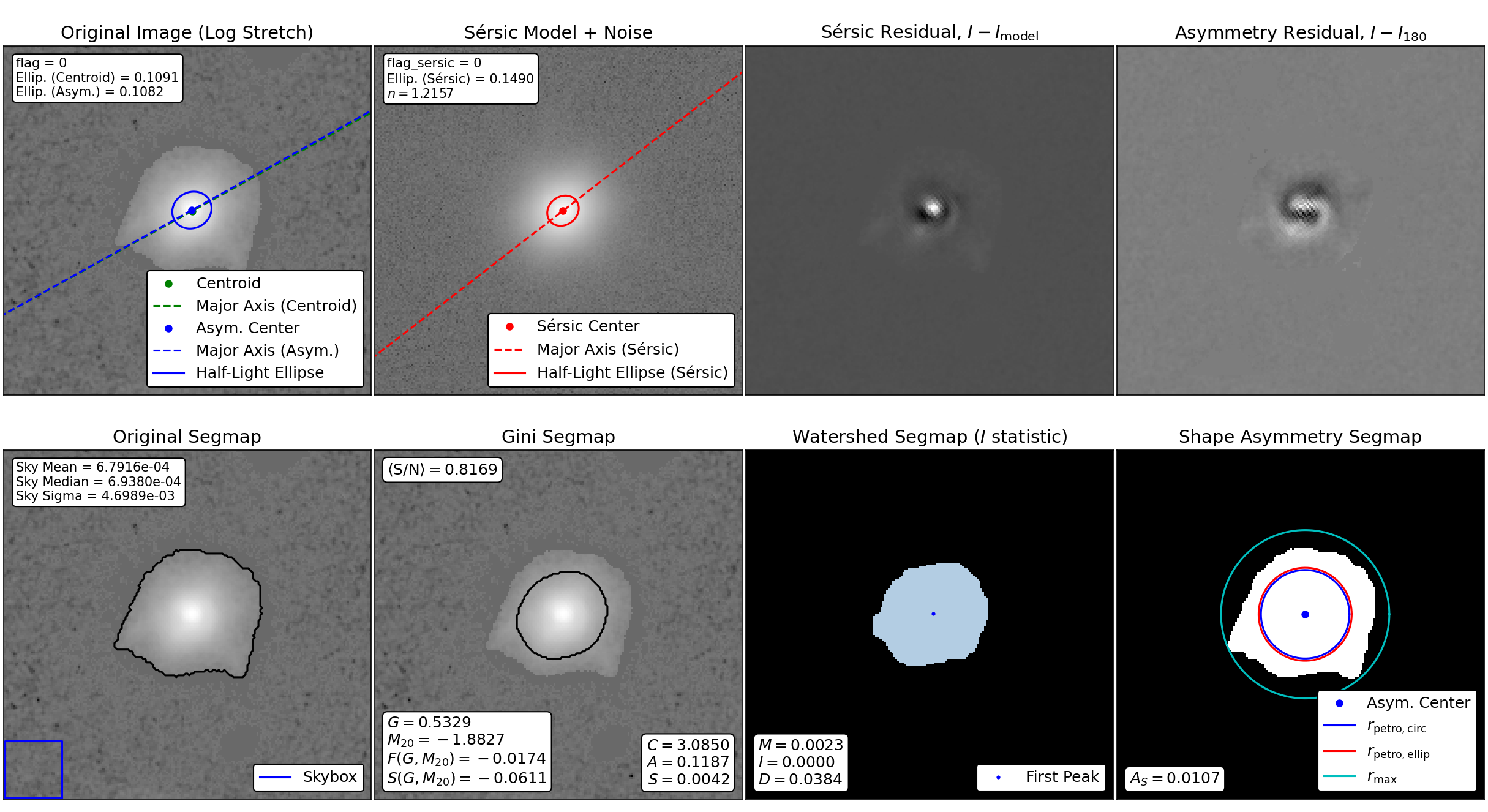}
\includegraphics[width=0.75\textwidth]{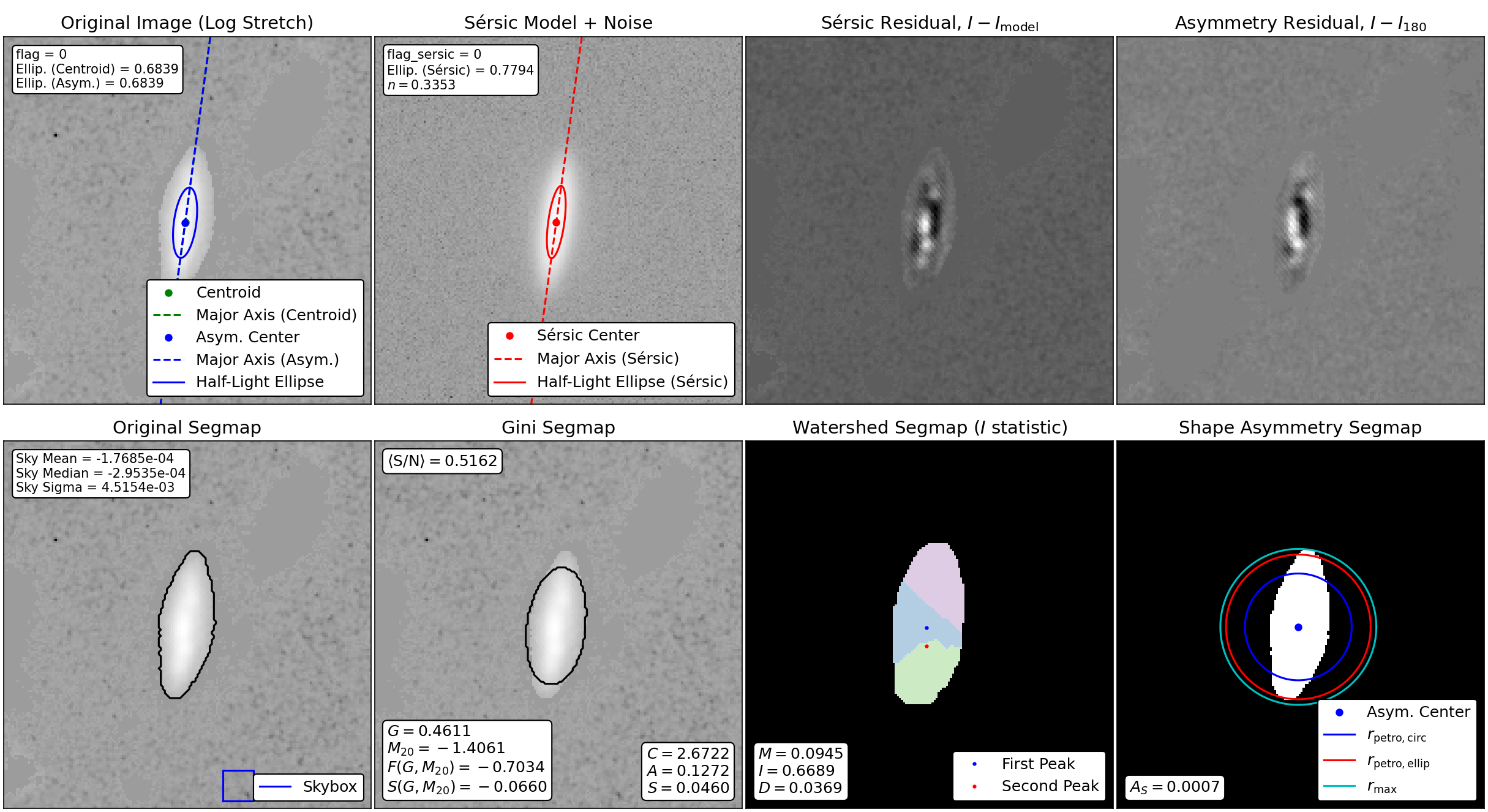}
\caption{\small Input and output F444W images from \texttt{statmorph}. Pairs of rows from top to bottom show the three targets 01a, 01b, and 01c. Images within each set of six are labeled, and the 01a images show the areas masked because of sources A1 and A2. All images are 6\arcsec\ on a side.}
\label{fig:statmorph} 
\end{figure*}

\begin{figure*}[h!]
\centering
\includegraphics[width=\textwidth]{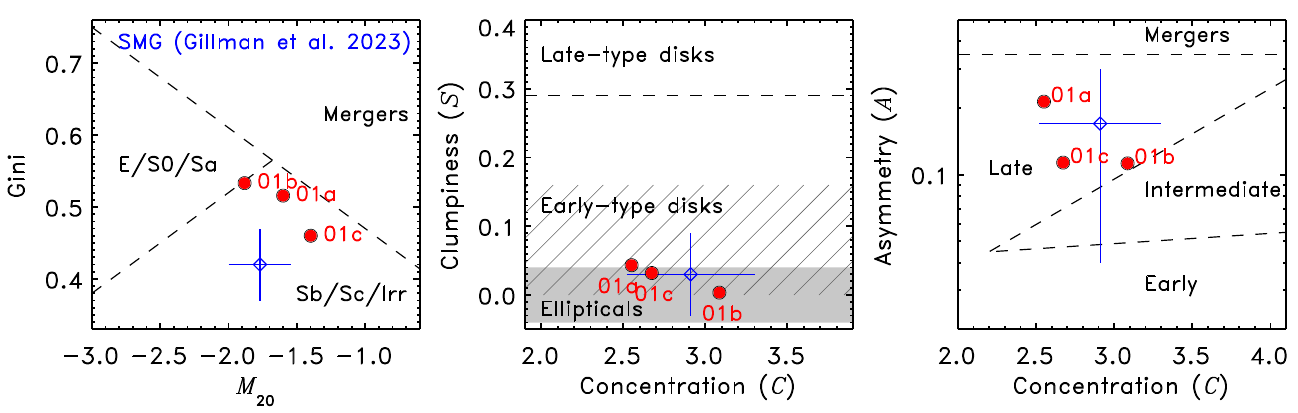}
\caption{\small Morphological parameters of the three CO emitters (filled red circles). Dashed lines and shaded regions show the division between different morphological types as defined by \citet{lotz08b} in the Gini-$M_{\rm 20}$ plane (left panel), by \citet{conselice03} in the $S-C$ plane (middle panel), and by \citet{bershady00} in the $A-C$ plane (right panel). The blue diamond represents the median value measured on  \jwst\ NIR images for a sample of SMGs \citep{gillman23} with error bars indicating the sample standard deviation of each parameter.}
\label{fig:morph_class} 
\end{figure*}

Fig.~\ref{fig:statmorph}  shows the input masked F444W images and the output
images obtained with \texttt{statmorph} along with key morphological
parameters. Table~\ref{tab:statmorph_params} gives the full list of derived morphological parameters.
Fig.~\ref{fig:morph_class} compares the \texttt{statmorph} parameters with the classification regions defined by \citet{lotz08b,conselice03,bershady00}. 
This morphological classification is based on nearby galaxies and on images at 5500\AA\ in the rest frame, and therefore it might not work as well for high-$z$ sources because of the different rest-frame wavelength (i.e., $\sim$1.3\,$\mu$m), spatial sampling and surface brightness limits. Some works find that galaxies become more compact at increasing redshifts, and accordingly the average concentration $C$ increases from 3.0 to 3.7 and the asymmetry $A$ from 0.22 to 0.30 from $z\simeq 1.25$ to $z\simeq 2.75$ \citep{whitney21,tohill21}, regardless of stellar mass.  The separation between disks and spheroids in the $A$--$C$ plane becomes less evident with increasing redshift and disappears for $z\gtrsim 2$, which corresponds to our sources. Other studies \citep{ren24}, using a large sample of simulated galaxies at $0.5<z<3$, find no significant variations in the morphological indicators for SFGs when considering rest-frame wavelengths $>1.1\mu$m. Since we carry out this analysis at rest-frame $\lambda\sim1.3\mu$m, the standard criteria can be safely applied.

\begin{figure*}[h!]
\centering
\includegraphics[width=\textwidth]{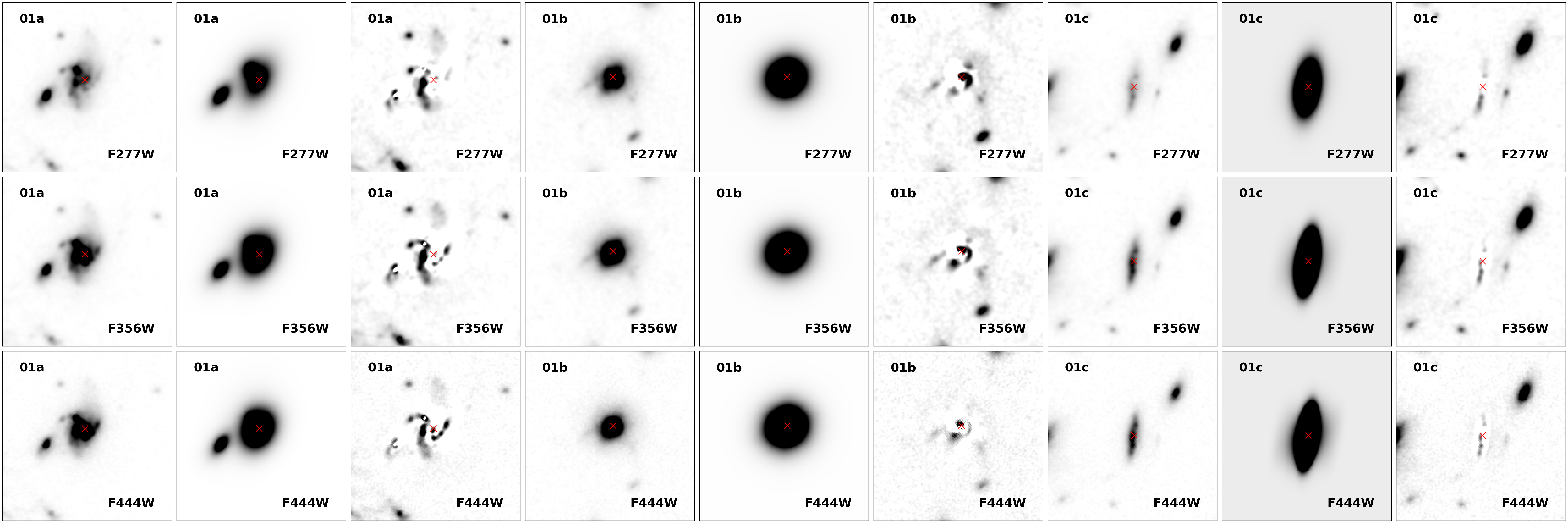}
\caption{\small \textsf{Galfit} input image (6\arcsec$\times$6\arcsec),
S\'ersic model, and residuals in three bands, F277W (top), F356W (central),
and F444W (bottom) and for sources 01a (left), 01b (middle), and 01c
(right). The source center is marked by a red cross.}
\label{fig:galfit} 
\end{figure*}

\textsf{Galfit} was run on all seven wide NIRCam bands, but only the LW images, F277W, F356W, and F444W, gave valid results for all three sources.  For 01a, we modeled simultaneously the bright nearby clump A1, and the nearby galaxy A3 (App.~\ref{app:01a_fgd_srcs}).  Fig.~\ref{fig:galfit}, shows the results.  For 01a, the residuals show several clumps that seem to be aligned along three spiral arms,  and some diffuse emission to the north.  The residual map of 01b shows emission in the center, an arc or spiral arm to the west, a clump to the southeast, and a straight feature, probably another galaxy, to the east. Source 01c contains four clumps along the galaxy major axis.  A faint feature, similar to a tidal tail, seems to connect the southern edge of the galaxy with a nearby, faint source to the southeast.

\end{appendix}

\end{document}